\documentclass[prx,aps,twocolumn,reprint,amsmath,floatfix,amssymb,showpacs,superscriptaddress]{revtex4-1}
\usepackage{xcolor}
\usepackage{subfigure}
\usepackage{pict2e}
\usepackage{multirow}
\newcommand{\be}{\begin{equation}}
\newcommand{\ee}{\end{equation}}
\newcommand{\bea}{\begin{eqnarray}}
\newcommand{\eea}{\end{eqnarray}}
\newcommand{\sbe}{\small\begin{equation}}
\newcommand{\see}{\end{equation}\normalsize}
\newcommand{\sbea}{\small\begin{eqnarray}}
\newcommand{\seea}{\end{eqnarray}\normalsize}

\usepackage{graphicx}
\usepackage{dcolumn}
\usepackage{bm}
\usepackage{color}

\usepackage{epstopdf}

\usepackage{amsmath, amssymb, graphics, setspace}

\newcommand{\mathsym}[1]{{}}
\newcommand{\unicode}[1]{{}}

\usepackage{titlesec}
\usepackage{bm}
\usepackage{comment}
\usepackage{verbatim}

\usepackage{graphicx}
\usepackage{subfigure}
\usepackage{tabularx}

\usepackage{color}
\usepackage[colorlinks,bookmarks=false,citecolor=blue,linkcolor=red,urlcolor=blue]{hyperref}

\definecolor{darkred}{rgb}{0.7,0.0,0.0}

\definecolor{darkblue}{rgb}{0,0.02,0.45}

\definecolor{darkgreen}{rgb}{0.02,0.45,0.0}

\definecolor{violet}{rgb}{0.8,0.2,0.6}

\def\be{\begin{equation}}
\def\ee{\end{equation}}
\def\bea{\begin{eqnarray}}
\def\eea{\end{eqnarray}}

\def\vec{\mathbf}

\def\mc{\mathcal}

\usepackage{times}

\begin{document}

\preprint{APS/123-QED}

\title{Magnetic Excitations and Interactions in the Kitaev Hyperhoneycomb Iridate $\beta$-Li$_2$IrO$_3$ }
\author{Thomas Halloran}
\affiliation{Institute for Quantum Matter and Department of Physics and Astronomy, Johns Hopkins University, Baltimore MD 21218, USA}
\author{Yishu Wang}
\affiliation{Institute for Quantum Matter and Department of Physics and Astronomy, Johns Hopkins University, Baltimore MD 21218, USA}
\affiliation{NIST Center for Neutron Research, Gaithersburg, Maryland\ 20899, USA}
\author{Mengqun Li}
\affiliation{School of Physics and Astronomy, University of Minnesota, Minneapolis, Minnesota 55455, USA}
\author{Ioannis Rousochatzakis}
\affiliation{Department of Physics, Loughborough University, Loughborough LE11 3TU, United Kingdom}
\author{Prashant Chauhan}
\affiliation{Institute for Quantum Matter and Department of Physics and Astronomy, Johns Hopkins University, Baltimore MD 21218, USA}
\author{M.B. Stone}
\affiliation{Neutron Scattering Division, Oak\ Ridge\ National\ Laboratory,\ Oak\ Ridge,\ Tennessee\ 37831, USA}
\author{Tomohiro Takayama}
\affiliation{Max Planck Institute for Solid State Research, Heisenbergstrasse 1, 70569 Stuttgart, Germany}
\author{Hidenori Takagi}
\affiliation{Max Planck Institute for Solid State Research, Heisenbergstrasse 1, 70569 Stuttgart, Germany}
\affiliation{Department of Physics, The University of Tokyo, Bunkyo, Japan}
\author{N. P. Armitage}
\affiliation{Institute for Quantum Matter and Department of Physics and Astronomy, Johns Hopkins University, Baltimore MD 21218, USA}
\author{Natalia B. Perkins}
\affiliation{School of Physics and Astronomy, University of Minnesota, Minneapolis, Minnesota 55455, USA}
\author{Collin Broholm}
\affiliation{Institute for Quantum Matter and Department of Physics and Astronomy, Johns Hopkins University, Baltimore MD 21218, USA}
\affiliation{Department of Materials Science and Engineering,
The\ Johns\ Hopkins\ University, Baltimore, Maryland\ 21218, USA}
\affiliation{NIST Center for Neutron Research, Gaithersburg, Maryland\ 20899, USA}

\date{\today}

\begin{abstract}

We present a thorough experimental study of the three-dimensional hyperhoneycomb Kitaev magnet $\beta$-Li$_2$IrO$_3$, using a combination of inelastic neutron scattering (INS), time-domain THz spectroscopy, and heat capacity measurements. The main results include a massive low-temperature reorganization of the INS spectral weight that evolves into a broad peak centered around 12 meV, and a distinctive peak in the THz data at 2.8(1) meV. A detailed comparison to powder-averaged spin-wave theory calculations reveals that the positions of these two features are controlled by the anisotropic $\Gamma$ coupling and the Heisenberg exchange $J$, respectively. The refined microscopic spin model places $\beta$-Li$_2$IrO$_3$ in  close proximity to the Kitaev spin liquid phase.
\end{abstract}

\maketitle


\section{Introduction}

In contrast to 3$d$ transition metal oxides with half filled orbital levels, where spins interact mainly through the isotropic Heisenberg interactions, strong spin-orbit coupling (SOC) in heavier 4$d$ and 5$d$ systems introduces anisotropic exchange interactions that may give rise to exotic forms of magnetism \cite{Winter2017ModelsMagnetism,Rau2016Spin-OrbitMaterials}. One prominent example is the bond-dependent Ising interaction between $S=1/2$ spins on a honeycomb lattice, which forms an exactly solvable quantum spin liquid known as the Kitaev spin liquid (KSL) \cite{Kitaev2006AnyonsBeyond}. A honeycomb lattice of spin-orbital $J_{\rm eff}=1/2$ degrees of freedom, formed by Ir$^{4+}$ and Ru$^{3+}$ coordinated by edge-sharing octahedra was shown to have the potential to realize this important model~\cite{Jackeli2009MottModels}. Experimental exploration of such materials has revealed spin-liquid-like features in antiferromagnetically ordered $\alpha-$RuCl$_3$ \cite{Kasahara2018Unusual-RuCl3,Banerjee2018Excitations-RuCl3} and also in H$_3$LiIr$_2$O$_6$ - albeit with inter-layer disorder - which shows no conventional magnetic order \cite{Kitagawa2018,Pei2020MagneticO6,Geirhos2020QuantumO6}. The demonstrative features of a KSL state however, remain elusive and the search for more ideal compounds to realize the KSL continues.

As part of this effort there is a need for a better understanding of factors that influence the strength of the various microscopic interactions
in insulating magnetic materials. Thus, experiments in magnetically ordered systems where magnetic interactions can be accurately determined by measuring and analyzing spin wave excitations can provide an important experimental reference point. Magnetic ordering has been observed in most KSL candidates, either due to the presence of inter-layer interactions or non-Kitaev interactions that destabilize the KSL. Particularly interesting is a family of such materials with  the chemical formula Li$_2$IrO$_3$. Ir$^{4+}$ ions in the $\alpha-$Li$_2$IrO$_3$ form 2D honeycomb lattices, in $\beta$-Li$_2$IrO$_3$ form a 3D hyperhoneycomb lattice~\cite{Takayama2015HyperhoneycombMagnetism,Majumder2019AnisotropicStudy,Ducatman2018Magnetic-Li2IrO3}, while in $\gamma$-Li$_2$IrO$_3$ form a stripy-hyperhoneycomb lattice~\cite{Modic2014RealizationIridate,Biffin2014Noncoplanar-li2iro3}. All members of the family share similar coordination and connectivity for the $J_{\rm eff}=1/2$ Ir$^{4+}$ ions. 

In this work we are interested in $\beta$-Li$_2$IrO$_3$, which develops long range magnetic order at $T_{\rm N}=38$~ K~\cite{Majumder2019AnisotropicStudy,Biffin2014Noncoplanar-li2iro3} with counter rotating spins with an incommensurate propagation wave vector ${\bf Q}=(0.57,0,0)$ indexed in the orthorhombic reciprocal lattice~\cite{Biffin2014a,Takayama2015HyperhoneycombMagnetism,Ruiz2017CorrelatedFields}. The application of a magnetic field along the $b-$axis can suppress this order with a critical field of $\mu_0H^\ast=2.8$~T~\cite{Ruiz2017CorrelatedFields}. The minimal spin Hamiltonian for the Li$_2$IrO$_3$ compounds has been suggested to contain Kitaev ($K$), Heisenberg ($J$), and off-diagonal anisotropic ($\Gamma$) exchange terms~\cite{ Tsirlin2021,Winter2017,Kimchi2015UnifiedLi2IrO3,Kim2016RevealingLiquid,Singh2012Relevance3,Lee2015TheoryIridates,Lee2016TwoMaterials,Katukuri2016TheState,Ducatman2018Magnetic-Li2IrO3,Rousochatzakis2018Magnetic-Li2IrO3}. In this $J$-$K$-$\Gamma$ model, the zero-field spin structure for $\beta$-Li$_2$IrO$_3$ is stabilized  in the regime where $|J|\leq K/3$, K$<$0, and $\Gamma<$0~ \cite{Rau2014GenericLimit}. Furthermore, it has been found that $H^\ast$ depends only on $J$, and that the value of $\mu_0H^\ast$ corresponds to a small value of $J\sim0.35$~meV
~\cite{Rousochatzakis2018Magnetic-Li2IrO3}. Still, the ground state order suggests a significant $\Gamma$ interaction.  The field-induced state for $H>H^*$ has been suggested to be a ``quantum correlated paramagnet" instead of a fully polarized state~\cite{Ruiz2017CorrelatedFields,Majumder2018Breakdown-Li2IrO3,Ruiz2017CorrelatedFields,Ducatman2018Magnetic-Li2IrO3,Rousochatzakis2018Magnetic-Li2IrO3}, with similarities to $\alpha$-RuCl$_3$~ \cite{Yadav2016Kitaev-RuCl3,Banerjee2018Excitations-RuCl3}. The fragility of the zero-field ground state order when introduced to a magnetic field as well as the low Néel temperature indicated relatively close proximity to the realization of the Kitaev model in three dimensions. This was recently supported by Resonant Inelastic X-ray (RIXS) studies \cite{Ruiz2021Magnon-spinon-Li2IrO3}, where signatures of long-lived fractionalized excitations separating the low-T ordered phase from the high-T paramagnetic regime were observed. 

The goal of this study is to investigate the magnetic excitations and refine the spin Hamiltonian of $\beta$-Li$_2$IrO$_3$ using a combination of inelastic neutron scattering , time domain THz spectroscopy, and heat capacity measurements on a powder sample. A close comparison is made to spin-wave theory calculations to connect the microscopic exchange parameters to the experiments. 

The main results can be summarised as follows: 
i) The specific heat and powder INS data are consistent with long-range magnetic ordering below $T_N=38.5(5)$ K, with a strong momentum dependence and distinct intensity around the incommensurate wavevector ${\bf Q}=0.57(1)~{\bf a}^\ast$. 
ii) The integrated INS intensity reveals a spin gap of $\Delta=2.1(1)$ meV in the excitation spectrum at the incommensurate wavevector, consistent with the anisotropic nature of this magnet. 
iii) At $T<T_N$, the powder INS spectra show a massive spectral weight transfer into a broad peak centered around 12 meV, fully consistent with powder-averaged, dynamical spin structure factor calculations based on linear spin-wave theory. The position of this peak depends very strongly on the value of $\Gamma$, providing a strong constraint on this parameter.
iv) The THz data reveal a distinctive, $Q=0$ peak at 2.8(1) meV that develops gradually below $T_N$, again in full agreement with spin wave theory. The position of this peak depends strongly on $J$, again providing a strong constraint on this parameter.
v) The best fit of the data to linear spin wave theory yields $\Gamma=-9.3(1)$ meV and  $K=-24(3)$ meV. The Heisenberg term $J$, which controls the gap at the magnetic Brillouin zone center, is further refined to $J=0.40(2)$~meV based on time-domain THz spectroscopy and heat capacity measurements. 
This set of exchange parameters refines previous estimates from various experimental and theoretical works (see Table~\ref{tab:literature_table} below) and places $\beta-$Li$_2$IrO$_3$ in the regime of dominant Kitaev interactions. Based on our results, the system is in close proximity to the ideal KSL point.

\section{Experimental methods}

\subsection{Materials synthesis}
The $\beta-$Li$_2$IrO$_3$ powder investigated in this work was synthesized by a solid-state reaction using powders of Li$_2$CO$_3$ and metallic Ir. The reagents were 99\% enriched $^7$Li and $^{193}$Ir to reduce the neutron absorption cross section and the incoherent scattering cross section \cite{Takayama2019Pressure-induced-Li2IrO3}. The obtained powder was confirmed to be a single phase of $\beta-$Li$_2$IrO$_3$ by powder x-ray diffraction.
\subsection{Neutron scattering}
Inelastic neutron scattering directly measures the dynamic spin structure factor in momentum-energy space. Specifically, the  normalized scattering intensity is given by
\begin{eqnarray}\label{eq:INS}
    \mc{I}({\bf Q},\omega)&=&r_0^2 |\frac{g}{2}F({\bf Q})|^2e^{-2W({\bf Q})}\nonumber\\
    &&\times \sum_{\alpha,\beta}(\delta_{\alpha\beta}-\hat{Q}_\alpha\hat{Q}_\beta)\mc{S}^{\alpha\beta}({\bf Q},\omega),
\end{eqnarray}
Here $r_0=5.3906$~fm, $g$ is the gyromagnetic ratio for the magnetic ion, $F({\bf Q})$ is the magnetic form factor, $2W({\bf Q})=\langle ({\bf Q \cdot u})^2\rangle $ is the Debye-Waller factor associated with displacements, $\bf u$ of the magnetic ion. The dynamic correlation function $\mc{S}^{\alpha\beta}({\bf Q},\omega)$ is given by 
\begin{align}
\label{eq:Sqw}
\mc{S}^{\alpha\beta}(\vec{Q},\omega)&=\frac{1}{2\pi\hbar}\!\!\int\!\! \text{d}t \, e^{-i\omega t}\langle S^\alpha(-\vec{Q},0)S^\beta(\vec{Q},t)\rangle
\end{align}
where $\hbar\vec{Q}$ is momentum transfer and $\hbar\omega$ is energy transfer. 
\sbe\label{eq:SofQ}
{ S}^\alpha({\bf Q},t) =\frac{1}{\mc{N}} \sum_{{\bf r}} e^{-i {\bf Q}\cdot{\bf r}} S^\alpha_{\bf r}(t)\,,
\see
is the Fourier transform of the spin operator at time $t$, with $\mc{N}$ being the total number of spins and ${\bf r}$ denoting the physical positions of the spins, with superscripts $\alpha$ and $\beta$ labelling the orthorhombic axes $a$, $b$, and $c$.

For a powder sample with an isotropic grain orientation distribution, neutron scattering probes a spherical average of $\mc{I}({\bf Q},\omega)$ in momentum space: 
\begin{align}
\mc{I}(Q,\omega)=\int\frac{d\Omega_{\bf Q}}{4\pi}\mc{I}({\bf Q},\omega).
\end{align}

This is the physical quantity presented in Figures \ref{fig:lio_main} and \ref{fig:lio_main_cuts}. Probing magnetic excitations in a powder sample of $\beta$-Li$_2$IrO$_3$ by INS presents three significant challenges. Stable isotopes of iridium all have significant neutron absorption. To minimize absorption we utilized the least absorbing  $^{193}$Ir isotope where $\sigma_a=111(5)$~barn for 25 meV neutrons. Secondly, for a powder sample, scattering at the important high symmetry $\Gamma$ point is accessible, but only near $Q=0$ in the first Brillouin zone. Beyond $Q=0$ the spherical averaging takes effect making it impossible to distinguish the scattering between points in the Brillouin zone of equivalent $Q$. The highest energy transfer that can be accessed for a given wave vector transfer $Q$ is $\hbar\omega_{max}=\bar{v}\hbar Q$ where $\bar{v}=(\hbar/2m)(k_i+k_f)$ is the average neutron velocity.  (Fig. \ref{fig:lio_main} and Appendix \ref{appendix:multi_ei}). Thirdly, the magnetic form factor for Ir$^{4+}$ decreases sharply for increasing momentum transfer $Q$, leading to a reduction in the magnetic scattering cross section by 50\% for $Q$=1.7 $\AA^{-1}$ and a 90\% reduction for $Q$=3.0 $\AA^{-1}$.

The neutron scattering experiment was conducted on the SEQUOIA spectrometer \cite{Granroth2010SEQUOIA:SNS} at the ORNL spallation neutron source. The powder was loaded under 1 atm $^4$He at room temperature in an annular aluminum can with an outer diameter of 20 mm and annulus thickness 0.5 mm. The height of the corresponding annular powder sample was 34 mm. The total mass was 3.8 g and the packing density was 3.6 g/cm$^3$ which is 50\% of the nominal density. The can was attached to the cold finger of a low-background closed cycle refrigeration cryostat. 

Data were acquired with fixed incident energy $E_{\rm i}$ = 18 meV, 30 meV, and 60 meV for each of the temperatures $T=4.0(1)$~K, $45.0(1)$~K, and $200.0(1)$~K. In addition,  $E_{\rm i}$=120 meV data were taken at $T=4.0(1)$~K and $300.0(1)$~K. For $E_{\rm i}$=18 meV and 30 meV we used the fine chopper configuration, while for $E_{\rm i}$=60 meV and 120 meV the high flux chopper was used. Energy and momentum transfer dependent absorption corrections determined by a Monte-Carlo method were applied to the data \cite{Schmitt1998AbsorptionDiffraction}. A temperature independent background was remove from each measurement using the principle of detailed balance, the details of which may be found in \ref{appendix:DB}. The one-phonon scattering was estimated from high temperature measurements where it dominates  and then subtracted from the lower temperature data. Elastic incoherent scattering from a vanadium standard sample was used to normalize count rates to absolute units in accordance with Eq.~\ref{eq:INS}. To cover a broad range of momentum-energy space with the right compromise between resolution and count rate, we combined data acquired for the four different incident energies. See Appendix \ref{appendix:neutron_corrections} for further details about the neutron scattering data analysis. 

Additional measurements were acquired with fixed incident energy $E_{\rm i}$ = 22 meV in the high flux configuration to examine the temperature dependence of the magnetic Bragg Peaks along with the inelastic spectrum. These data were corrected for absorption,  normalized to vanadium, and binned in 2.5 K temperature steps.

\subsection{THz spectroscopy}
Time-domain THz spectroscopy was performed using a custom-built system with frequency range 0.2-2~THz \cite{Laurita2017LowMagnets} at zero magnetic field. The measurement was performed on a dry pressed powder pellet of $\beta-$Li$_2$IrO$_3$ with diameter 5 mm, thickness 0.6 mm and mass 5.0(1) mg. Transmission spectra were collected at temperatures from $T=3$~K to $40$~K. The $T=50$~K spectrum was used as an approximately nonmagnetic reference. 

The energy dependent complex THz transmission through a slab of material with thickness $d$ may be written as
\begin{equation}
    \Tilde{{\cal T}}(\omega) = \frac{4 n_s}{(1+n_s)^2} \exp{(i\frac{\omega d}{c} (n_s -1))}.
    \label{eq:ref_transmission}
\end{equation}
Here, $n_s=n - ik$ is the complex refractive index of the material. Note, that the real part of the refractive index $n$ indicates the phase velocity, while the imaginary part $k$ is called the absorption coefficient and measures the attenuation of the electromagnetic wave while propagating through the material.  We solve for $n_s$ numerically for all temperatures. The index of refraction at temperature $T$ is given by $n_T=\sqrt{\epsilon(1+\chi_M)}$, where $\epsilon$ is the generalized permitivity and $\chi_M$ is the magnetic susceptibility. For a sample that has a magnetic response below a reference temperature $T_{ref}$ and no magnetic response above $T_{ref}$, we derive an expression for magnetic susceptibility by taking the ratio of the refractive indices above and below the transition temperature. 
\begin{equation}
    \frac{n_T}{n_{T_{ref}}} = \sqrt{1+\chi_M},
    \label{eq:thz_n_ratio}
\end{equation}
from which we obtain 
\begin{equation}
    \chi_M(\omega) = (\frac{n_T}{n_{T_{ref}}})^2 -1
    \label{eq:THz_chim}
\end{equation}
See Ref. \cite{Laurita2017LowMagnets, Chauhan2020TunableChain} for full details of the derivation and use of this technique.

\subsection{Heat capacity}
Heat capacity measurements were performed in a Quantum Design  physical properties measurement system (PPMS). To enhance thermal conduction, we used a pellet pressed of equal parts by mass of silver and $\beta$-Li$_2$IrO$_3$. Measurements were taken at zero field for the range of $T=2-300$~K  and at $\mu_0H=14$~T for the range of $T=2-100$~K. The silver contribution to the specific heat was subtracted based on tabulated values in Refs. \cite{ErPH5K,MeadsThe300K,BldgmiLibraryLiterature}. Low temperature heat capacity measurements for $T=0.1-3.5$~K were performed using the PPMS dilution refrigerator option. This measurement was done on a 1.50(1) mg piece of a pressed pellet of pure $\beta$-Li$_2$IrO$_3$ with no silver.

\section{Experimental Results}
\label{resultsection}

\begin{figure}[t]
    \centering
    \includegraphics[width=1.0\columnwidth]{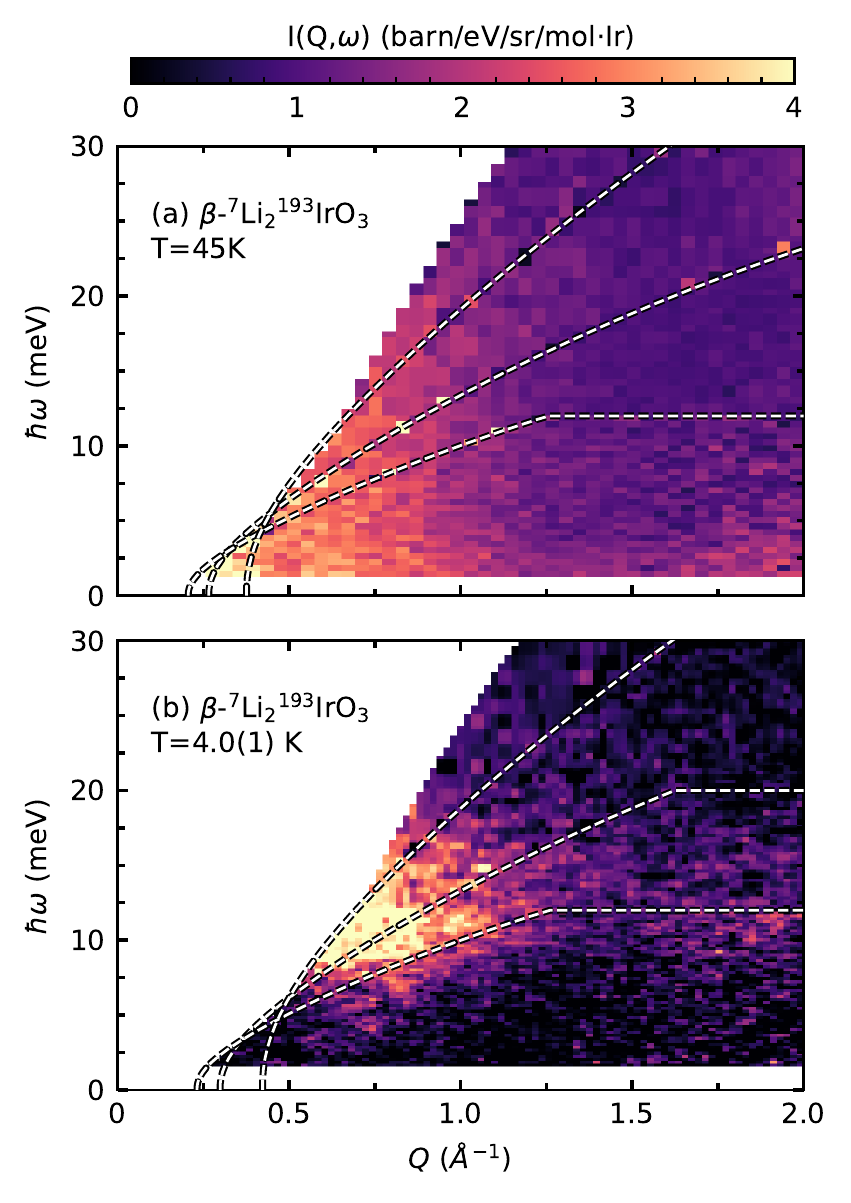}
    \caption{Magnetic excitation spectrum of $\beta$-$^7$Li$_2^{193}$IrO$_3$ probed by inelastic neutron scattering at (a) $T=45$~K  and (b) $T=4$~K. The data combines scattering from neutrons of incident energies 10.5 meV, 30 meV, 60 meV, and 120 meV. Data acquired at $T=200$~K and $300$~K were used to determine the temperature dependent one-phonon scattering, which was subtracted to isolate the  magnetic scattering (see Appendix \ref{appendix:neutron_corrections}). The dashed lines show the kinematic limit for each incident neutron energy employed.}
    \label{fig:lio_main}
\end{figure}
\begin{figure}[t]
    \centering
    \includegraphics[width=1.0\columnwidth]{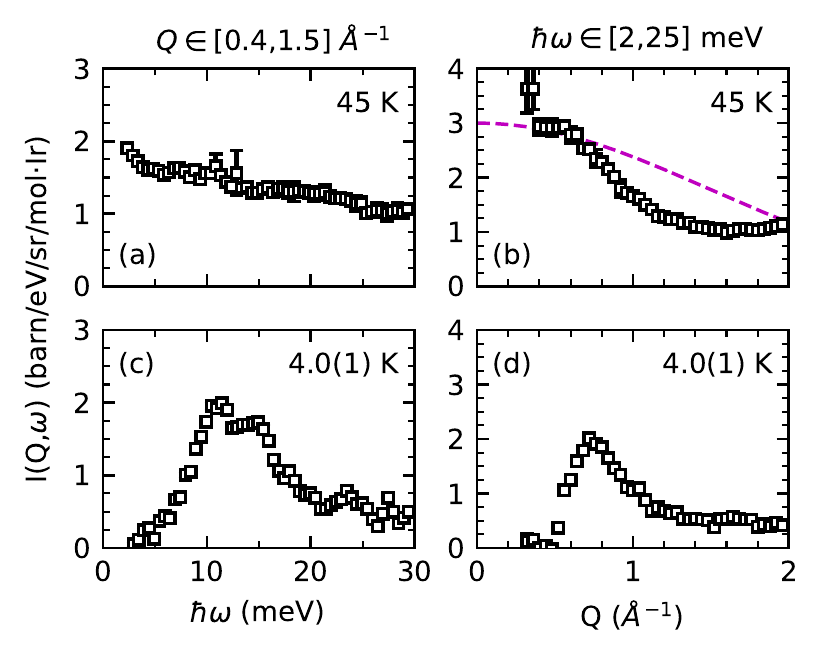}
    \caption{Cuts across experimental data shown in Fig.~\ref{fig:lio_main}. The spectra in (a) and (c) represent averages over $Q\in[0.4,1.5]$~\AA$^{-1}$ while the $Q-$dependent scattering intensity in (b) and (d) average over $\hbar\omega\in[2,25]$~meV. $Q-$averaging is weighted by $Q^2$ to represent the average of $I(Q,\omega)$ throughout a spherical shell of momentum space. The kinematic limits indicated in Fig.~\ref{fig:lio_main} impact these cuts as the averages can only be extended over kinematically accessible regimes of $Q$ and $\hbar\omega$. The dashed purple line in (b) shows the scaled magnetic form factor $|F(\textbf{Q})|^2$ for Ir$^{4+}$.}
    \label{fig:lio_main_cuts}
\end{figure}
\subsection{Magnetic neutron scattering}

Fig. \ref{fig:lio_main}(a,b) show color images of the inelastic magnetic neutron scattering from a powder sample of $\beta$-Li$_2$IrO$_3$ at T=45 K and T=4 K, which are above and below the magnetic ordering temperature respectively. Representative energy and momentum cuts through the same data are shown in Fig \ref{fig:lio_main_cuts}. In the paramagnetic phase at $T=45.0(1)$~K (Fig. \ref{fig:lio_main}(a) and Fig. \ref{fig:lio_main_cuts}(a,b)), the spectrum extends from the lowest accessible energy transfer of 2 meV to beyond the kinematic limit of the experiment (near 30 meV). 

The scattering cross section is further attenuated with increasing $Q$ than the squared iridium form factor (dashed line, Fig. \ref{fig:lio_main_cuts}(b)). This indicates short ranged inter-site spin correlations.  The spectrum of fluctuations (Fig. \ref{fig:lio_main_cuts}(a)) extends to energies well beyond $k_BT$, which is characteristic of frustrated magnetic materials where competing interactions do not favor a state with long range spin order. In particular the $T=0$ dynamic spin correlation function of the Kitaev quantum spin liquid is virtually Q-independent while the spin flip excitation spectrum is broad and featureless above a gap \cite{Knolle2014DynamicsFluxes,Smith2015NeutronLiquid}. 

\begin{figure}
    \centering
    \includegraphics[width=0.9\columnwidth]{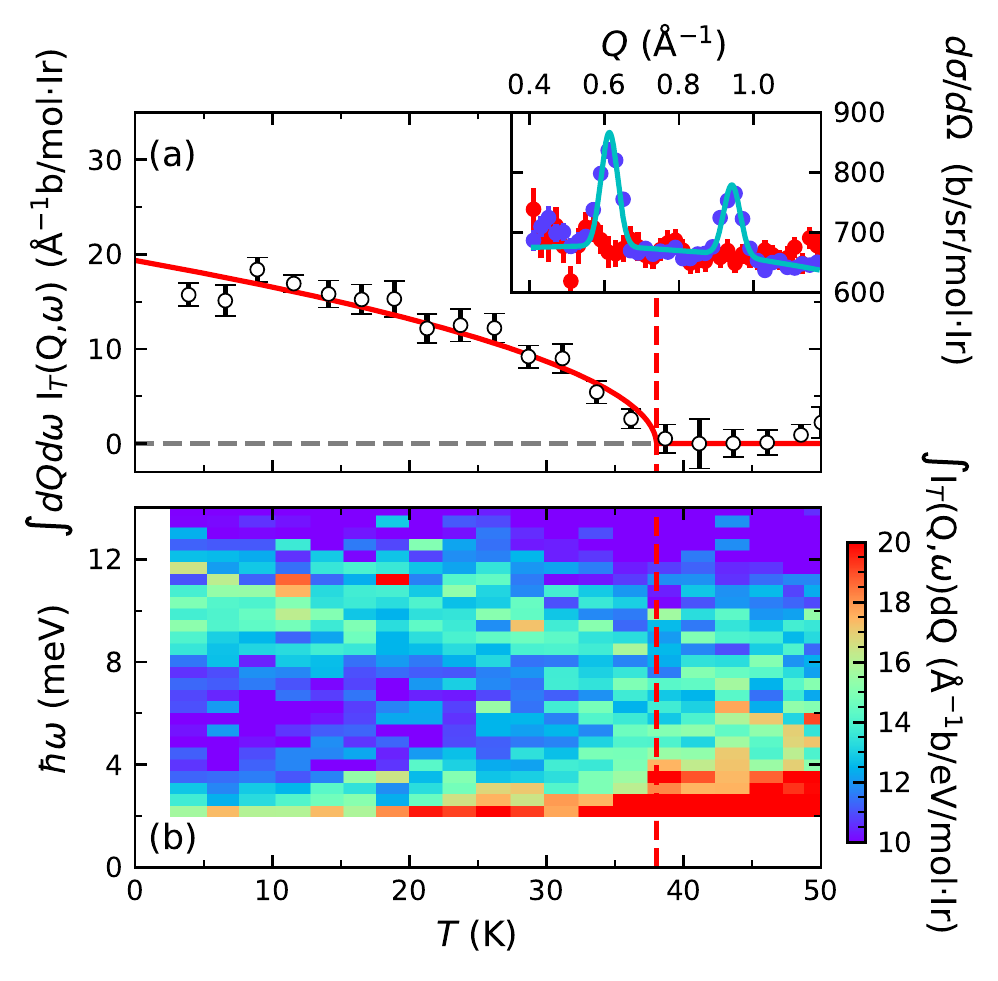}
    \caption{Temperature dependence of magnetic neutron scattering from $\beta$-$^7$Li$_2^{193}$IrO$_3$. (a) Temperature dependence of the $Q$-Integrated intensity of the magnetic Bragg peaks \textcolor{black}{(1,1,1)-$\textbf{k}$ and (0,0,0)$\pm\textbf{k}$} shown in the inset. The line through the data is that of an order parameter squared with critical exponent $\beta=0.42(6)$ and the critical temperature $T_N=38$~K (dashed red line) determined from the specific heat capacity data in Fig.~\ref{fig:lio_hc}(a). The inset depicts the integrated elastic scattering around the magnetic Bragg peaks at $T=3.95$~K (blue points) and $T=50.0$~K (red points). the cyan line depicts an example of the gaussian fits used to extract magnetic diffraction intensity which is a measure of the  staggered magnetization squared. (b) Temperature dependent inelastic magnetic neutron scattering integrated over $Q\in[0.5,1.0]$~\AA$^{-1}$. }
    \label{fig:order_param}
\end{figure}

A massive rearrangement of spectral weight occurs upon cooling to $T=4.0(1)$~K $\ll T_N$ (Fig. \ref{fig:lio_main}(b)).  Lower energy magnetic neutron scattering shifts to a $\hbar\omega\approx12$ meV intensity maximum. 
Strong momentum dependence develops in the low energy regime (Fig. \ref{fig:lio_main_cuts}) with a distinct intensity near the incommensurate magnetic wavevector $Q_m=0.57(1)a^*$ \cite{Williams2016}. There is an apparent gap $\Delta = 2.1(1)$ meV in the magnetic excitation spectrum (Fig. \ref{fig:lio_main_cuts}(c)). 

The detailed temperature dependence of magnetic neutron scattering from $\beta$-Li$_2$IrO$_3$ is in Fig.~\ref{fig:order_param}. Panel (a) shows the development of elastic magnetic Bragg peaks as an order parameter (squared). The Néel  temperature is consistent with a peak in the magnetic specific heat plotted as $\Delta C(T)/T$ versus $T$ in the inset to Fig.~\ref{fig:lio_hc}(a).  Fixing the critical temperature to $T_N=38.5(5)$~K inferred from $\Delta C(T)/T$ to be discussed later, a fit to the temperature dependent Bragg intensity yields a rough estimate of the critical exponent $\beta=0.42(6)$. This value is consistent with the value for the 3D Ising model ($\beta_{\rm Ising}=0.326$) but also indistinguishable from the Heisenberg ($\beta_{\rm Heisenberg}=0.365$) and XY models ($\beta_{\rm XY}=0.345$). Fig. \ref{fig:order_param}(b) displays the temperature dependence of the magnetic excitation spectrum as a color image. The data illustrate depletion of low energy inelastic scattering as in an Ising-like phase transition and the transfer of  spectral weight into a broad peak centered at 10 meV (Fig. 2(c)). 
 
\begin{figure}[]
    \centering
    \includegraphics[width=1.0\columnwidth]{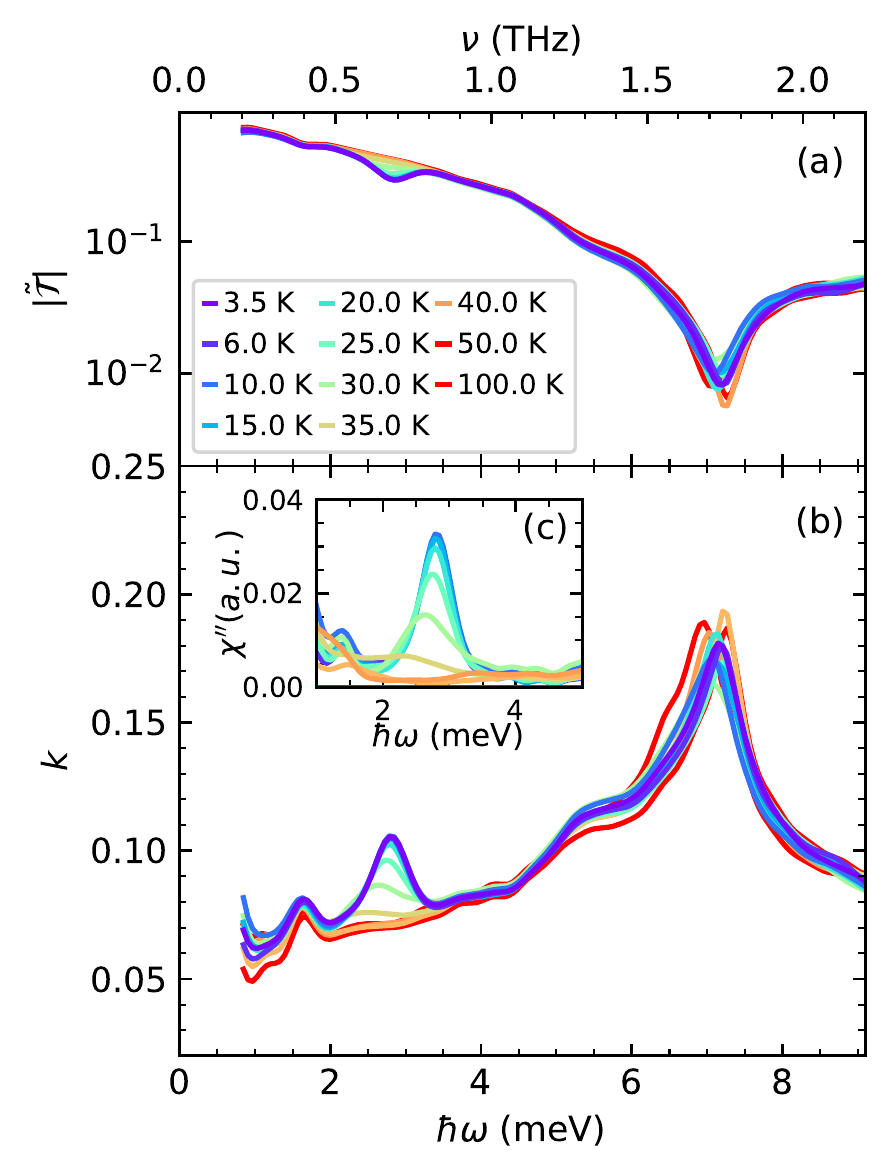}
    \caption{Time-domain THz spectroscopy of $\beta$-$^7$Li$_2^{193}$IrO$_3$. (a) Raw THz transmission data at temperatures from $3$~K to $40$~K. (b) Absorption coefficient $k$ inferred from the transmission data in (a). The sharply defined temperature dependent peak at 2.8(1) meV is associated with a magnetic excitation from the long range ordered state. Based on Eq.\ref{eq:THz_chim}, this peak is shown as $\chi^{\prime\prime}(\omega)$ in the inset. A second excitation is seen around 7.0(2) meV but its temperature dependence is quite different from the 2.8 meV peak and its origin is unclear. The quality of the higher energy data is impacted by the reduced transmission (see frame (a)). The temperature independent peak at 1.6 meV arises from an instrumental interference effect.}
    \label{fig:lio_thz}
\end{figure}
\subsection{THz spectroscopy}
\label{sec:TDTS}
The $Q=0$ spectrum is of particular interest as it reflects spin-space anisotropy, which is central to the Kitaev model. 
To measure this, we use THz spectroscopy. The THz transmission spectrum for $\beta-\rm Li_2IrO_3$ is presented in Fig. \ref{fig:lio_thz}(a), and the absorption coefficient $k$ is shown in Fig. \ref{fig:lio_thz}(b). Low THz transmission at higher energy transfers precluded reliable measurements of spin-wave excitations above 5 meV. The higher energy feature seen in THz transmission near 7 meV is not temperature dependent and thus we do not attribute it to magnons. Another feature at $\approx$1.6 meV is an instrumental interference effect that is not associated with the electromagnetic response of the sample. However, the peak at $2.8(1)$~meV has a Lorentzian shape and gradually forms at temperatures below $T_{\rm N}=38$~K. The corresponding $T-$ dependent imaginary part of the dynamic magnetic susceptibility $\chi^{\prime\prime}(\omega)$ at $Q=0$ is shown in Fig. \ref{fig:lio_thz}(c). We associate this peak with a zone center gap in  magnetic excitations from the ordered state. The finite gap in the excitation spectrum is direct evidence of anisotropic magnetic interactions.  


\begin{figure}
    \centering
    \includegraphics[width=1.0\columnwidth]{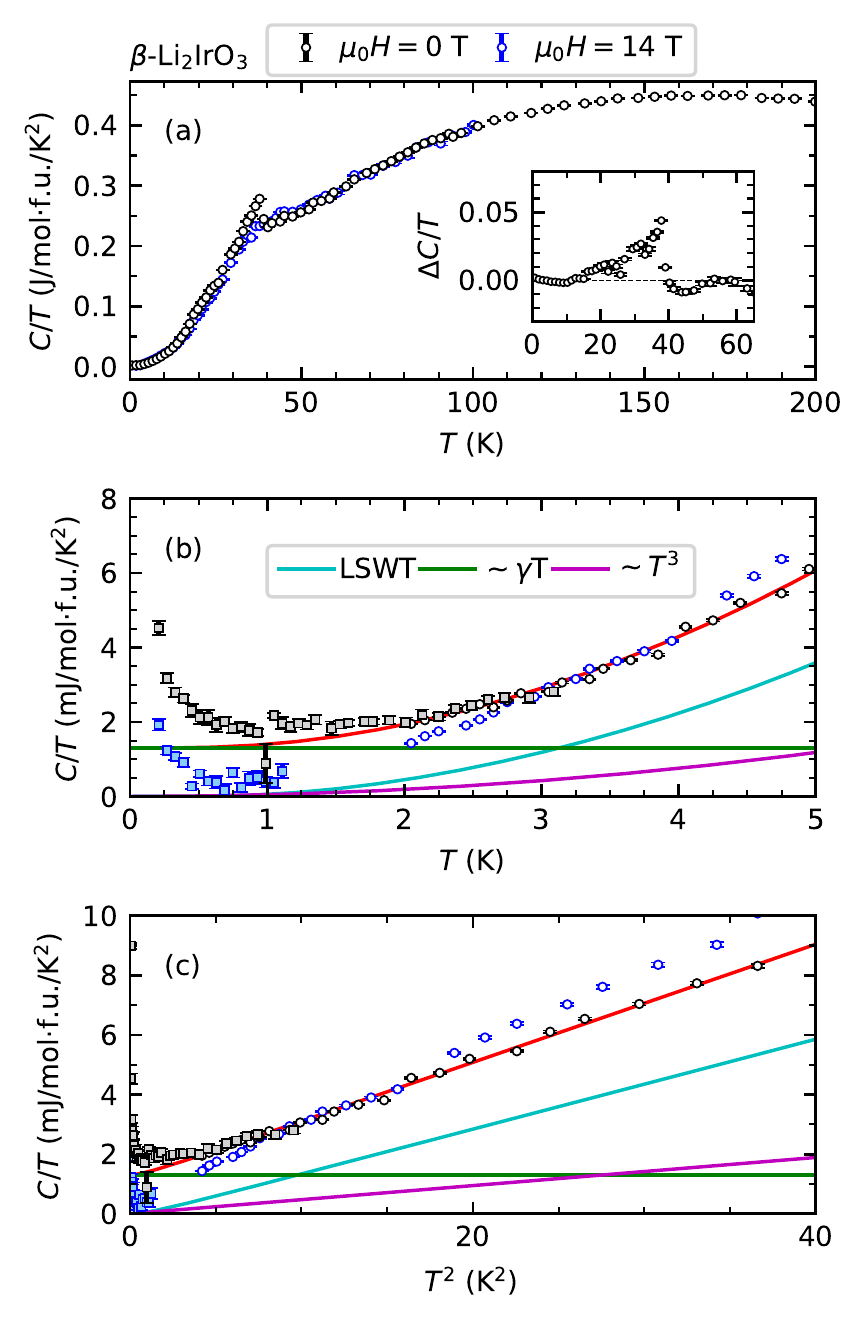}
    \caption{Specific heat capacity of $\beta$-$^7$Li$_2^{193}$IrO$_3$ plotted as $C(T)/T$. In all subplots black points represent zero field measurements, blue points represent 14 T measurements. Square symbols denote data acquired using the dilution refrigerator configuration of a physical properties measurement system. Circle symbols denote data from the high temperature configuration of the PPMS. (a) Specific heat at zero-field and $\mu_0H=14$~T.  (Inset) View around $T_{\rm N}=38$~K highlighting the peak associated with the magnetic phase transition. The plotted quantity $\Delta C(T)/T$ is the difference between the zero field and 14 T measurements. (b) Detailed view of the low $T$ regime with model fitting. The red line is the sum of three terms: (1) The calculated contribution from 3D antiferromagnetic spin waves above a 2.8 meV gap in the excitation spectrum (Fig. 4), (2) a $T-$ linear term that vanishes at high fields, and (3) a separate $T^3$ term that we associate with acoustic phonons.   
    (c) $C/T$ vs $T^2$. The $T-$linear term in $C(T)$ is clearly visible here as the intercept with the y-axis. The 14 T data (blue points) show a zero intercept so that the $T-$linear term appears to be associated with magnetic excitations. }
    \label{fig:lio_hc}
\end{figure}

\subsection{Heat Capacity}
To access the $Q-$averaged magnetic excitation spectrum to the lowest energies, and to probe the critical regime near the magnetic phase transition, we measured the specific heat capacity $C(T)$. The result is shown as a plot of $C/T$ in Fig.~\ref{fig:lio_hc}(a). The zero field data set (black symbols) features a sharp peak at $T_{\rm N}=38$~K that is suppressed and broadened by the application of a 14 T magnetic field (inset to Fig.~\ref{fig:lio_hc}(a)).  This indicates a magnetic phase transition in an anisotropic material as detailed in previous studies on single crystalline samples \cite{Majumder2019AnisotropicStudy,Takayama2015HyperhoneycombMagnetism,Ruiz2017CorrelatedFields}. The change in entropy associated with the temperature regime in the immediate vicinity of the phase transition (10 K to 40 K) is estimated to $\Delta S_m=0.41(1){\rm J mol}^{-1} {\rm K}^{-1}$. Representing  just 7\% of the $R$ln2 total spin entropy per iridium, this small peak indicates the magnetic order is incomplete and/or develops from a strongly correlated state. This observation is consistent with the small ordered moment ($0.47(1)~\mu_B$ \cite{Biffin2014Noncoplanar-li2iro3}) and the fact that the magnetic excitation spectrum extends well beyond $k_BT$ for $T=45$~K $>T_N$ (Fig. \ref{fig:lio_main}(a) and Fig. \ref{fig:lio_main_cuts}(a)).  Correspondingly, the peak in $C(T)$ marking the phase transition rides on a broad maximum, which in addition to contributions from phonons, is associated with the development of short range spin correlations. 

Fig.~\ref{fig:lio_hc}(b,c) display $C(T)/T$ in the low temperature regime. While the upturn for $T<1$~K is associated with hyperfine splitting of the nuclear spin-3/2 of iridium, there is also a Sommerfeld-like term $C(T)=\gamma T$ that is unusual for a long range ordered insulating magnetic material. $\gamma$ is driven to zero in a field of 14 T, which indicates it is associated with gapless  electronic excitations. Further analysis and discussion of the heat capacity data is in sections~\ref{C_analysis} and \ref{discussionandconclusion}.

\begin{figure*}[t]
    \centering
    \includegraphics[width=1.0\textwidth]{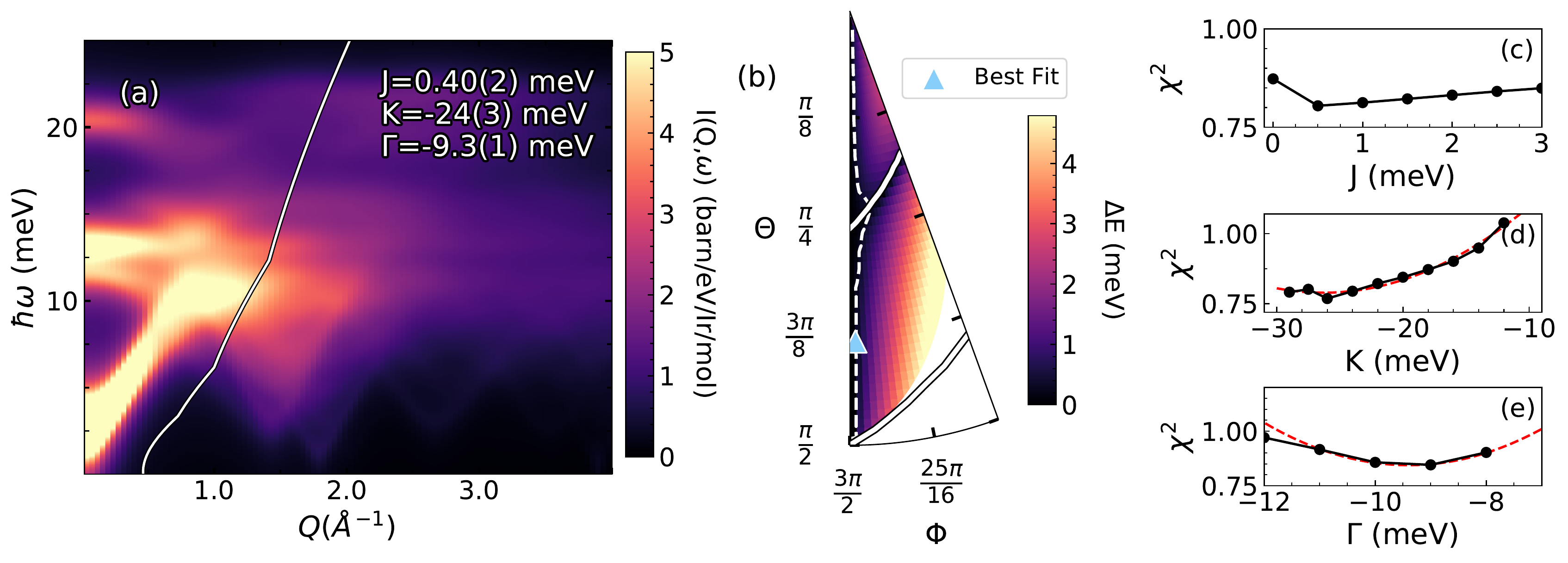}
    \caption{Results from fitting linear spin wave theory to magnetic neutron scattering data for $\beta$-$^7$Li$_2^{193}$IrO$_3$. (a) The calculated spectrum of inelastic neutron scattering for the best fit parameters. The white line represents the kinematic limit of the experiment which combined four different incident neutron energies. (b) The calculated gap at the $\Gamma$ point in the phase space of $J$-$K$-$\Gamma$ model, as described by parameters $\theta$ and $\phi$ in Eqn.~(\ref{eq:parameterizeHamiltonian}).The solid white lines depict phase boundaries between dominant $K$ and $\Gamma$ interactions as described in Ref. \cite{Ducatman2018Magnetic-Li2IrO3} and the triangle represents the best fit result from our experiment which is used to calculate the spectrum in panel (a). The white space represents the region of phase space in which the observed magnetic structure is not stabilized as the lowest energy ground state. The dashed white line shows the path through phase space where $\Delta E=0.4$~meV. (c-e) The $\chi^2$ goodness of fit versus the free parameters $J, K$, and $\Gamma$. The red dashed lines in (d) and (e) are parabolic fits from which the minima and their uncertainties were extracted. The experimental uncertainty for each parameter is determined by the range of the parameter over which $\chi^2$ is less than $\chi^2_{min} / (1+1/(N_{d} - N_{p}))$, with $N_{d}$ being the number of pixels in our measurement and $N_{p}$ being the number of free parameters in the fit. This fitting process is not able to determine the Heisenberg exchange parameter $J$ because the kinematically accessible part of the magnetic neutron scattering cross section is insensitive to $J$. }
    \label{fig:lio_theory_composite}
\end{figure*}

\section{Analysis}

\subsection{Spin Hamiltonian and spin wave theory}
To account for the counter-rotating spin structure, the minimal spin Hamiltonian for $\beta$-Li$_2$IrO$_3$ is the $J$-$K$-$\Gamma$ model~\cite{Lee2015TheoryIridates,Lee2016TwoMaterials,Ducatman2018Magnetic-Li2IrO3, Rousochatzakis2018Magnetic-Li2IrO3, Liu2020KitaevCompounds,Li2021ModifiedMagnets} 
\sbe\label{eq:Hamiltonian0}
\mc{H}\!=\!\sum_t\sum_{\langle ij\rangle\in t}\mc{H}_{ij}^t,
\see 
where
\sbe\label{eq:Hamiltonian}
\mc{H}_{ij}^t=J\vec{S}_i\cdot\vec{S}_j+K S_i^{\alpha_t}S_j^{\alpha_t}+\sigma_t\Gamma(S_i^{\beta_t}S_j^{\gamma_t}+S_i^{\gamma_t}S_j^{\beta_t}).
\see
Following the nomenclature of Ref.~\cite{Li2020Torque}, ${\bf S}_i$ denotes the pseudo-spin $j_{\text{eff}}\!=\!1/2$ operator at site $i$. The five different types of NN Ir-Ir bond are labeled $t\in\{x,y,z,x',y'\}$ with associated cartesian components $(\alpha_t,\beta_t,\gamma_t)\!=\!(x,y,z)$, $(y,z,x)$, and $(z,x,y)$ for $t\in\{x,x'\}$, $\{y,y'\}$, and $\{z\}$, respectively. For simplicity, we take $K$ to be bond independent though the $z$ type bond by symmetry is distinguishable from the $(x,x')$ and ${y,y'}$ type bonds. The prefactor $\sigma_t\!=\!\pm1$ determines the sign of the $\Gamma$ interactions which is bond-dependent and prescribed by lattice symmetry~\cite{Lee2015TheoryIridates}. Fixing the overall energy scale to be $J^2+K^2+\Gamma^2\equiv 1$, the Hamiltonian in Eq.(\ref{eq:Hamiltonian}) can be parameterized in terms of polar angles $\theta$ and $\phi$ as 
\sbe\label{eq:parameterizeHamiltonian}
J=\sin{\theta}\cos{\phi},\ K=\sin{\theta}\sin{\phi},\ \Gamma=\text{sgn}({\Gamma})\cos{\theta}.
 \see 

Within this parameterization, Luttinger-Tisza (LT) analysis indicates the counter-rotating order is stabilized in the approximate range of $(\theta,\phi)\in[(0,\frac{\pi}{2}),(\frac{3\pi}{2},\frac{13\pi}{8})]$ \cite{Ducatman2018,Lee2016TwoMaterials}. Varying mostly with $\phi$, the incommensurate wave vector $\textbf{Q}=h{\bf a^*}$ takes on values in the range $0.53 \lesssim h \lesssim 0.80$ \cite{Lee2015TheoryIridates}, which may be compared to the experimental value of $h=0.57(1)$\cite{Biffin2014a}. The LT method does not include the higher harmonic components that  are required to ensure a fixed spin length for a general incommensurate order. Instead of using an incommensurate wavevector, we take the magnetic structure to be a long-wavelength deformation of the closest $\textbf{Q}=(2/3,0,0)$ commensurate approximation of the true order~\cite{Ducatman2018}. Within the region of phase space with dominant Kitaev interactions, this description accounts for most experimental findings reported so far, including the observed static spin structure factor, the irreducible representation, magnetic structure of two counter-rotating spin sublattices, the response under a magnetic field, as well as Raman scattering ~\cite{Ruiz2017CorrelatedFields,Williams2016,Li2020,Li2020Torque,Rousochatzakis2018Magnetic-Li2IrO3,Ruiz2021Magnon-spinon-Li2IrO3,Yang2022Signaturesbeta-Li_2IrO_3}.

Given the Hamlitonian (\ref{eq:Hamiltonian}) and the parameterization (\ref{eq:parameterizeHamiltonian}), we employ the standard $1/S$ semiclassical expansion (see details in  Refs.~\cite{Ducatman2018,Li2020Torque}) and compute the dynamical spin structure factor (DSF) given by  Eq.(\ref{eq:Sqw}).
To accommodate the zero-field spiral magnetic order with propagation wavevector  $\bf Q\parallel a^*$, we use an enlarged magnetic unit cell composed of three orthorhombic unit cells along the ${\bf a}$-axis with 48 magnetic sites \cite{Ducatman2018}.

\begin{figure}[t]
    \centering
    \includegraphics[width=1.0\columnwidth]{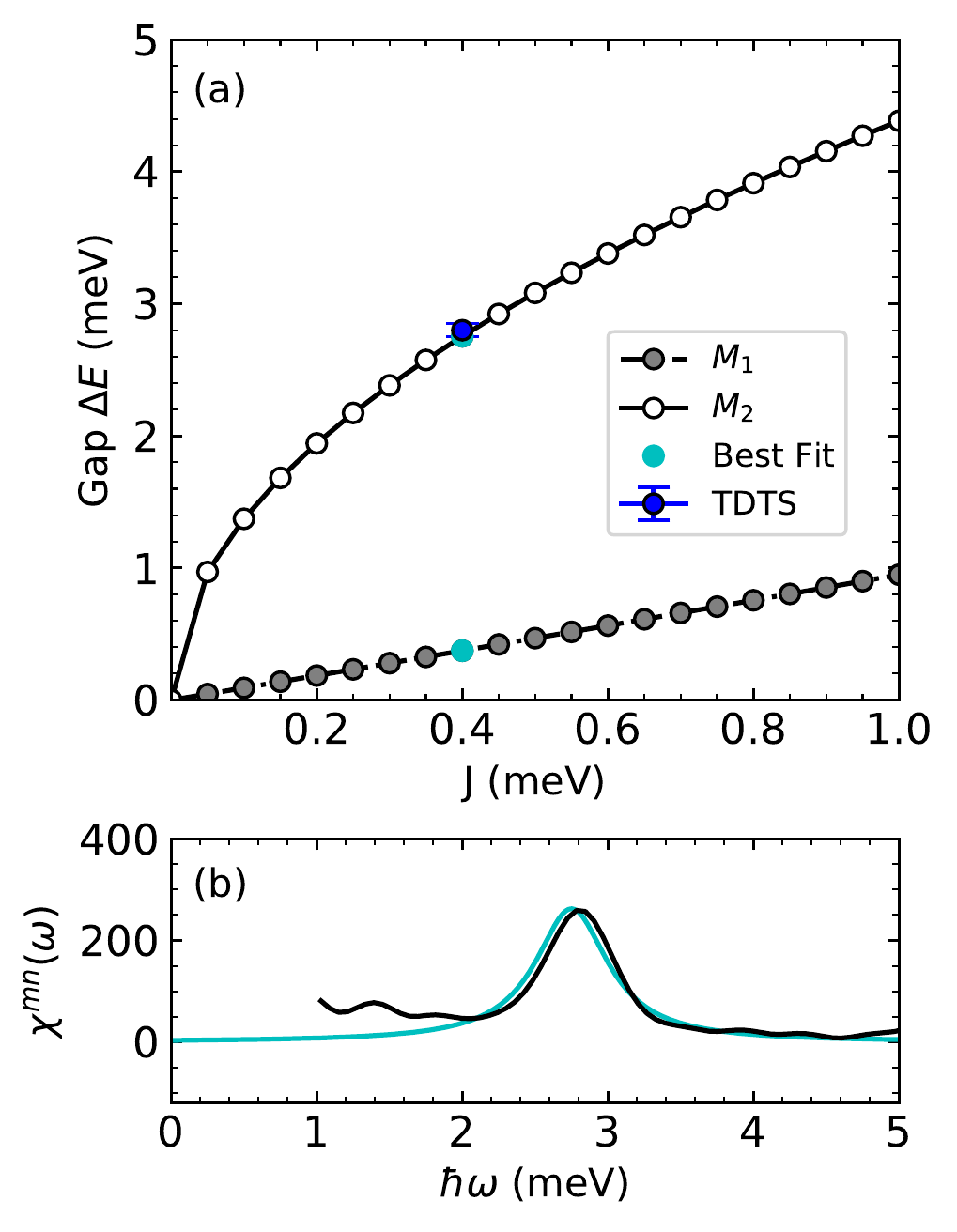}
    \caption{(a) Calculated energies at the $\Gamma$ point for spin wave modes $M_1$ and $M_2$ in $\beta$-$^7$Li$_2^{193}$IrO$_3$ as described in the text versus the Heisenberg exchange constant $J$. Cyan points highlight the overall best fit value of $J=0.4$~meV. The blue point shows the observed $M_2$ mode energy determined by THz spectroscopy (TDTS). (b) Calculated THz response for $\beta-$Li$_2$IrO$_3$ (blue) and the measured TDTS at $T=3$~K (black). The main peak is clearly described by the model in cyan using the best fit parameters $J=0.4$~meV, $K=-24$~meV, and $\Gamma=-9.3$~meV. }
    \label{fig:THz_response_theory}
\end{figure}

\subsection{Extraction of Exchange Parameters}

We describe the excitation in the antiferromagnetic state of $\beta-$Li$_2$IrO$_3$ by linear spin wave theory and refine the exchange parameters $J, K$ and $\Gamma$ in Eq.(\ref{eq:Hamiltonian}) by comparison to inelastic neutron scattering, THz spectroscopy, and heat capacity measurements presented in section~\ref{resultsection}.  

Fig.~\ref{fig:lio_theory_composite}(a) presents the calculated ${\cal S}(Q,\omega)$ with the best fit INS spectra (Fig.~\ref{fig:lio_main}, Appendix  \ref{appendix:jkg}), with $K=-24(3)$~meV, $\Gamma=-9.3(1)$~meV, and $J\approx 0.40(2)$~ meV. For $Q$ within the kinematic limits set by the experimental conditions and indicated by the white line in Fig.~\ref{fig:lio_theory_composite}(a), the model calculation is consistent with the experimental data in Fig. 1(b). In the $J$-$K$-$\Gamma$ model, the effects of each parameter on the calculated spectra are shown in Appendix \ref{appendix:jkg}, from which it is immediately clear that the part of ${\cal S}(Q,\omega)$ accessed in our neutron scattering experiment is sensitive to $\Gamma$ and $K$. In contrast, the Heisenberg interaction $J$ barely affects the spectrum that is visible in INS, and therefore is not well constrained by the neutron data. The fidelity in refining $J, K, \Gamma$ from INS is quantitatively reflected in the variation of $\chi^2$ with each parameter as shown in Fig.~\ref{fig:lio_theory_composite} (c-e). While $\chi^2$ v.s. $K$ and $\Gamma$ have a parabolic shape, Fig.~\ref{fig:lio_theory_composite} (c) shows the value of $J$ is not well constrained by the INS data. The magnitude of the $Q=0$ gap $\Delta E$ is however, closely linked to $J$. To determine  $J$, we therefore turn to  THz spectroscopy, which probes the $\Gamma$ point spectrum and is sensitive to $\Delta E$. The result of calculating $\Delta E$ using LSWT over the phase space consistent with the incommensurate order is shown in Fig.~\ref{fig:lio_theory_composite} (b). Near our best fit point from INS, $\Delta E$ increases strongly with increasing values of $\phi$ or $J$. This is shown explicitly in Fig. \ref{fig:THz_response_theory} (a). 

Using the values of $K$ and $\Gamma$ and the range of $J$ determined from INS, the LSWT predicts the lowest two modes at $Q=0$ to be at $0-1$~meV and $1-5$~meV.  We designate these modes as $M_1$ and $M_2$ respectively and their dependence on $J$ within linear spin wave theory is presented in Fig.~\ref{fig:THz_response_theory}. Assigning the $2.8$~meV peak observed in THz spectroscopy (Fig.~\ref{fig:lio_thz}) to the higher energy mode leads to the refinement of $J=0.40(2)$~meV and the prediction of the lower energy mode at 0.4 meV. Here we do not consider the effects of magnon anharmonicity, which may renormalize $M_1$ and $M_2$  to be much closer together in energy. The $Q=0$ mode energy estimates obtained 
by treating the magnon interactions at the level of a mean-field decoupling of the quartic terms (and disregarding the magnon decay processes driven by the cubic terms) appear to be able to bring $M_1$ and $M_2$ much closer  to each other with renormalized energies slightly below and slightly above 3 meV~\cite{Yang2022Signaturesbeta-Li_2IrO_3}.

\begin{table*}
\centering
 \noindent
 \begin{tabular}{|p{2cm} |p{10cm}|p{5cm} |} 
 \hline
 Source & $JK\Gamma$ Estimate & Method \\
 \hline\hline
 Ref. \cite{Ducatman2018Magnetic-Li2IrO3} & $|K|>|\Gamma|\gg|J|$ & LLG using magnetic structure  \\ 
 \hline
 Ref. \cite{Majumder2019} & $\Gamma=-15(11)$ meV, $3J+K=-11(4)$ meV& Single crystal Magnetization\\
 \hline
 Ref.~\cite{Rousochatzakis2018Magnetic-Li2IrO3} & $J\approx 0.3$ meV, $|K|\gg|J|, |G|\gg|J|$ & $H_c$ and Magnetic Structure  \\
 \hline
 Ref. \cite{Stavropoulos2018Counter-rotatingModel} & $K\approx\Gamma <0$, $J\ll|K|, J>0$ & Diagonalization  \\
 \hline
 Ref. \cite{Katukuri2016} & $K=[-15,-12]$ meV, $\Gamma=[-3.9,-2.1]$ meV, $J=[0,1.5]$ meV & Diagonalization and MRCI+SOC   \\ [1ex] 
 \hline
 Refs. \cite{Li2020Torque,Ruiz2021Magnon-spinon-Li2IrO3} & $K=-18$ meV, $\Gamma=-10$ meV, $J=0.4$ meV & Raman Scattering, RIXS  \\
 \hline
 This work &$K=-24(3)$ meV, $\Gamma=-9.3(1)$ meV, $J=0.40(2)$ meV & INS, TDTS, Heat capacity \\
 \hline
\end{tabular}
\caption{Summary of literature estimates of exchange parameters in $\beta$-Li$_2$IrO$_3$.}
\label{tab:literature_table}
\end{table*}

We now use the extracted parameters to calculate the full spectrum of magnetic excitations for comparison with future INS or RIXS experiments on single crystals. The spin wave dispersion relation along a high symmetry trajectory through the magnetic Brillouin zone has previously been reported in Ref.~\cite{Yang2022Signaturesbeta-Li_2IrO_3}. Fig.~\ref{fig:DSF_1} shows the diagonal components of the dynamic spin structure factor, $\mc{S}^{aa}(\vec{Q},\omega)$, $\mc{S}^{bb}(\vec{Q},\omega)$, $\mc{S}^{cc}(\vec{Q},\omega)$ for wave vector transfer along  the  orthorhombic (100) direction. The off-diagonal elements $\mc{S}^{ab}(Q,\omega)$ and $\mc{S}^{bc}(Q,\omega)$ are zero by the $C_{2b}$ symmetry (two-fold rotation around the ${\bf b}$ axis) of the zero-field ground state and the ground state when the magnetic field is applied along the b axis. The off-diagonal component $\mc{S}_{ac}(Q,\omega)$ is much smaller than the diagonal components.

\subsection{Application of LSWT to Heat Capacity}
\label{C_analysis}
The temperature-dependent specific heat capacity $C_{\rm mag}(T)\approx\partial U/\partial T$ can be modeled by bosonic magnons starting from
\begin{equation}
    U(T) = \int_\Delta^{\infty} \frac{\epsilon~ g(\epsilon)}{e^{\beta \epsilon} -1} d\epsilon.
    \label{eq:hc_boson}
\end{equation}
Here, $\epsilon$ is the spin wave excitation energy, $\Delta$ is the excitation gap, $\beta=1/k_BT$, and $g(\epsilon) $ is the magnon density of states\cite{Scheie2019HomogeneousAntiferromagnet}. For a three dimensional antiferromagnet with a dispersion relation $\epsilon(q)=\sqrt{\Delta^2+(cq)^2}$ (where $\Delta$ is the spin gap and $c$ is the spin-wave velocity), the density of states is 
\begin{equation}
    g(\epsilon) = \frac{V\eta}{2c^3 \pi^2}\epsilon\sqrt{\epsilon^2 - \Delta^2}\,,
    \label{eq:disp_2D}
\end{equation}
where $V$ is the unit cell volume and $\eta$ is a constant. The 3D nature of the low energy magnons is reflected in the linearity of $C_{\rm mag}/T$ versus $T^2$ for temperatures above $\Delta/k_B$ (Fig. \ref{fig:lio_hc}(c)). The values for $c$, $\Delta$, and $\eta=4$ were fixed by the calculated linear spin-wave dispersion for the lowest energy band at the $\Gamma$ point. Because $c$ and $\Delta$ are closely coupled, the specific heat data do not provide an independent estimate for $\Delta$. To fit the measured specific heat, we must add a $T-$linear term with $\gamma=1.28(2)$ mJ/mol/K$^2$, an additional $T^3$ term that may be associated with acoustic phonons, as well as a contribution at very low temperatures from the nuclear specific heat ($C(T)\propto 1/T^2$). A $T-$linear term in the specific heat capacity was previously associated with fermionic quasi-particles in spin-1/2 chains \cite{Oshikawa1999CharacterizationPresented} and quantum spin liquid candidates \cite{Yamashita2008ThermodynamicSalt} though it can also be associated with localized possibly disorder related excitations \cite{taillefer}.  


\section{Discussion and Conclusions}
\label{discussionandconclusion}

Using a double isotope powder sample and the high intensity time of flight spectrometer SEQUOIA at the SNS, we have acquired inelastic magnetic neutron scattering data within the paramagnetic and within the incommensurate magnetically ordered state of the spin-1/2 hyper-honeycomb lattice of $\beta-^{7}$Li$_2^{193}$IrO$_3$. Complementary information about electronic excitations was obtained through time domain THz spectroscopy and specific heat capacity measurements.

To quantitatively establish a spin Hamiltonian for this material we compared the multiple inelastic neutron scattering measurements
 to detailed spin-wave theory of the  anisotropic nearest neighbor $J$-$K$-$\Gamma$ model. This comparison 
constrained the parameters to $J=0.40(2)$~ meV, $K = -24(3)$~meV, and $\Gamma=-9.3(1)$~meV. 
The same set of parameters also predicts the excitations at the $\Gamma$ point observed by TDTS and is consistent with the heat capacity measurements.

\begin{figure}[t]
\includegraphics[width=0.9\columnwidth]{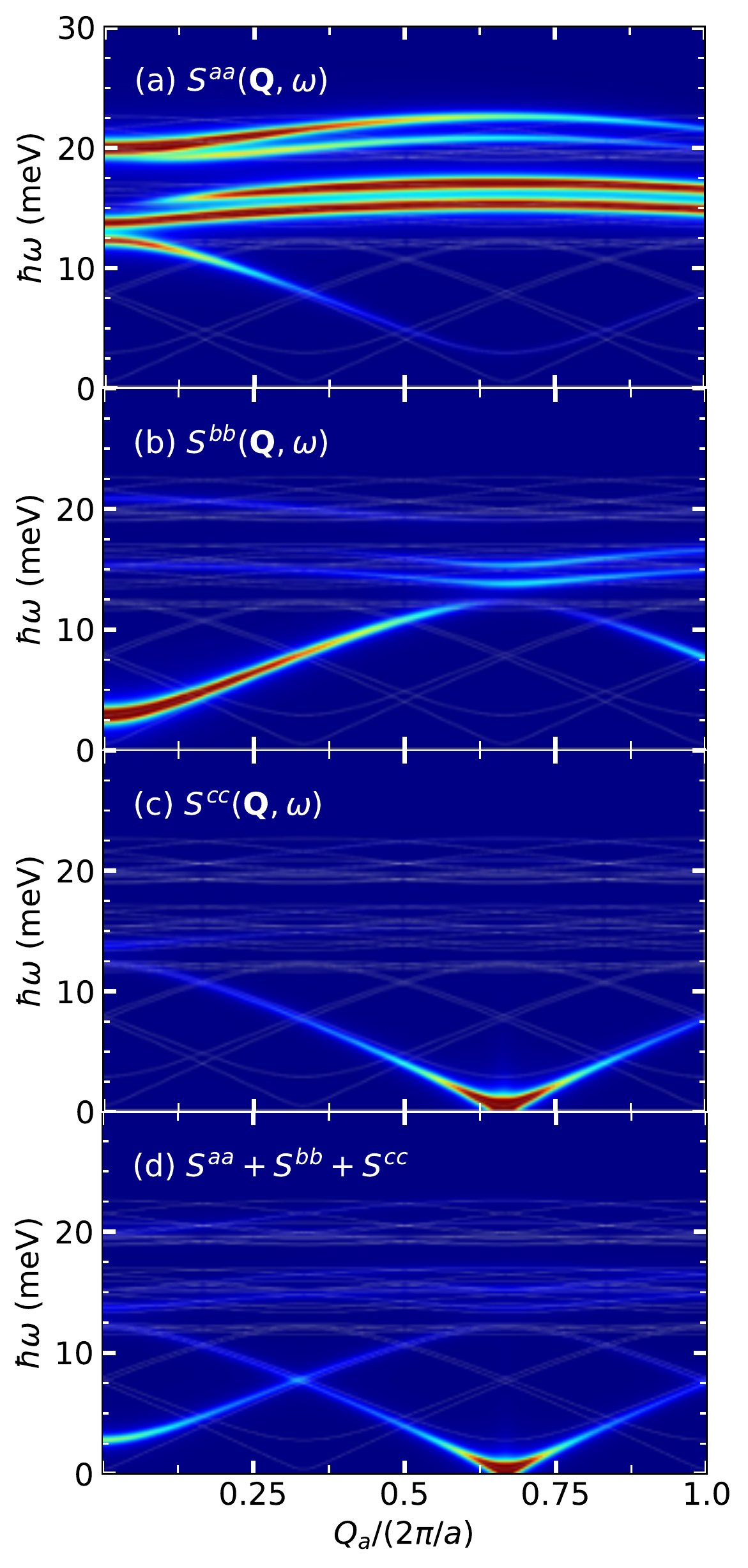}
\caption{The  diagonal components  of the spin dynamical  structure factor (a-c) and their sum (d). In the background, we also plot the linear spin wave spectrum (shown by thin white lines). The color scale is arbitrary and different for each panel. Calculations were performed for the parameter set of $J$=0.4 meV, K=-24 meV, and $\Gamma$=-9.3 meV.}
\label{fig:DSF_1}
\end{figure}

Inelastic neutron scattering response shows a characteristic broad feature around 10 meV that is consistent with powder-averaged dynamical spin structure factor calculations within linear spin-wave theory. This feature appears to arise from the combination of: i) the particularly strong response in the $\mc{S}^{aa}$ channel in that regime (see Fig.~\ref{fig:DSF_1}(a), ii) the otherwise large number of excitation modes in the same regime (reflecting, in part, the large magnetic unit cell of the multi-sublattice ordered state), iii) the result of powder averaging in a  system with a highly anisotropic response (see different panels in Fig.~\ref{fig:DSF_1}). To these we should also add the effect of higher-order spin-wave corrections, which, although not calculated explicitly here, are known to be significant especially in highly anisotropic systems~\cite{Winter2017BreakdownMagnet}. 

Furthermore, the extracted Heisenberg exchange $J$ is consistent with the  critical field for the ordered state, which is predicted to vary as $\mu_B \mu_0H_c=0.46J$~\cite{Rousochatzakis2018Magnetic-Li2IrO3}. Using our result for $J$ we find $\mu_0 H_c\approx 3.2(2)$~T, which is consistent with the observed value of 2.8 T. Secondly, the strong absorption feature near $\hbar \omega=7$ meV in our THz data is close to the prediction of the third-lowest excitation at $\hbar\omega=7.5$~meV at the $\Gamma$ point though low THz transmission precludes definite identification of the 7.5 meV anomaly (Fig.~\ref{fig:lio_thz}(b)) with a magnetic excitation. The feature also exists for $T>T_N$ and appears temperature independent even at high temperatures, making its nature rather unclear. However, the main feature in the THz spectrocsopy at $\hbar\omega$~meV is consistent with the two peaks seen in recent Raman scattering work. These peaks lie at 2.5 and 3.0 meV and are signatures of non-Loudon-Fleury scattering processes, generated by magnetic-dipole-like terms in the Raman vertex, which are of a similar nature as the excitations probed by THz and INS \cite{Yang2022Signaturesbeta-Li_2IrO_3}. Our results are generally consistent with previous theoretical and experimental works that have attempted to estimate these parameters as shown in Table~ \ref{tab:literature_table}.

The ratios of $J/|K|=0.017(2)$ and $|\Gamma|/|K|=0.39(5)$ suggest that  $\beta-$Li$_2$IrO$_3$ is in relative close proximity to the ideal KSL phase.
While the precise details of the quantum phase diagram of the hyperhoneycomb $J$-$K$-$\Gamma$ model have not yet been established, a comparison to the closely related two-dimensional honeycomb case is rather instructive, as the above ratios would place the two-dimensional model {\it within} the KSL phase~\cite{Wang2019OneLattice}.
Though we are dealing with a three-dimensional material, the proximity of $\beta$-Li$_2$IrO$_3$ to a highly frustrated point in parameter space is further corroborated by the relatively low value of $T_N$ compared to the dominant energy scale of the problem ($T_N/|K|\sim 0.13$), the small amount of entropy associated with the magnetic phase transition, and by the reported fragility of the low-$T$ incommensurate phase under modest external perturbations. Indeed, both a modest external magnetic field and hydrostatic pressure suppress the order at $\mu_0H_c=2.8$~T \cite{Ruiz2017CorrelatedFields,Majumder2019} and 1.4 GPa~\cite{Majumder2018Breakdown-Li2IrO3,Takayama2019Pressure-induced-Li2IrO3}, respectively. 
Moreover, one should also keep in mind that, at the classical level, the $\Gamma$ interaction fails to lift the infinite ground state degeneracy of the Kitaev model completely, and an infinite submanifold of ground states still remains along the $J=0$ line for negative $K$ and $\Gamma$~\cite{Ducatman2018Magnetic-Li2IrO3}.

Finally, we discuss the linear term $\gamma T$ in the zero-field heat capacity (Fig.~\ref{fig:lio_hc}). 
The presence of linear term in the specific heat of a magnetic origin in $\beta-$Li$_2$IrO$_3$ is interesting because it cannot be accounted for by the spin wave theory that was used here to successfully account for the ordered state as well and the inelastic magnetic neutron scattering spectrum. A $T-$linear term can arise from fermionic quasi-particles with the Sommerfeld constant $\gamma$ proportional to the density of states at the Fermi level. $\gamma$ ranges from mJ/mole/K$^2$ in uncorrelated metals such as copper to J/mol/K$^2$ in heavy-fermion systems  \cite{Aoki2019UnconventionalUte2,Chen2002Observation2CoIn8,Bruning2008CeFePO:Correlations}. With no fermionic charge carriers, a linear term is generally unexpected for insulators. The complete suppression of $\gamma$ for $\beta-$Li$_2$IrO$_3$  under a 14 T magnetic field  (Fig.~\ref{fig:lio_hc}) indicates its magnetic origin and is consistent with prior NMR and THz data \cite{PhysRevB.101.214417}. While a field dependent Sommerfeld term could arise from a metallic impurity phase, powder XRD studies as well as low temperature susceptibility measurements place an upper limit on the concentration of impurities at the percent level (see Appendix \ref{appendix:sample_quality}). If such impurities are the origin of the $\gamma T$ term they would need to be heavy fermion like. 

Intrinsic $T-$linear specific heat terms have been reported in materials with quasi-one-dimensional spin-1/2 chains such as copper pyrazine di-nitrate\cite{PhysRevB.59.1008}. There it is quantitatively accounted for by fermionic spinons, which form a Luttinger liquid with $\gamma=2/3R(k_B/J)$ in zero field. A linear term was also reported in  QSL candidate materials including the hyperkagome system Na$_4$Ir$_3$O$_8$ \cite{Singh2013Spin8}. In our measurement, $\gamma = 1.20(1)$ mJ/mol/K$^2$ is comparable to Na$_4$Ir$_3$O$_8$ where $\gamma\approx 2$mJ/mol/K$^2$ \cite{Singh2013Spin8}, but less than other insulating QSL candidates like $\kappa-$(BEDT-TTF)$_2$Cu$_2$(CN)$_3$ where $\gamma=12$ mJ/mol/K$^2$ \cite{Yamashita2008ThermodynamicSalt}. For spin-1/2 chains two different field effects on $\gamma$ have been demonstrated. In copper pyrazine dinitrate the $T-$linear term is enhanced in an applied field. This is fully accounted for theoretically as resulting from a field induced suppression of the spinon-velocity, which enhances the spinon density of states at the fermi level. The opposite effect is found in copper benzoate, which has two magnetic sites per 1D unit cell with symmetry-related g-tensors. As a result  a uniform applied field produces an effective staggered field that opens a gap in the magnetic excitation spectrum and replaces the $T-$linear specific heat capacity by exponentially activated behavior\cite{PhysRevLett.79.1750,PhysRevLett.93.017204,doi:10.1143/JPSJS.72SB.36}. It seems unlikely that exotic fermionic quasi-particles associated with a proximate spin liquid phase would contribute to the very lowest energy part of the excitation spectrum within a symmetry-breaking long range ordered state of $\beta$-Li$_2$IrO$_3$. The incommensurate nature of the order could be relevant as it admits a gapless phason mode as well as complex domain wall structures both of which have the potential to contribute to the low temperature heat capacity. A detailed study of the field dependence of $\gamma$ to determine for example whether its disappearance coincides with the 2.8 T critical field~\cite{Ruiz2017CorrelatedFields} would be informative.

\section{Acknowledgements}

This work was supported as part of the Institute for Quantum Matter, an Energy Frontier Research Center funded by the U.S. Department of Energy, Office of Science, Basic Energy Sciences under Award No. DE-SC0019331. CB was supported by the Gordon and Betty Moore foundation EPIQS program under GBMF9456.  The research at the ORNL Spallation Neutron Source was sponsored by the U.S. Department of Energy, Office of Basic Energy Sciences. We would also like to acknowledge the help of Sayak Dasgupta for heat capacity calculations. Natalia B. Perkins and Mengqun Li were supported by the U.S. Department of Energy, Office of Science, Basic Energy Sciences under Award No. DE-SC0018056. I.R. was supported by the Engineering and Physical Sciences Research Council [grant number EP/V038281/1]. A portion of this research used resources at the Spallation Neutron Source, a DOE Office of Science User Facility operated by the Oak Ridge National Laboratory.

\section{Appendices}

\appendix

\section{Calculated scattering $J,K,\Gamma$ Dependence}
\label{appendix:jkg}
\begin{figure*}[]
    \centering
    \includegraphics[width=0.9\textwidth]{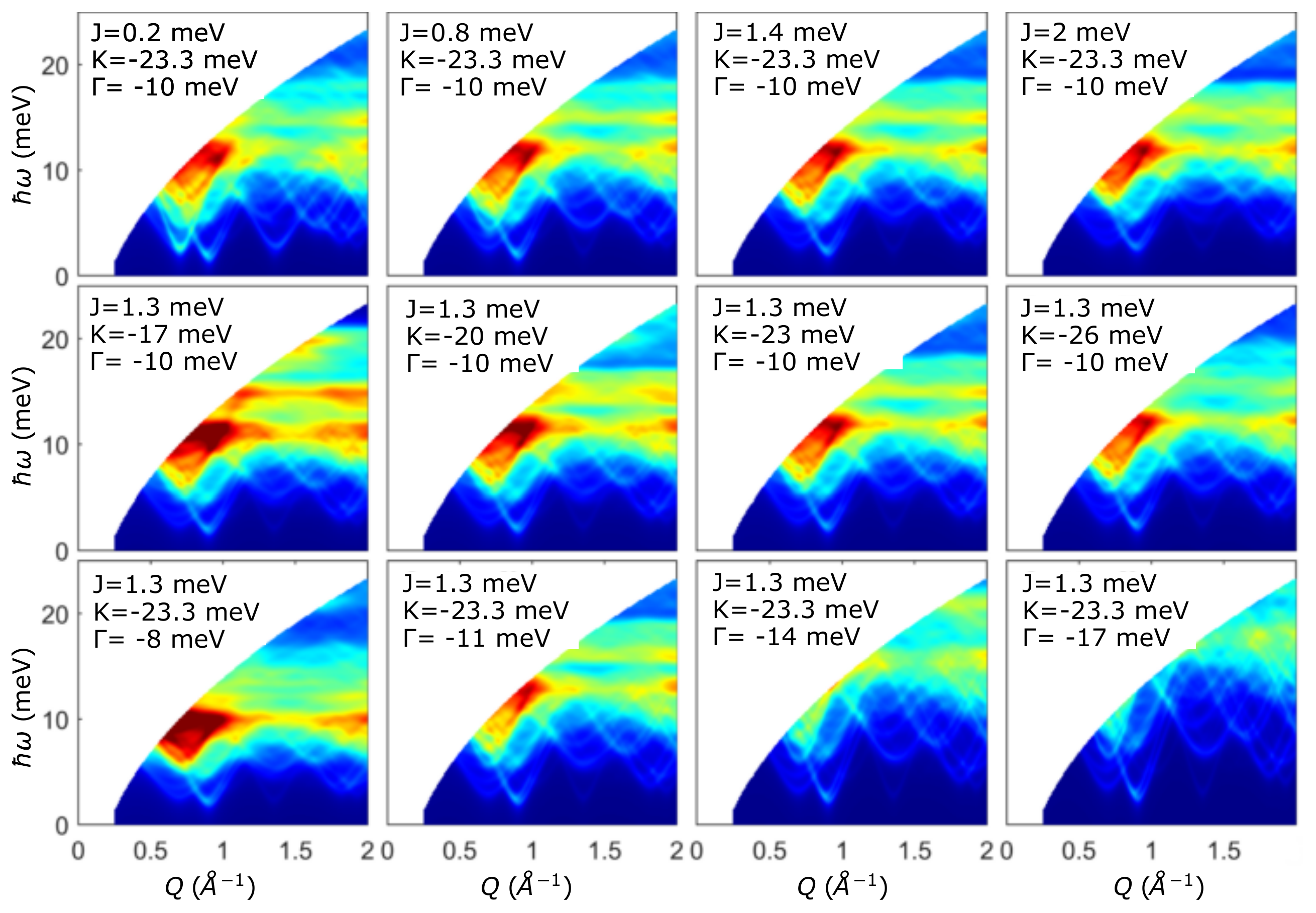}
    \caption{Calculated powder averaged INS intensities for $E_i=30$ meV with variations in $J$, $K$, and $\Gamma$. The magnetic form factor of Ir$^{4+}$ is not considered in the calculation, and a constant broadening in energy of width 1 meV is used to approximate instrumental resolution effects. While $\Gamma$ and $K$ are well constrained by this method, $J$ has little influence on the scattering data that is accessible for $E_i=30$ meV. The intensity scale is arbitrary but consistent across the calculations.}
    \label{fig:ins_evolution}
\end{figure*}
The fits to the measured INS data were performed by the systematic calculation of neutron scattering spectra for many sets of exchange parameters. Spectra were first calculated globally in the region of 
$J-K-\Gamma$ parameter space that stabilizes the incommensurate magnetic order \cite{Ducatman2018Magnetic-Li2IrO3}. Then the calculations were refined  for the smaller range of parameters that best reproduce the observed spectra.

The pattern of scattering varies differently with each of the exchange parameters, as illustrated in Fig. \ref{fig:ins_evolution}.  $J$ has very little effect on the spectra. The value of $\Gamma$ has the greatest impact on the excitation spectra, controlling the overall energy scale of the spin-waves. The influence of $K$ is more subtle. It impacts the energies and intensities of select bands at energies higher than $\Gamma$. This is seen in the second row of spectra in Fig. \ref{fig:ins_evolution}. Thus, inelastic magnetic neutron scattering scattering from a powder sample of $^7$Li and $^{193}$Ir can be used to extract values for $K$ and $\Gamma$ but not $J$. 

\section{Neutron Data Analysis}
\label{appendix:neutron_corrections}
\subsection{Annular Absorption Correction}
Although we have taken care to minimize absorption by means of a custom annular aluminum sample can and isotopic enrichment of both $^{193}$Ir and $^{7}$Li, significant energy transfer dependent neutron absorption still affects the measurement. To account for this, we use a Monte-Carlo method implemented in the Mantid software which takes into account the full neutron path for each pixel as a function of scattering angle and energy transfer \cite{Arnold2014MantidExperiments}. Further details may be found in the Mantid documentation for the AnnularRingAbsorption method. Finally, the measurements were normalized to units of barn/eV/sr/ mol Ir by comparing to the scattering intensity measured for a known quantity of vanadium.

\subsection{Detailed Balance Correction}
\label{appendix:DB}
To isolate the true inelastic neutron scattering from the sample from various types of background we used the fact that  inelastic scattering from the sample at a given temperature $T$ must obey the detailed balance principle $S(-\omega)=\exp(-\hbar\omega/k_{\rm B}T)S(\omega)$. The component of the scattering that does not obey detailed balance between positive and negative energy transfer processes can be considered an approximately temperature-independent background component. Assuming that there exists a true inelastic signal $S(Q,\omega)$ and a temperature-independent background $I_{\rm bkg}(Q,|\omega|)$, the scattering intensity for a particular point in $Q-\omega$ space at temperature $T$, should obey the following relationship:

\begin{gather}
    I(T,\delta)=I_{bkg}(\delta) + e^{-\beta\omega (1-\delta)/2}S(T)
    \label{eq:db_lsq}
\end{gather}

Here, $\delta=[1,-1]$ and denotes positive or negative energy transfer. Due to kinematic constraints, some values of $Q$ and $|\omega|$ do not permit $(Q,|\omega|)$ and $(Q,-|\omega|)$ to be simultaneously accessed. These points are therefore excluded from the analysis. For every pixel in $Q-\omega$ space the value of $I_{\rm bkg}(\delta)$, $I_{\rm bkg}(-\delta)$, and $S(T)$ are  independent free parameters. $S(T)$ can furthermore take on different values for each temperature. This leads to $2N_T$ equations with $2+N_T$ unknowns, where $N_T$ is the number of temperatures where full $Q-\omega$ dependent data sets have been acquired. So long as $N_T\geq2$, a solution can be found for the $Q-\omega$ dependent inelastic scattering cross section at each temperature. 

The system of equations was solved using a weighted ordinary least-squares method \cite{Seabold2010Statsmodels:EconometricPython}, resulting in full solutions of $S(T)$, $I_{bkg}(\delta)$, and $I_{bkg}(-\delta)$ for every value of $Q, \omega$, and $T$ where scattering data are available for positive and negative values of $\omega$ Estimates for the standard error are also obtained. This detailed balance correction was performed separately for each incident neutron energy. We note that the method requires accurate correction for energy-dependent neutron absorption and detector efficiency. Also note that temperature-dependent elastic scattering is not treated correctly with this method.

\begin{figure*}
    \centering
    \includegraphics[width=0.9\textwidth]{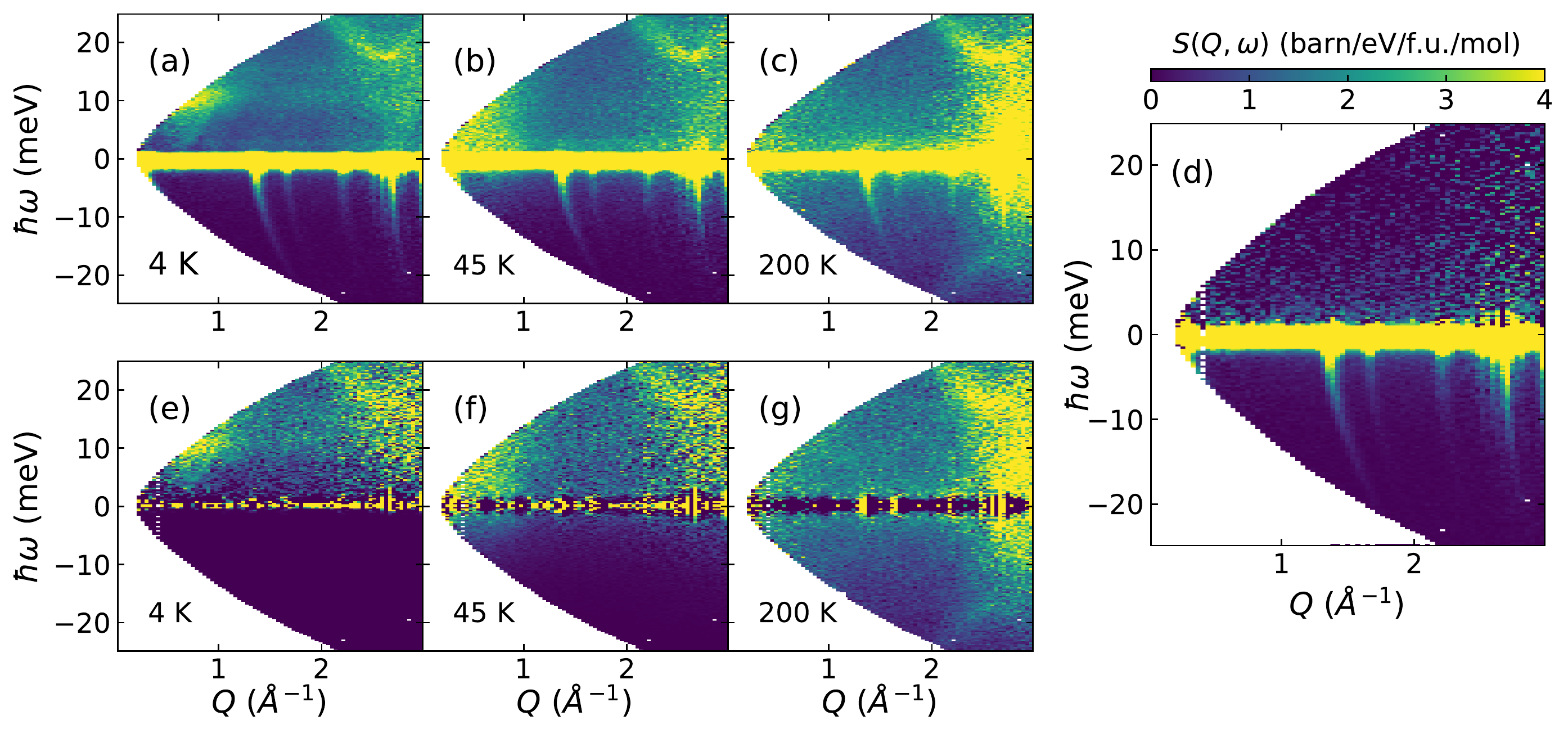}
    \caption{Inelastic neutron scattering spectra for $\beta$-$^7$Li$_2^{193}$IrO$_3$ before and after background subtraction as described in Appendix \ref{appendix:DB}. (a-c) Spectra measured at $T=4$~K, 45 K, and 200 K with  incident neutron energy $E_{\rm i}=30.0$~meV. The detailed balance correction was performed for measurements with all $E_i$ configurations. (d) The temperature-independent background $I_{\rm bkg}(Q,\omega)$ (Eqn.~\ref{eq:db_lsq}).}
    \label{fig:db_app_fig}
\end{figure*}

\begin{figure}
    \centering
    \includegraphics[width=0.8\columnwidth]{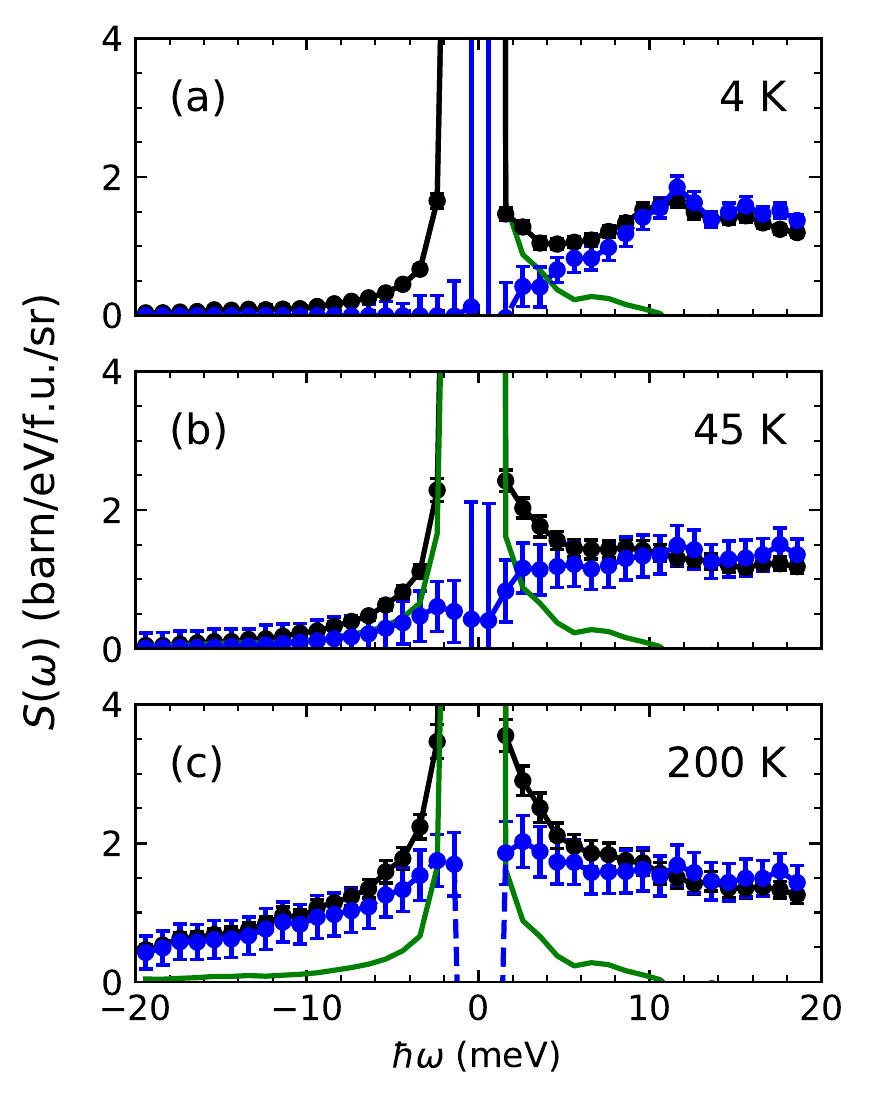}
    \caption{Energy cuts with integration over $Q\in[1.5,2]$~\AA$^{-1}$ for $\beta$-$^7$Li$_2^{193}$IrO$_3$ extracted from the data sets shown in Fig \ref{fig:db_app_fig}. The black symbols represent raw data before the detailed balance correction was applied. The blue symbols represent the extracted inelastic scattering $I_T(Q,\omega)$ and the green line represents the inferred temperature-independent background $I_{\rm bkg}(Q,\omega)$.}
    \label{fig:db_cut_fig}
\end{figure}

Fig. \ref{fig:db_app_fig} shows how inelastic scattering with contributions from magnetism and phonons is separated from temperature independent backgrounds and elastic scattering.  Note the successful removal of the tails of Bragg scattering for $\hbar\omega<0$. These are clearly seen in the raw data (Fig. \ref{fig:db_app_fig} (a-c)). They are identified as a temperature-independent background (Fig. \ref{fig:db_app_fig}(d)), and are no longer present in the inelastic scattering spectra after the procedure (Fig. \ref{fig:db_app_fig} (e-g)). In Fig. \ref{fig:db_cut_fig} (a-c), we presented the energy cuts integrating over $1.5\AA^{-1}<Q<2.0\AA^{-1}$ for both raw data and the inelastic spectra after detailed balance correction. We note that at high energy transfers in panels (a-c), the scattering intensity in the extracted inelastic spectra (blue) exceeds that in the raw data (black) at some $\omega$ values. This is because the detailed balance method is not employing a direct point-by-point subtraction but carries out a least-square fit to the full temperature dependent data set under the assumption that the non-inelastic scattering (backgrounds and elastic scattering) is temperature-independent. Both statistical uncertainty and any   temperature-dependence of non-inelastic contributions to the measured count rates can cause the intensity after this background correction process to exceed the raw count rate within error bars (Fig. \ref{fig:db_cut_fig}). 

\subsection{Subtracting the contribution from phonons}
\begin{figure}
    \centering
    \includegraphics[width=1\columnwidth]{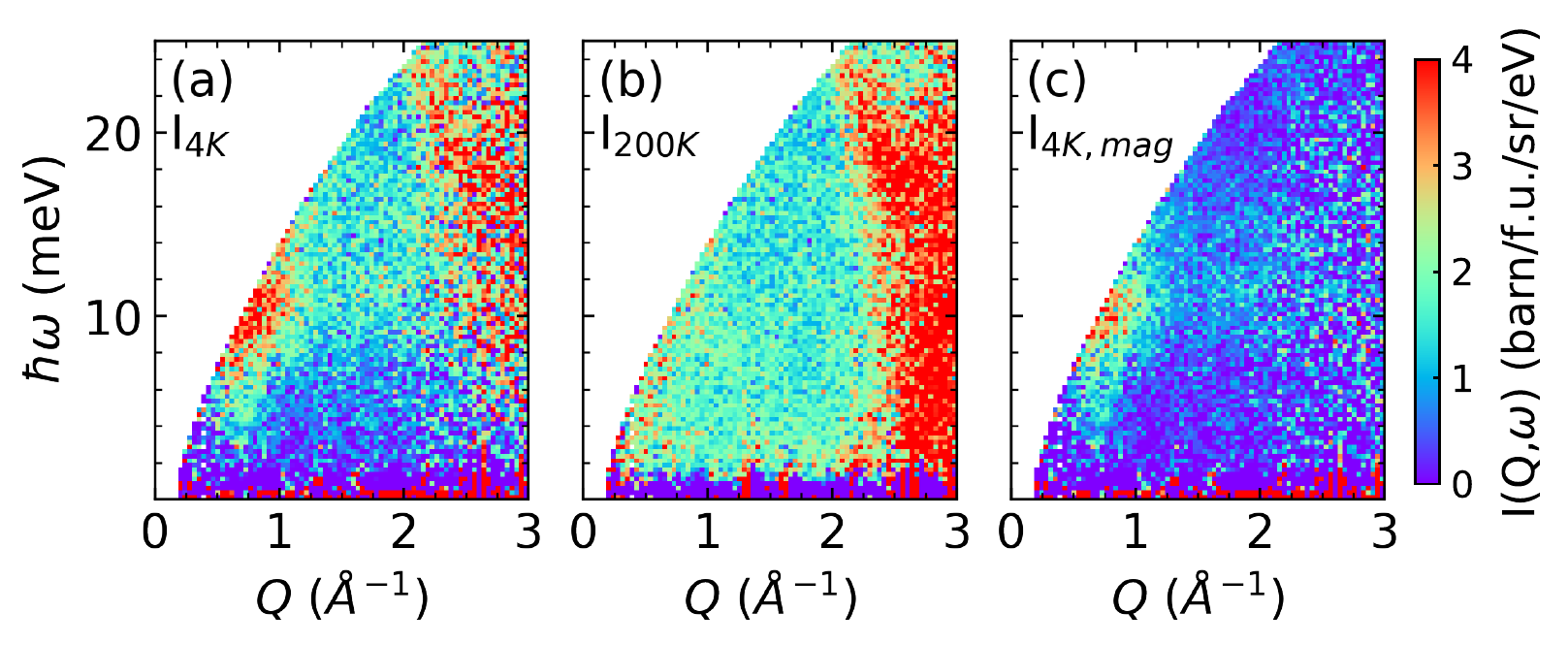}
    \caption{Demonstration of the subtraction of phonon scattering from inelastic neutron scattering data for $\beta$-$^7$Li$_2^{193}$IrO$_3$  using Eq. \ref{eq:bose_fact}. $T=4$~K data (a) is subtracted by the appropriately scaled $T=200$~K data (b) to produce the reported magnetic scattering (c).}
    \label{fig:phonon_fig}
\end{figure}

The contribution from phonons can be removed by comparing scattering spectra taken at low- and high-temperatures. Specifically in our experiment, measurements at $T=4$~K and $T=200$~K have been used. Because phonon scattering increases with $T$ while magnetic scattering is capped by the total moment sum-rule, the high-temperature scattering cross section can be regarded as dominated by scattering from phonons. Using the expression for one-phonon scattering and neglecting the temperature dependence of the Debye Waller factor, the magnetic scattering $I_{\rm m}(Q,\omega)$ can be approximated by: 
\begin{eqnarray}
I_{\rm m}(Q,\omega) &=& I_{L}(Q,\omega) \nonumber\\
 &&- \frac{1-e^{-\beta_H \hbar\omega}}{1-e^{-\beta_L \hbar\omega}}\nonumber\\
&&\times e^{(\langle u^2\rangle_H-\langle u^2\rangle_L)Q^2}\nonumber\\
&&\times I_H (Q,\omega),
\label{eq:bose_fact}
\end{eqnarray}
where the subscripts $L$ and $H$ represent low and high temperatures respectively, and $\beta_i = 1/k_BT_i (i=L, H)$. The ratio of Debye Waller factors on the third line may be approximated as 1 for small $Q$ and moderate $T_H$. Note that this procedure does not remove multi-phonon scattering. An example of the results of this process are shown for $E_i$=30 meV in Fig. \ref{fig:phonon_fig}.
\subsection{Combination of Measurements of Differing Incident Neutron Energy}
\label{appendix:multi_ei}
For a given incident neutron energy $E_{\rm i}$, access to the $(Q,\omega)$ space is constrained by the kinematic limit and the elastic resolution of the instrument. By combining spectra acquired for several incident energies it is possible to extended the accessible range of $Q-\omega$ space. 

The lowest accessible $Q$ is given by: 
\begin{equation}
    Q_{\rm min}(k_{\rm i},k_{\rm f}) = \sqrt{k_{\rm i}^2 + k_{\rm f}^2 - 2k_{\rm i}k_{\rm f} \cos{(2\theta_{\rm min})}}.
    \label{eq:kin_limit}
\end{equation}
Here $\hbar k_{\rm i}$ and $\hbar k_{\rm f}$ are the initial and final momentum of the neutrons and $2\theta_{\rm min}$ denotes the minimum accessible scattering angle of the spectrometer. In the forward scattering limit ($2\theta_{min}=0$), this expression simplifies to 
\begin{equation}
    Q_{\rm min}=|k_{\rm i}-k_{\rm f}|=\frac{|k_{\rm i}^2-k_{\rm f}^2|}{k_{\rm i}+k_{\rm f}}=\frac{|\omega|}{\bar{v}},
\end{equation}
where $\bar{v}=(v_{\rm i}+v_{\rm f})/2$ is the average neutron speed.
The elastic full-width-half-maximum energy resolution $\Delta E\propto E_i$ increases in proportion to $E_{\rm i}$. While the constant of proportionality can be reduced for example by reducing the opening time of the monochromating chopper on a direct geometry spectrometer, better energy resolution generally implies more restricted access to Q-space. 

To optimize resolution and coverage that we combine data acquired for multiple values of $E_{rm i}$.  In the main Fig \ref{fig:lio_main}, for each pixel in the $Q-\omega$ space, we present the magnetic scattering intensity as measured in the configuration with highest instrument resolution in our experiment. As different configurations contribute to different regions in the $Q-\omega$ space, the boundaries separating them are marked by white dashed lines. 

\subsection{Simultaneous fitting of Multiple Incident Neutron Energies}
\label{appendix:multi_ei_chisqr}
Now that the scattering data taken at each incident energy configuration have been corrected for absorption, background, and phonon contributions, they are ready to be compared with the calculated magnetic scattering cross section associated with spin wave excitation. In this procedure, We used all the data acquired with different $E_{\rm i}$ configurations. For a particular set of parameters ($J,K,\Gamma$), we first performed a high resolution calculation of the LSWT over the $Q-\omega$ space probed in our neutron scattering measurements. Then, for each $E_{\rm i}$, the calculated spectrum is convoluted with the energy- and momentum-dependent resolution function, which is different for each instrumental configuration. We also masked out the elastic line based on the elastic resolution of the particular instrumental configuration. A pixel by pixel fit of $J,K,\Gamma$ was then performed by minimizing  $\chi^2$ defined as follows:
\begin{gather}
    \chi^2_{E_{\rm i}} = \frac{1}{N_{E_{\rm i}}}\sum_{j=1}^{N_{E_{\rm i}}} \frac{(I^{E_{\rm i}}_{\rm mag}(Q_j,\omega_j) - A^{E_{\rm i}}I^{E_{\rm i}}_{\rm LSWT}(Q_j,\omega_j))^2}{\sigma I^{E_{\rm i}}_{\rm mag}(Q_j,\omega_j)^2}\nonumber\\
    \chi^2 = \sum_{E_{\rm i}}\chi^2_{E_{\rm i}}.
\end{gather}
In the first equation, $j$ labels the pixels in the entire $Q-\omega$ space probed by neutrons with a certain incident energy, $N_{E_{\rm i}}$ represent the total number of pixels, $I_{\rm mag}^{E_{\rm i}}$ represents the measured magnetic scattering in the configuration of incident energy $E_{\rm i}$, $I_{\rm LSWT}^{E_{\rm i}}$ represents the calculated spin-wave spectra convoluted with the resolution function at that incident energy, and $\sigma I^{E_{\rm i}}_{\rm mag}$ is the standard deviation associated with the measured magnetic scattering intensity $I_{\rm mag}^{E_{\rm i}}$. An overall scale factor $A^{E_{\rm i}}$ was introduced to minimize $\chi^2_{E_{\rm i}}$ for each $E_{\rm i}$. 
The values of $\chi^2 $ calculated in this way are presented in Fig. \ref{fig:lio_theory_composite} (c-e).

\subsection{Verification of sample quality}
\label{appendix:sample_quality}

\begin{figure}
    \centering
    \includegraphics[width=1\columnwidth]{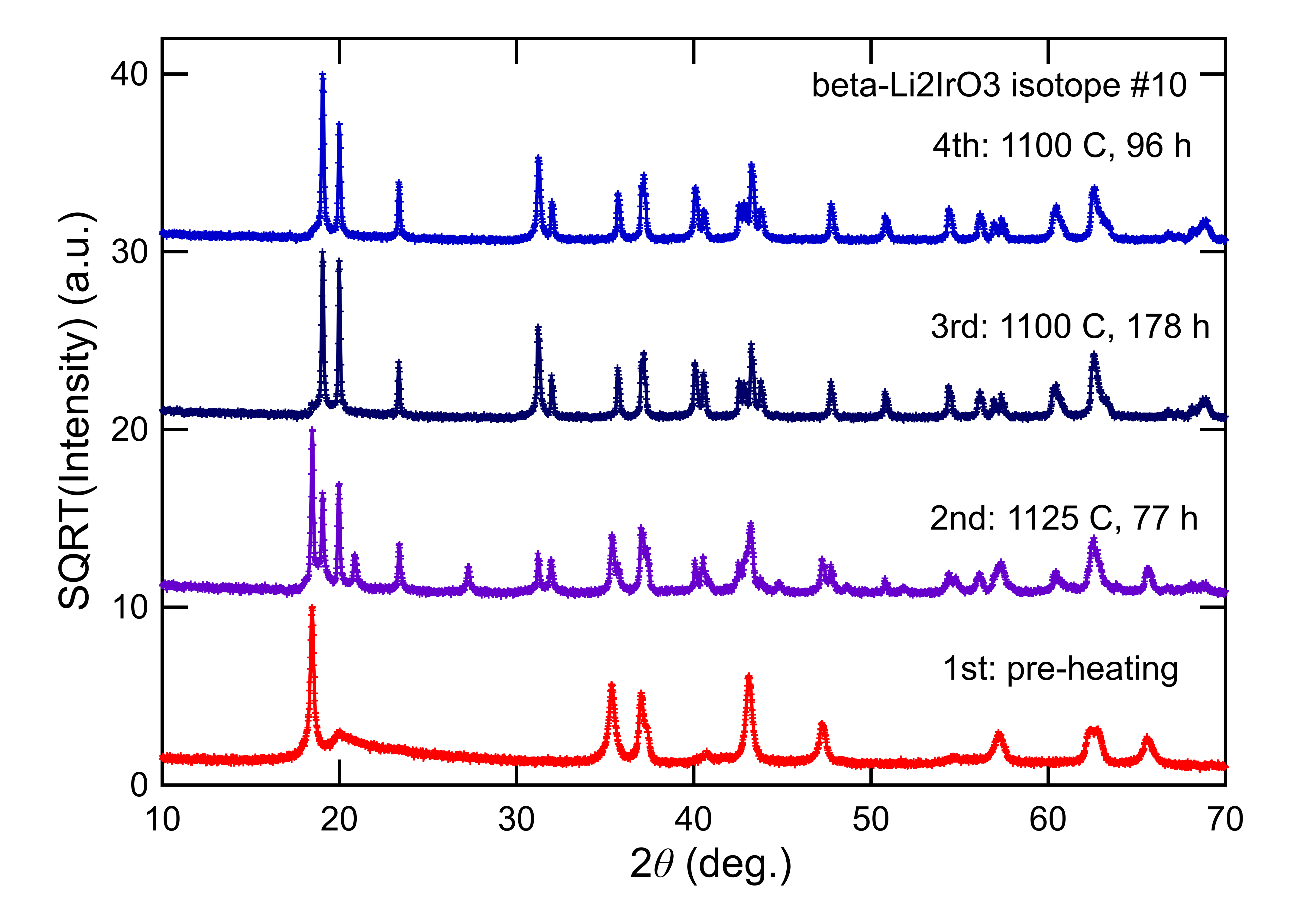}
    \includegraphics[width=1\columnwidth]{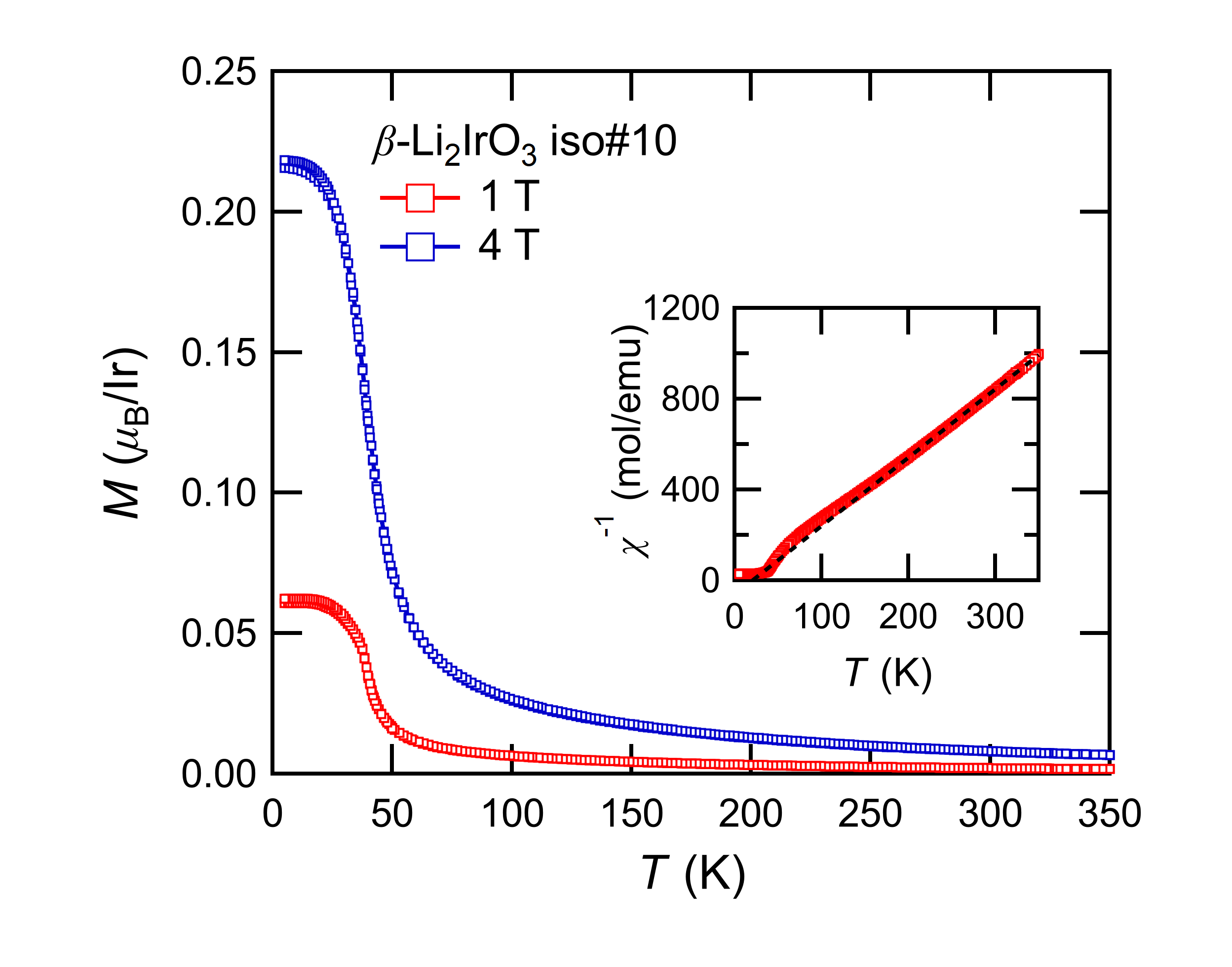}
    \caption{ (a) Powder x-ray diffraction patterns from one batch of $\beta-^{7}$Li$_2^{193}$IrO$_3$ showing the evolution of the diffraction pattern through various annealing steps. No impurity phases were observed in the final product for any of the batches used in this study. (b) Magnetization and susceptibility data for the same powder sample. The main feature in the data indicates the magnetic phase transition expected to occur at $T_N=38$~K.}
    \label{fig:impurity fig}
\end{figure}

A $\gamma T$ term in the specific heat capacity can arise from various types of impurities in the sample. To look for such impurities powder samples used for this work were characterized through x-ray diffraction and magnetic susceptibility measurements. Identical measurements were performed on each batch of powder used for INS, heat capacity, and THz experiments. In the final products used for the experiments no impurities were detected by these methods indicating phase pure materials with minority phases below the 1\% level. One batch that was discarded contained IrO$_2$ impurities, which therefore may be present in trace amounts in our samples. 

We present the effective moments and Curie-Weiss temperatures extracted from fits to $\chi^{-1}$ curves as seen in the inset to Fig. \ref{fig:impurity fig}(b). Note that we only include batches that were used in the experiments, which is why indexing does not start from one.
\begin{table}[!h]
 \centering
 \noindent
 \begin{tabular}{|p{0.2\columnwidth} |p{0.2\columnwidth}|p{0.25\columnwidth}|p{0.2\columnwidth} |} 
 \hline
 Batch No. & $\Theta_{CW}$ (K) & $\mu_{eff}$ ($\mu_B/$Ir) & Mass (g) \\
 \hline\hline
 4 & 58.6 & 1.31 & 0.24  \\ 
 5 & 93.6 & 1.22 & 0.22  \\ 
 6 & 22.4 & 1.65 & 0.56 \\
 7 & 56.4 & 1.41 & 0.55 \\
 8 & 31.0 & 1.55 & 0.51 \\
 9 & 26.2 & 1.61 & 0.58 \\
 10 & 20.2 & 1.63 & 0.59 \\
 11 & 10.2 & 1.63 & 0.68 \\
 \hline
\end{tabular}
\caption{Magnetic properties of each batch of $^7$Li and $^{193}$Ir powder used in this work. }
\label{tab:impurity_table}
\end{table}

\vspace{1cm}
\bibliography{bLIO,sorsamp,reference}

\providecommand{\noopsort}[1]{}\providecommand{\singleletter}[1]{#1}%
\begin{thebibliography}{66}%
\makeatletter
\providecommand \@ifxundefined [1]{%
 \@ifx{#1\undefined}
}%
\providecommand \@ifnum [1]{%
 \ifnum #1\expandafter \@firstoftwo
 \else \expandafter \@secondoftwo
 \fi
}%
\providecommand \@ifx [1]{%
 \ifx #1\expandafter \@firstoftwo
 \else \expandafter \@secondoftwo
 \fi
}%
\providecommand \natexlab [1]{#1}%
\providecommand \enquote  [1]{``#1''}%
\providecommand \bibnamefont  [1]{#1}%
\providecommand \bibfnamefont [1]{#1}%
\providecommand \citenamefont [1]{#1}%
\providecommand \href@noop [0]{\@secondoftwo}%
\providecommand \href [0]{\begingroup \@sanitize@url \@href}%
\providecommand \@href[1]{\@@startlink{#1}\@@href}%
\providecommand \@@href[1]{\endgroup#1\@@endlink}%
\providecommand \@sanitize@url [0]{\catcode `\\12\catcode `\$12\catcode
  `\&12\catcode `\#12\catcode `\^12\catcode `\_12\catcode `\%12\relax}%
\providecommand \@@startlink[1]{}%
\providecommand \@@endlink[0]{}%
\providecommand \url  [0]{\begingroup\@sanitize@url \@url }%
\providecommand \@url [1]{\endgroup\@href {#1}{\urlprefix }}%
\providecommand \urlprefix  [0]{URL }%
\providecommand \Eprint [0]{\href }%
\providecommand \doibase [0]{http://dx.doi.org/}%
\providecommand \selectlanguage [0]{\@gobble}%
\providecommand \bibinfo  [0]{\@secondoftwo}%
\providecommand \bibfield  [0]{\@secondoftwo}%
\providecommand \translation [1]{[#1]}%
\providecommand \BibitemOpen [0]{}%
\providecommand \bibitemStop [0]{}%
\providecommand \bibitemNoStop [0]{.\EOS\space}%
\providecommand \EOS [0]{\spacefactor3000\relax}%
\providecommand \BibitemShut  [1]{\csname bibitem#1\endcsname}%
\let\auto@bib@innerbib\@empty
\bibitem [{\citenamefont {Winter}\ \emph
  {et~al.}(2017{\natexlab{a}})\citenamefont {Winter}, \citenamefont {Tsirlin},
  \citenamefont {Daghofer}, \citenamefont {Van Den~Brink}, \citenamefont
  {Singh}, \citenamefont {Gegenwart},\ and\ \citenamefont
  {Valent{\'{i}}}}]{Winter2017ModelsMagnetism}%
  \BibitemOpen
  \bibfield  {author} {\bibinfo {author} {\bibfnamefont {S.~M.}\ \bibnamefont
  {Winter}}, \bibinfo {author} {\bibfnamefont {A.~A.}\ \bibnamefont {Tsirlin}},
  \bibinfo {author} {\bibfnamefont {M.}~\bibnamefont {Daghofer}}, \bibinfo
  {author} {\bibfnamefont {J.}~\bibnamefont {Van Den~Brink}}, \bibinfo {author}
  {\bibfnamefont {Y.}~\bibnamefont {Singh}}, \bibinfo {author} {\bibfnamefont
  {P.}~\bibnamefont {Gegenwart}}, \ and\ \bibinfo {author} {\bibfnamefont
  {R.}~\bibnamefont {Valent{\'{i}}}},\ }\href {\doibase
  10.1088/1361-648X/aa8cf5} {\enquote {\bibinfo {title} {{Models and materials
  for generalized Kitaev magnetism}},}\ } (\bibinfo {year}
  {2017}{\natexlab{a}})\BibitemShut {NoStop}%
\bibitem [{\citenamefont {Rau}\ \emph {et~al.}(2016)\citenamefont {Rau},
  \citenamefont {Lee},\ and\ \citenamefont {Kee}}]{Rau2016Spin-OrbitMaterials}%
  \BibitemOpen
  \bibfield  {author} {\bibinfo {author} {\bibfnamefont {J.~G.}\ \bibnamefont
  {Rau}}, \bibinfo {author} {\bibfnamefont {E.~K.-H.}\ \bibnamefont {Lee}}, \
  and\ \bibinfo {author} {\bibfnamefont {H.-Y.}\ \bibnamefont {Kee}},\ }\href
  {\doibase 10.1146/annurev-conmatphys-031115-011319} {\bibfield  {journal}
  {\bibinfo  {journal} {Annual Review of Condensed Matter Physics}\ }\textbf
  {\bibinfo {volume} {7}},\ \bibinfo {pages} {195} (\bibinfo {year}
  {2016})}\BibitemShut {NoStop}%
\bibitem [{\citenamefont {Kitaev}(2006)}]{Kitaev2006AnyonsBeyond}%
  \BibitemOpen
  \bibfield  {author} {\bibinfo {author} {\bibfnamefont {A.}~\bibnamefont
  {Kitaev}},\ }\href {\doibase 10.1016/j.aop.2005.10.005} {\bibfield  {journal}
  {\bibinfo  {journal} {Annals of Physics}\ }\textbf {\bibinfo {volume}
  {321}},\ \bibinfo {pages} {2} (\bibinfo {year} {2006})}\BibitemShut {NoStop}%
\bibitem [{\citenamefont {Jackeli}\ and\ \citenamefont
  {Khaliullin}(2009)}]{Jackeli2009MottModels}%
  \BibitemOpen
  \bibfield  {author} {\bibinfo {author} {\bibfnamefont {G.}~\bibnamefont
  {Jackeli}}\ and\ \bibinfo {author} {\bibfnamefont {G.}~\bibnamefont
  {Khaliullin}},\ }\href {\doibase 10.1103/PhysRevLett.102.017205} {\bibfield
  {journal} {\bibinfo  {journal} {Physical Review Letters}\ }\textbf {\bibinfo
  {volume} {102}} (\bibinfo {year} {2009}),\
  10.1103/PhysRevLett.102.017205}\BibitemShut {NoStop}%
\bibitem [{\citenamefont {Kasahara}\ \emph {et~al.}(2018)\citenamefont
  {Kasahara}, \citenamefont {Sugii}, \citenamefont {Ohnishi}, \citenamefont
  {Shimozawa}, \citenamefont {Yamashita}, \citenamefont {Kurita}, \citenamefont
  {Tanaka}, \citenamefont {Nasu}, \citenamefont {Motome}, \citenamefont
  {Shibauchi},\ and\ \citenamefont {Matsuda}}]{Kasahara2018Unusual-RuCl3}%
  \BibitemOpen
  \bibfield  {author} {\bibinfo {author} {\bibfnamefont {Y.}~\bibnamefont
  {Kasahara}}, \bibinfo {author} {\bibfnamefont {K.}~\bibnamefont {Sugii}},
  \bibinfo {author} {\bibfnamefont {T.}~\bibnamefont {Ohnishi}}, \bibinfo
  {author} {\bibfnamefont {M.}~\bibnamefont {Shimozawa}}, \bibinfo {author}
  {\bibfnamefont {M.}~\bibnamefont {Yamashita}}, \bibinfo {author}
  {\bibfnamefont {N.}~\bibnamefont {Kurita}}, \bibinfo {author} {\bibfnamefont
  {H.}~\bibnamefont {Tanaka}}, \bibinfo {author} {\bibfnamefont
  {J.}~\bibnamefont {Nasu}}, \bibinfo {author} {\bibfnamefont {Y.}~\bibnamefont
  {Motome}}, \bibinfo {author} {\bibfnamefont {T.}~\bibnamefont {Shibauchi}}, \
  and\ \bibinfo {author} {\bibfnamefont {Y.}~\bibnamefont {Matsuda}},\ }\href
  {\doibase 10.1103/PhysRevLett.120.217205} {\bibfield  {journal} {\bibinfo
  {journal} {Physical Review Letters}\ }\textbf {\bibinfo {volume} {120}}
  (\bibinfo {year} {2018}),\ 10.1103/PhysRevLett.120.217205}\BibitemShut
  {NoStop}%
\bibitem [{\citenamefont {Banerjee}\ \emph {et~al.}(2018)\citenamefont
  {Banerjee}, \citenamefont {Lampen-Kelley}, \citenamefont {Knolle},
  \citenamefont {Balz}, \citenamefont {Aczel}, \citenamefont {Winn},
  \citenamefont {Liu}, \citenamefont {Pajerowski}, \citenamefont {Yan},
  \citenamefont {Bridges}, \citenamefont {Savici}, \citenamefont {Chakoumakos},
  \citenamefont {Lumsden}, \citenamefont {Tennant}, \citenamefont {Moessner},
  \citenamefont {Mandrus},\ and\ \citenamefont
  {Nagler}}]{Banerjee2018Excitations-RuCl3}%
  \BibitemOpen
  \bibfield  {author} {\bibinfo {author} {\bibfnamefont {A.}~\bibnamefont
  {Banerjee}}, \bibinfo {author} {\bibfnamefont {P.}~\bibnamefont
  {Lampen-Kelley}}, \bibinfo {author} {\bibfnamefont {J.}~\bibnamefont
  {Knolle}}, \bibinfo {author} {\bibfnamefont {C.}~\bibnamefont {Balz}},
  \bibinfo {author} {\bibfnamefont {A.~A.}\ \bibnamefont {Aczel}}, \bibinfo
  {author} {\bibfnamefont {B.}~\bibnamefont {Winn}}, \bibinfo {author}
  {\bibfnamefont {Y.}~\bibnamefont {Liu}}, \bibinfo {author} {\bibfnamefont
  {D.}~\bibnamefont {Pajerowski}}, \bibinfo {author} {\bibfnamefont
  {J.}~\bibnamefont {Yan}}, \bibinfo {author} {\bibfnamefont {C.~A.}\
  \bibnamefont {Bridges}}, \bibinfo {author} {\bibfnamefont {A.~T.}\
  \bibnamefont {Savici}}, \bibinfo {author} {\bibfnamefont {B.~C.}\
  \bibnamefont {Chakoumakos}}, \bibinfo {author} {\bibfnamefont {M.~D.}\
  \bibnamefont {Lumsden}}, \bibinfo {author} {\bibfnamefont {D.~A.}\
  \bibnamefont {Tennant}}, \bibinfo {author} {\bibfnamefont {R.}~\bibnamefont
  {Moessner}}, \bibinfo {author} {\bibfnamefont {D.~G.}\ \bibnamefont
  {Mandrus}}, \ and\ \bibinfo {author} {\bibfnamefont {S.~E.}\ \bibnamefont
  {Nagler}},\ }\href {\doibase 10.1038/s41535-018-0079-2} {\bibfield  {journal}
  {\bibinfo  {journal} {npj Quantum Materials}\ }\textbf {\bibinfo {volume}
  {3}} (\bibinfo {year} {2018}),\ 10.1038/s41535-018-0079-2}\BibitemShut
  {NoStop}%
\bibitem [{\citenamefont {Kitagawa}\ \emph {et~al.}(2018)\citenamefont
  {Kitagawa}, \citenamefont {Takayama}, \citenamefont {Matsumoto},
  \citenamefont {Kato}, \citenamefont {Takano}, \citenamefont {Kishimoto},
  \citenamefont {Dinnebier}, \citenamefont {Jackeli},\ and\ \citenamefont
  {Takagi}}]{Kitagawa2018}%
  \BibitemOpen
  \bibfield  {author} {\bibinfo {author} {\bibfnamefont {K.}~\bibnamefont
  {Kitagawa}}, \bibinfo {author} {\bibfnamefont {T.}~\bibnamefont {Takayama}},
  \bibinfo {author} {\bibfnamefont {Y.}~\bibnamefont {Matsumoto}}, \bibinfo
  {author} {\bibfnamefont {A.}~\bibnamefont {Kato}}, \bibinfo {author}
  {\bibfnamefont {R.}~\bibnamefont {Takano}}, \bibinfo {author} {\bibfnamefont
  {Y.}~\bibnamefont {Kishimoto}}, \bibinfo {author} {\bibfnamefont
  {R.}~\bibnamefont {Dinnebier}}, \bibinfo {author} {\bibfnamefont
  {G.}~\bibnamefont {Jackeli}}, \ and\ \bibinfo {author} {\bibfnamefont
  {H.}~\bibnamefont {Takagi}},\ }\href {\doibase
  http://dx.doi.org/10.1038/nature25482} {\bibfield  {journal} {\bibinfo
  {journal} {Nature}\ }\textbf {\bibinfo {volume} {554}},\ \bibinfo {pages}
  {341} (\bibinfo {year} {2018})}\BibitemShut {NoStop}%
\bibitem [{\citenamefont {Pei}\ \emph {et~al.}(2020)\citenamefont {Pei},
  \citenamefont {Huang}, \citenamefont {Li}, \citenamefont {Chen},
  \citenamefont {Xi}, \citenamefont {Wang}, \citenamefont {Shi}, \citenamefont
  {Yu}, \citenamefont {Liu}, \citenamefont {Wang}, \citenamefont {Ye},
  \citenamefont {Huang},\ and\ \citenamefont {Mei}}]{Pei2020MagneticO6}%
  \BibitemOpen
  \bibfield  {author} {\bibinfo {author} {\bibfnamefont {S.}~\bibnamefont
  {Pei}}, \bibinfo {author} {\bibfnamefont {L.~L.}\ \bibnamefont {Huang}},
  \bibinfo {author} {\bibfnamefont {G.}~\bibnamefont {Li}}, \bibinfo {author}
  {\bibfnamefont {X.}~\bibnamefont {Chen}}, \bibinfo {author} {\bibfnamefont
  {B.}~\bibnamefont {Xi}}, \bibinfo {author} {\bibfnamefont {X.~W.}\
  \bibnamefont {Wang}}, \bibinfo {author} {\bibfnamefont {Y.}~\bibnamefont
  {Shi}}, \bibinfo {author} {\bibfnamefont {D.}~\bibnamefont {Yu}}, \bibinfo
  {author} {\bibfnamefont {C.}~\bibnamefont {Liu}}, \bibinfo {author}
  {\bibfnamefont {L.}~\bibnamefont {Wang}}, \bibinfo {author} {\bibfnamefont
  {F.}~\bibnamefont {Ye}}, \bibinfo {author} {\bibfnamefont {M.}~\bibnamefont
  {Huang}}, \ and\ \bibinfo {author} {\bibfnamefont {J.~W.}\ \bibnamefont
  {Mei}},\ }\href {\doibase 10.1103/PhysRevB.101.201101} {\bibfield  {journal}
  {\bibinfo  {journal} {Physical Review B}\ }\textbf {\bibinfo {volume}
  {101}},\ \bibinfo {pages} {201101} (\bibinfo {year} {2020})}\BibitemShut
  {NoStop}%
\bibitem [{\citenamefont {Geirhos}\ \emph {et~al.}(2020)\citenamefont
  {Geirhos}, \citenamefont {Lunkenheimer}, \citenamefont {Blankenhorn},
  \citenamefont {Claus}, \citenamefont {Matsumoto}, \citenamefont {Kitagawa},
  \citenamefont {Takayama}, \citenamefont {Takagi}, \citenamefont
  {K{\'{e}}zsm{\'{a}}rki},\ and\ \citenamefont {Loidl}}]{Geirhos2020QuantumO6}%
  \BibitemOpen
  \bibfield  {author} {\bibinfo {author} {\bibfnamefont {K.}~\bibnamefont
  {Geirhos}}, \bibinfo {author} {\bibfnamefont {P.}~\bibnamefont
  {Lunkenheimer}}, \bibinfo {author} {\bibfnamefont {M.}~\bibnamefont
  {Blankenhorn}}, \bibinfo {author} {\bibfnamefont {R.}~\bibnamefont {Claus}},
  \bibinfo {author} {\bibfnamefont {Y.}~\bibnamefont {Matsumoto}}, \bibinfo
  {author} {\bibfnamefont {K.}~\bibnamefont {Kitagawa}}, \bibinfo {author}
  {\bibfnamefont {T.}~\bibnamefont {Takayama}}, \bibinfo {author}
  {\bibfnamefont {H.}~\bibnamefont {Takagi}}, \bibinfo {author} {\bibfnamefont
  {I.}~\bibnamefont {K{\'{e}}zsm{\'{a}}rki}}, \ and\ \bibinfo {author}
  {\bibfnamefont {A.}~\bibnamefont {Loidl}},\ }\href {\doibase
  10.1103/PhysRevB.101.184410} {\bibfield  {journal} {\bibinfo  {journal}
  {Physical Review B}\ }\textbf {\bibinfo {volume} {101}},\ \bibinfo {pages}
  {184410} (\bibinfo {year} {2020})}\BibitemShut {NoStop}%
\bibitem [{\citenamefont {Takayama}\ \emph {et~al.}(2015)\citenamefont
  {Takayama}, \citenamefont {Kato}, \citenamefont {Dinnebier}, \citenamefont
  {Nuss}, \citenamefont {Kono}, \citenamefont {Veiga}, \citenamefont {Fabbris},
  \citenamefont {Haskel},\ and\ \citenamefont
  {Takagi}}]{Takayama2015HyperhoneycombMagnetism}%
  \BibitemOpen
  \bibfield  {author} {\bibinfo {author} {\bibfnamefont {T.}~\bibnamefont
  {Takayama}}, \bibinfo {author} {\bibfnamefont {A.}~\bibnamefont {Kato}},
  \bibinfo {author} {\bibfnamefont {R.}~\bibnamefont {Dinnebier}}, \bibinfo
  {author} {\bibfnamefont {J.}~\bibnamefont {Nuss}}, \bibinfo {author}
  {\bibfnamefont {H.}~\bibnamefont {Kono}}, \bibinfo {author} {\bibfnamefont
  {L.~S.}\ \bibnamefont {Veiga}}, \bibinfo {author} {\bibfnamefont
  {G.}~\bibnamefont {Fabbris}}, \bibinfo {author} {\bibfnamefont
  {D.}~\bibnamefont {Haskel}}, \ and\ \bibinfo {author} {\bibfnamefont
  {H.}~\bibnamefont {Takagi}},\ }\href {\doibase
  10.1103/PhysRevLett.114.077202} {\bibfield  {journal} {\bibinfo  {journal}
  {Physical Review Letters}\ }\textbf {\bibinfo {volume} {114}} (\bibinfo
  {year} {2015}),\ 10.1103/PhysRevLett.114.077202}\BibitemShut {NoStop}%
\bibitem [{\citenamefont {Majumder}\ \emph
  {et~al.}(2019{\natexlab{a}})\citenamefont {Majumder}, \citenamefont {Freund},
  \citenamefont {Dey}, \citenamefont {Prinz-Zwick}, \citenamefont
  {B{\"{u}}ttgen}, \citenamefont {Skourski}, \citenamefont {Jesche},
  \citenamefont {Tsirlin},\ and\ \citenamefont
  {Gegenwart}}]{Majumder2019AnisotropicStudy}%
  \BibitemOpen
  \bibfield  {author} {\bibinfo {author} {\bibfnamefont {M.}~\bibnamefont
  {Majumder}}, \bibinfo {author} {\bibfnamefont {F.}~\bibnamefont {Freund}},
  \bibinfo {author} {\bibfnamefont {T.}~\bibnamefont {Dey}}, \bibinfo {author}
  {\bibfnamefont {M.}~\bibnamefont {Prinz-Zwick}}, \bibinfo {author}
  {\bibfnamefont {N.}~\bibnamefont {B{\"{u}}ttgen}}, \bibinfo {author}
  {\bibfnamefont {Y.}~\bibnamefont {Skourski}}, \bibinfo {author}
  {\bibfnamefont {A.}~\bibnamefont {Jesche}}, \bibinfo {author} {\bibfnamefont
  {A.~A.}\ \bibnamefont {Tsirlin}}, \ and\ \bibinfo {author} {\bibfnamefont
  {P.}~\bibnamefont {Gegenwart}},\ }\href {\doibase
  10.1103/PhysRevMaterials.3.074408} {\bibfield  {journal} {\bibinfo  {journal}
  {Physical Review Materials}\ }\textbf {\bibinfo {volume} {3}} (\bibinfo
  {year} {2019}{\natexlab{a}}),\ 10.1103/PhysRevMaterials.3.074408}\BibitemShut
  {NoStop}%
\bibitem [{\citenamefont {Ducatman}\ \emph
  {et~al.}(2018{\natexlab{a}})\citenamefont {Ducatman}, \citenamefont
  {Rousochatzakis},\ and\ \citenamefont
  {Perkins}}]{Ducatman2018Magnetic-Li2IrO3}%
  \BibitemOpen
  \bibfield  {author} {\bibinfo {author} {\bibfnamefont {S.}~\bibnamefont
  {Ducatman}}, \bibinfo {author} {\bibfnamefont {I.}~\bibnamefont
  {Rousochatzakis}}, \ and\ \bibinfo {author} {\bibfnamefont {N.~B.}\
  \bibnamefont {Perkins}},\ }\href {\doibase 10.1103/PhysRevB.97.125125}
  {\bibfield  {journal} {\bibinfo  {journal} {Physical Review B}\ }\textbf
  {\bibinfo {volume} {97}},\ \bibinfo {pages} {125125} (\bibinfo {year}
  {2018}{\natexlab{a}})}\BibitemShut {NoStop}%
\bibitem [{\citenamefont {Modic}\ \emph {et~al.}(2014)\citenamefont {Modic},
  \citenamefont {Smidt}, \citenamefont {Kimchi}, \citenamefont {Breznay},
  \citenamefont {Biffin}, \citenamefont {Choi}, \citenamefont {Johnson},
  \citenamefont {Coldea}, \citenamefont {Watkins-Curry}, \citenamefont
  {McCandless}, \citenamefont {Chan}, \citenamefont {Gandara}, \citenamefont
  {Islam}, \citenamefont {Vishwanath}, \citenamefont {Shekhter}, \citenamefont
  {McDonald},\ and\ \citenamefont {Analytis}}]{Modic2014RealizationIridate}%
  \BibitemOpen
  \bibfield  {author} {\bibinfo {author} {\bibfnamefont {K.~A.}\ \bibnamefont
  {Modic}}, \bibinfo {author} {\bibfnamefont {T.~E.}\ \bibnamefont {Smidt}},
  \bibinfo {author} {\bibfnamefont {I.}~\bibnamefont {Kimchi}}, \bibinfo
  {author} {\bibfnamefont {N.~P.}\ \bibnamefont {Breznay}}, \bibinfo {author}
  {\bibfnamefont {A.}~\bibnamefont {Biffin}}, \bibinfo {author} {\bibfnamefont
  {S.}~\bibnamefont {Choi}}, \bibinfo {author} {\bibfnamefont {R.~D.}\
  \bibnamefont {Johnson}}, \bibinfo {author} {\bibfnamefont {R.}~\bibnamefont
  {Coldea}}, \bibinfo {author} {\bibfnamefont {P.}~\bibnamefont
  {Watkins-Curry}}, \bibinfo {author} {\bibfnamefont {G.~T.}\ \bibnamefont
  {McCandless}}, \bibinfo {author} {\bibfnamefont {J.~Y.}\ \bibnamefont
  {Chan}}, \bibinfo {author} {\bibfnamefont {F.}~\bibnamefont {Gandara}},
  \bibinfo {author} {\bibfnamefont {Z.}~\bibnamefont {Islam}}, \bibinfo
  {author} {\bibfnamefont {A.}~\bibnamefont {Vishwanath}}, \bibinfo {author}
  {\bibfnamefont {A.}~\bibnamefont {Shekhter}}, \bibinfo {author}
  {\bibfnamefont {R.~D.}\ \bibnamefont {McDonald}}, \ and\ \bibinfo {author}
  {\bibfnamefont {J.~G.}\ \bibnamefont {Analytis}},\ }\href {\doibase
  10.1038/ncomms5203} {\bibfield  {journal} {\bibinfo  {journal} {Nature
  Communications}\ }\textbf {\bibinfo {volume} {5}} (\bibinfo {year} {2014}),\
  10.1038/ncomms5203}\BibitemShut {NoStop}%
\bibitem [{\citenamefont {Biffin}\ \emph
  {et~al.}(2014{\natexlab{a}})\citenamefont {Biffin}, \citenamefont {Johnson},
  \citenamefont {Kimchi}, \citenamefont {Morris}, \citenamefont {Bombardi},
  \citenamefont {Analytis}, \citenamefont {Vishwanath},\ and\ \citenamefont
  {Coldea}}]{Biffin2014Noncoplanar-li2iro3}%
  \BibitemOpen
  \bibfield  {author} {\bibinfo {author} {\bibfnamefont {A.}~\bibnamefont
  {Biffin}}, \bibinfo {author} {\bibfnamefont {R.~D.}\ \bibnamefont {Johnson}},
  \bibinfo {author} {\bibfnamefont {I.}~\bibnamefont {Kimchi}}, \bibinfo
  {author} {\bibfnamefont {R.}~\bibnamefont {Morris}}, \bibinfo {author}
  {\bibfnamefont {A.}~\bibnamefont {Bombardi}}, \bibinfo {author}
  {\bibfnamefont {J.~G.}\ \bibnamefont {Analytis}}, \bibinfo {author}
  {\bibfnamefont {A.}~\bibnamefont {Vishwanath}}, \ and\ \bibinfo {author}
  {\bibfnamefont {R.}~\bibnamefont {Coldea}},\ }\href {\doibase
  10.1103/PhysRevLett.113.197201} {\bibfield  {journal} {\bibinfo  {journal}
  {Physical Review Letters}\ }\textbf {\bibinfo {volume} {113}} (\bibinfo
  {year} {2014}{\natexlab{a}}),\ 10.1103/PhysRevLett.113.197201}\BibitemShut
  {NoStop}%
\bibitem [{\citenamefont {Biffin}\ \emph
  {et~al.}(2014{\natexlab{b}})\citenamefont {Biffin}, \citenamefont {Johnson},
  \citenamefont {Choi}, \citenamefont {Freund}, \citenamefont {Manni},
  \citenamefont {Bombardi}, \citenamefont {Manuel}, \citenamefont {Gegenwart},\
  and\ \citenamefont {Coldea}}]{Biffin2014a}%
  \BibitemOpen
  \bibfield  {author} {\bibinfo {author} {\bibfnamefont {A.}~\bibnamefont
  {Biffin}}, \bibinfo {author} {\bibfnamefont {R.~D.}\ \bibnamefont {Johnson}},
  \bibinfo {author} {\bibfnamefont {S.}~\bibnamefont {Choi}}, \bibinfo {author}
  {\bibfnamefont {F.}~\bibnamefont {Freund}}, \bibinfo {author} {\bibfnamefont
  {S.}~\bibnamefont {Manni}}, \bibinfo {author} {\bibfnamefont
  {A.}~\bibnamefont {Bombardi}}, \bibinfo {author} {\bibfnamefont
  {P.}~\bibnamefont {Manuel}}, \bibinfo {author} {\bibfnamefont
  {P.}~\bibnamefont {Gegenwart}}, \ and\ \bibinfo {author} {\bibfnamefont
  {R.}~\bibnamefont {Coldea}},\ }\href {\doibase 10.1103/PhysRevB.90.205116}
  {\bibfield  {journal} {\bibinfo  {journal} {Phys. Rev. B}\ }\textbf {\bibinfo
  {volume} {90}},\ \bibinfo {pages} {205116} (\bibinfo {year}
  {2014}{\natexlab{b}})}\BibitemShut {NoStop}%
\bibitem [{\citenamefont {Ruiz}\ \emph {et~al.}(2017)\citenamefont {Ruiz},
  \citenamefont {Frano}, \citenamefont {Breznay}, \citenamefont {Kimchi},
  \citenamefont {Helm}, \citenamefont {Oswald}, \citenamefont {Chan},
  \citenamefont {Birgeneau}, \citenamefont {Islam},\ and\ \citenamefont
  {Analytis}}]{Ruiz2017CorrelatedFields}%
  \BibitemOpen
  \bibfield  {author} {\bibinfo {author} {\bibfnamefont {A.}~\bibnamefont
  {Ruiz}}, \bibinfo {author} {\bibfnamefont {A.}~\bibnamefont {Frano}},
  \bibinfo {author} {\bibfnamefont {N.~P.}\ \bibnamefont {Breznay}}, \bibinfo
  {author} {\bibfnamefont {I.}~\bibnamefont {Kimchi}}, \bibinfo {author}
  {\bibfnamefont {T.}~\bibnamefont {Helm}}, \bibinfo {author} {\bibfnamefont
  {I.}~\bibnamefont {Oswald}}, \bibinfo {author} {\bibfnamefont {J.~Y.}\
  \bibnamefont {Chan}}, \bibinfo {author} {\bibfnamefont {R.~J.}\ \bibnamefont
  {Birgeneau}}, \bibinfo {author} {\bibfnamefont {Z.}~\bibnamefont {Islam}}, \
  and\ \bibinfo {author} {\bibfnamefont {J.~G.}\ \bibnamefont {Analytis}},\
  }\href {\doibase 10.1038/s41467-017-01071-9} {\bibfield  {journal} {\bibinfo
  {journal} {Nature Communications}\ }\textbf {\bibinfo {volume} {8}},\
  \bibinfo {pages} {1} (\bibinfo {year} {2017})}\BibitemShut {NoStop}%
\bibitem [{\citenamefont {Tsirlin}\ and\ \citenamefont
  {Gegenwart}(2021)}]{Tsirlin2021}%
  \BibitemOpen
  \bibfield  {author} {\bibinfo {author} {\bibfnamefont {A.~A.}\ \bibnamefont
  {Tsirlin}}\ and\ \bibinfo {author} {\bibfnamefont {P.}~\bibnamefont
  {Gegenwart}},\ }\href {https://doi.org/10.1002/pssb.202100146} {\bibfield
  {journal} {\bibinfo  {journal} {Phys. Status Solidi B}\ ,\ \bibinfo {pages}
  {2100146}} (\bibinfo {year} {2021})}\BibitemShut {NoStop}%
\bibitem [{\citenamefont {Winter}\ \emph
  {et~al.}(2017{\natexlab{b}})\citenamefont {Winter}, \citenamefont {Tsirlin},
  \citenamefont {Daghofer}, \citenamefont {van~den Brink}, \citenamefont
  {Singh}, \citenamefont {Gegenwart},\ and\ \citenamefont
  {Valenti­}}]{Winter2017}%
  \BibitemOpen
  \bibfield  {author} {\bibinfo {author} {\bibfnamefont {S.~M.}\ \bibnamefont
  {Winter}}, \bibinfo {author} {\bibfnamefont {A.~A.}\ \bibnamefont {Tsirlin}},
  \bibinfo {author} {\bibfnamefont {M.}~\bibnamefont {Daghofer}}, \bibinfo
  {author} {\bibfnamefont {J.}~\bibnamefont {van~den Brink}}, \bibinfo {author}
  {\bibfnamefont {Y.}~\bibnamefont {Singh}}, \bibinfo {author} {\bibfnamefont
  {P.}~\bibnamefont {Gegenwart}}, \ and\ \bibinfo {author} {\bibfnamefont
  {R.}~\bibnamefont {Valenti­}},\ }\href
  {http://stacks.iop.org/0953-8984/29/i=49/a=493002} {\bibfield  {journal}
  {\bibinfo  {journal} {J. Phys.: Condens. Matter}\ }\textbf {\bibinfo {volume}
  {29}},\ \bibinfo {pages} {493002} (\bibinfo {year}
  {2017}{\natexlab{b}})}\BibitemShut {NoStop}%
\bibitem [{\citenamefont {Kimchi}\ \emph {et~al.}(2015)\citenamefont {Kimchi},
  \citenamefont {Coldea},\ and\ \citenamefont
  {Vishwanath}}]{Kimchi2015UnifiedLi2IrO3}%
  \BibitemOpen
  \bibfield  {author} {\bibinfo {author} {\bibfnamefont {I.}~\bibnamefont
  {Kimchi}}, \bibinfo {author} {\bibfnamefont {R.}~\bibnamefont {Coldea}}, \
  and\ \bibinfo {author} {\bibfnamefont {A.}~\bibnamefont {Vishwanath}},\
  }\href {\doibase 10.1103/PhysRevB.91.245134} {\bibfield  {journal} {\bibinfo
  {journal} {Physical Review B - Condensed Matter and Materials Physics}\
  }\textbf {\bibinfo {volume} {91}},\ \bibinfo {pages} {245134} (\bibinfo
  {year} {2015})}\BibitemShut {NoStop}%
\bibitem [{\citenamefont {Kim}\ \emph {et~al.}(2016)\citenamefont {Kim},
  \citenamefont {Kim},\ and\ \citenamefont {Kee}}]{Kim2016RevealingLiquid}%
  \BibitemOpen
  \bibfield  {author} {\bibinfo {author} {\bibfnamefont {H.~S.}\ \bibnamefont
  {Kim}}, \bibinfo {author} {\bibfnamefont {Y.~B.}\ \bibnamefont {Kim}}, \ and\
  \bibinfo {author} {\bibfnamefont {H.~Y.}\ \bibnamefont {Kee}},\ }\href
  {\doibase 10.1103/PhysRevB.94.245127} {\bibfield  {journal} {\bibinfo
  {journal} {Physical Review B}\ }\textbf {\bibinfo {volume} {94}},\ \bibinfo
  {pages} {245127} (\bibinfo {year} {2016})}\BibitemShut {NoStop}%
\bibitem [{\citenamefont {Singh}\ \emph {et~al.}(2012)\citenamefont {Singh},
  \citenamefont {Manni}, \citenamefont {Reuther}, \citenamefont {Berlijn},
  \citenamefont {Thomale}, \citenamefont {Ku}, \citenamefont {Trebst},\ and\
  \citenamefont {Gegenwart}}]{Singh2012Relevance3}%
  \BibitemOpen
  \bibfield  {author} {\bibinfo {author} {\bibfnamefont {Y.}~\bibnamefont
  {Singh}}, \bibinfo {author} {\bibfnamefont {S.}~\bibnamefont {Manni}},
  \bibinfo {author} {\bibfnamefont {J.}~\bibnamefont {Reuther}}, \bibinfo
  {author} {\bibfnamefont {T.}~\bibnamefont {Berlijn}}, \bibinfo {author}
  {\bibfnamefont {R.}~\bibnamefont {Thomale}}, \bibinfo {author} {\bibfnamefont
  {W.}~\bibnamefont {Ku}}, \bibinfo {author} {\bibfnamefont {S.}~\bibnamefont
  {Trebst}}, \ and\ \bibinfo {author} {\bibfnamefont {P.}~\bibnamefont
  {Gegenwart}},\ }\href {\doibase 10.1103/PhysRevLett.108.127203} {\bibfield
  {journal} {\bibinfo  {journal} {Physical Review Letters}\ }\textbf {\bibinfo
  {volume} {108}} (\bibinfo {year} {2012}),\
  10.1103/PhysRevLett.108.127203}\BibitemShut {NoStop}%
\bibitem [{\citenamefont {Lee}\ and\ \citenamefont
  {Kim}(2015)}]{Lee2015TheoryIridates}%
  \BibitemOpen
  \bibfield  {author} {\bibinfo {author} {\bibfnamefont {E.~K.~H.}\
  \bibnamefont {Lee}}\ and\ \bibinfo {author} {\bibfnamefont {Y.~B.}\
  \bibnamefont {Kim}},\ }\href {\doibase 10.1103/PhysRevB.91.064407} {\bibfield
   {journal} {\bibinfo  {journal} {Physical Review B}\ }\textbf {\bibinfo
  {volume} {91}},\ \bibinfo {pages} {64407} (\bibinfo {year}
  {2015})}\BibitemShut {NoStop}%
\bibitem [{\citenamefont {Lee}\ \emph {et~al.}(2016)\citenamefont {Lee},
  \citenamefont {Rau},\ and\ \citenamefont {Kim}}]{Lee2016TwoMaterials}%
  \BibitemOpen
  \bibfield  {author} {\bibinfo {author} {\bibfnamefont {E.~K.~H.}\
  \bibnamefont {Lee}}, \bibinfo {author} {\bibfnamefont {J.~G.}\ \bibnamefont
  {Rau}}, \ and\ \bibinfo {author} {\bibfnamefont {Y.~B.}\ \bibnamefont
  {Kim}},\ }\href {\doibase 10.1103/PhysRevB.93.184420} {\bibfield  {journal}
  {\bibinfo  {journal} {Physical Review B}\ }\textbf {\bibinfo {volume} {93}},\
  \bibinfo {pages} {184420} (\bibinfo {year} {2016})}\BibitemShut {NoStop}%
\bibitem [{\citenamefont {Katukuri}\ \emph
  {et~al.}(2016{\natexlab{a}})\citenamefont {Katukuri}, \citenamefont {Yadav},
  \citenamefont {Hozoi}, \citenamefont {Nishimoto},\ and\ \citenamefont {Van
  Den~Brink}}]{Katukuri2016TheState}%
  \BibitemOpen
  \bibfield  {author} {\bibinfo {author} {\bibfnamefont {V.~M.}\ \bibnamefont
  {Katukuri}}, \bibinfo {author} {\bibfnamefont {R.}~\bibnamefont {Yadav}},
  \bibinfo {author} {\bibfnamefont {L.}~\bibnamefont {Hozoi}}, \bibinfo
  {author} {\bibfnamefont {S.}~\bibnamefont {Nishimoto}}, \ and\ \bibinfo
  {author} {\bibfnamefont {J.}~\bibnamefont {Van Den~Brink}},\ }\href {\doibase
  10.1038/srep29585} {\bibfield  {journal} {\bibinfo  {journal} {Scientific
  Reports}\ }\textbf {\bibinfo {volume} {6}} (\bibinfo {year}
  {2016}{\natexlab{a}}),\ 10.1038/srep29585}\BibitemShut {NoStop}%
\bibitem [{\citenamefont {Rousochatzakis}\ and\ \citenamefont
  {Perkins}(2018)}]{Rousochatzakis2018Magnetic-Li2IrO3}%
  \BibitemOpen
  \bibfield  {author} {\bibinfo {author} {\bibfnamefont {I.}~\bibnamefont
  {Rousochatzakis}}\ and\ \bibinfo {author} {\bibfnamefont {N.~B.}\
  \bibnamefont {Perkins}},\ }\href {\doibase 10.1103/PhysRevB.97.174423}
  {\bibfield  {journal} {\bibinfo  {journal} {Physical Review B}\ }\textbf
  {\bibinfo {volume} {97}},\ \bibinfo {pages} {174423} (\bibinfo {year}
  {2018})}\BibitemShut {NoStop}%
\bibitem [{\citenamefont {Rau}\ \emph {et~al.}(2014)\citenamefont {Rau},
  \citenamefont {Lee},\ and\ \citenamefont {Kee}}]{Rau2014GenericLimit}%
  \BibitemOpen
  \bibfield  {author} {\bibinfo {author} {\bibfnamefont {J.~G.}\ \bibnamefont
  {Rau}}, \bibinfo {author} {\bibfnamefont {E.~K.~H.}\ \bibnamefont {Lee}}, \
  and\ \bibinfo {author} {\bibfnamefont {H.~Y.}\ \bibnamefont {Kee}},\ }\href
  {\doibase 10.1103/PhysRevLett.112.077204} {\bibfield  {journal} {\bibinfo
  {journal} {Physical Review Letters}\ }\textbf {\bibinfo {volume} {112}}
  (\bibinfo {year} {2014}),\ 10.1103/PhysRevLett.112.077204}\BibitemShut
  {NoStop}%
\bibitem [{\citenamefont {Majumder}\ \emph {et~al.}(2018)\citenamefont
  {Majumder}, \citenamefont {Manna}, \citenamefont {Simutis}, \citenamefont
  {Orain}, \citenamefont {Dey}, \citenamefont {Freund}, \citenamefont {Jesche},
  \citenamefont {Khasanov}, \citenamefont {Biswas}, \citenamefont {Bykova},
  \citenamefont {Dubrovinskaia}, \citenamefont {Dubrovinsky}, \citenamefont
  {Yadav}, \citenamefont {Hozoi}, \citenamefont {Nishimoto}, \citenamefont
  {Tsirlin},\ and\ \citenamefont {Gegenwart}}]{Majumder2018Breakdown-Li2IrO3}%
  \BibitemOpen
  \bibfield  {author} {\bibinfo {author} {\bibfnamefont {M.}~\bibnamefont
  {Majumder}}, \bibinfo {author} {\bibfnamefont {R.~S.}\ \bibnamefont {Manna}},
  \bibinfo {author} {\bibfnamefont {G.}~\bibnamefont {Simutis}}, \bibinfo
  {author} {\bibfnamefont {J.~C.}\ \bibnamefont {Orain}}, \bibinfo {author}
  {\bibfnamefont {T.}~\bibnamefont {Dey}}, \bibinfo {author} {\bibfnamefont
  {F.}~\bibnamefont {Freund}}, \bibinfo {author} {\bibfnamefont
  {A.}~\bibnamefont {Jesche}}, \bibinfo {author} {\bibfnamefont
  {R.}~\bibnamefont {Khasanov}}, \bibinfo {author} {\bibfnamefont {P.~K.}\
  \bibnamefont {Biswas}}, \bibinfo {author} {\bibfnamefont {E.}~\bibnamefont
  {Bykova}}, \bibinfo {author} {\bibfnamefont {N.}~\bibnamefont
  {Dubrovinskaia}}, \bibinfo {author} {\bibfnamefont {L.~S.}\ \bibnamefont
  {Dubrovinsky}}, \bibinfo {author} {\bibfnamefont {R.}~\bibnamefont {Yadav}},
  \bibinfo {author} {\bibfnamefont {L.}~\bibnamefont {Hozoi}}, \bibinfo
  {author} {\bibfnamefont {S.}~\bibnamefont {Nishimoto}}, \bibinfo {author}
  {\bibfnamefont {A.~A.}\ \bibnamefont {Tsirlin}}, \ and\ \bibinfo {author}
  {\bibfnamefont {P.}~\bibnamefont {Gegenwart}},\ }\href {\doibase
  10.1103/PhysRevLett.120.237202} {\bibfield  {journal} {\bibinfo  {journal}
  {Physical Review Letters}\ }\textbf {\bibinfo {volume} {120}} (\bibinfo
  {year} {2018}),\ 10.1103/PhysRevLett.120.237202}\BibitemShut {NoStop}%
\bibitem [{\citenamefont {Yadav}\ \emph {et~al.}(2016)\citenamefont {Yadav},
  \citenamefont {Bogdanov}, \citenamefont {Katukuri}, \citenamefont
  {Nishimoto}, \citenamefont {Van Den~Brink},\ and\ \citenamefont
  {Hozoi}}]{Yadav2016Kitaev-RuCl3}%
  \BibitemOpen
  \bibfield  {author} {\bibinfo {author} {\bibfnamefont {R.}~\bibnamefont
  {Yadav}}, \bibinfo {author} {\bibfnamefont {N.~A.}\ \bibnamefont {Bogdanov}},
  \bibinfo {author} {\bibfnamefont {V.~M.}\ \bibnamefont {Katukuri}}, \bibinfo
  {author} {\bibfnamefont {S.}~\bibnamefont {Nishimoto}}, \bibinfo {author}
  {\bibfnamefont {J.}~\bibnamefont {Van Den~Brink}}, \ and\ \bibinfo {author}
  {\bibfnamefont {L.}~\bibnamefont {Hozoi}},\ }\href {\doibase
  10.1038/srep37925} {\bibfield  {journal} {\bibinfo  {journal} {Scientific
  Reports}\ }\textbf {\bibinfo {volume} {6}} (\bibinfo {year} {2016}),\
  10.1038/srep37925}\BibitemShut {NoStop}%
\bibitem [{\citenamefont {Ruiz}\ \emph {et~al.}(2021)\citenamefont {Ruiz},
  \citenamefont {Breznay}, \citenamefont {Li}, \citenamefont {Rousochatzakis},
  \citenamefont {Allen}, \citenamefont {Zinda}, \citenamefont {Nagarajan},
  \citenamefont {Lopez}, \citenamefont {Islam}, \citenamefont {Upton},
  \citenamefont {Kim}, \citenamefont {Said}, \citenamefont {Huang},
  \citenamefont {Gog}, \citenamefont {Casa}, \citenamefont {Birgeneau},
  \citenamefont {Koralek}, \citenamefont {Analytis}, \citenamefont {Perkins},\
  and\ \citenamefont {Frano}}]{Ruiz2021Magnon-spinon-Li2IrO3}%
  \BibitemOpen
  \bibfield  {author} {\bibinfo {author} {\bibfnamefont {A.}~\bibnamefont
  {Ruiz}}, \bibinfo {author} {\bibfnamefont {N.~P.}\ \bibnamefont {Breznay}},
  \bibinfo {author} {\bibfnamefont {M.}~\bibnamefont {Li}}, \bibinfo {author}
  {\bibfnamefont {I.}~\bibnamefont {Rousochatzakis}}, \bibinfo {author}
  {\bibfnamefont {A.}~\bibnamefont {Allen}}, \bibinfo {author} {\bibfnamefont
  {I.}~\bibnamefont {Zinda}}, \bibinfo {author} {\bibfnamefont
  {V.}~\bibnamefont {Nagarajan}}, \bibinfo {author} {\bibfnamefont
  {G.}~\bibnamefont {Lopez}}, \bibinfo {author} {\bibfnamefont
  {Z.}~\bibnamefont {Islam}}, \bibinfo {author} {\bibfnamefont {M.~H.}\
  \bibnamefont {Upton}}, \bibinfo {author} {\bibfnamefont {J.}~\bibnamefont
  {Kim}}, \bibinfo {author} {\bibfnamefont {A.~H.}\ \bibnamefont {Said}},
  \bibinfo {author} {\bibfnamefont {X.~R.}\ \bibnamefont {Huang}}, \bibinfo
  {author} {\bibfnamefont {T.}~\bibnamefont {Gog}}, \bibinfo {author}
  {\bibfnamefont {D.}~\bibnamefont {Casa}}, \bibinfo {author} {\bibfnamefont
  {R.~J.}\ \bibnamefont {Birgeneau}}, \bibinfo {author} {\bibfnamefont {J.~D.}\
  \bibnamefont {Koralek}}, \bibinfo {author} {\bibfnamefont {J.~G.}\
  \bibnamefont {Analytis}}, \bibinfo {author} {\bibfnamefont {N.~B.}\
  \bibnamefont {Perkins}}, \ and\ \bibinfo {author} {\bibfnamefont
  {A.}~\bibnamefont {Frano}},\ }\href {\doibase 10.1103/PhysRevB.103.184404}
  {\bibfield  {journal} {\bibinfo  {journal} {Physical Review B}\ }\textbf
  {\bibinfo {volume} {103}},\ \bibinfo {pages} {184404} (\bibinfo {year}
  {2021})}\BibitemShut {NoStop}%
\bibitem [{\citenamefont {Takayama}\ \emph {et~al.}(2019)\citenamefont
  {Takayama}, \citenamefont {Krajewska}, \citenamefont {Gibbs}, \citenamefont
  {Yaresko}, \citenamefont {Ishii}, \citenamefont {Yamaoka}, \citenamefont
  {Ishii}, \citenamefont {Hiraoka}, \citenamefont {Funnell}, \citenamefont
  {Bull},\ and\ \citenamefont {Takagi}}]{Takayama2019Pressure-induced-Li2IrO3}%
  \BibitemOpen
  \bibfield  {author} {\bibinfo {author} {\bibfnamefont {T.}~\bibnamefont
  {Takayama}}, \bibinfo {author} {\bibfnamefont {A.}~\bibnamefont {Krajewska}},
  \bibinfo {author} {\bibfnamefont {A.~S.}\ \bibnamefont {Gibbs}}, \bibinfo
  {author} {\bibfnamefont {A.~N.}\ \bibnamefont {Yaresko}}, \bibinfo {author}
  {\bibfnamefont {H.}~\bibnamefont {Ishii}}, \bibinfo {author} {\bibfnamefont
  {H.}~\bibnamefont {Yamaoka}}, \bibinfo {author} {\bibfnamefont
  {K.}~\bibnamefont {Ishii}}, \bibinfo {author} {\bibfnamefont
  {N.}~\bibnamefont {Hiraoka}}, \bibinfo {author} {\bibfnamefont {N.~P.}\
  \bibnamefont {Funnell}}, \bibinfo {author} {\bibfnamefont {C.~L.}\
  \bibnamefont {Bull}}, \ and\ \bibinfo {author} {\bibfnamefont
  {H.}~\bibnamefont {Takagi}},\ }\href {\doibase 10.1103/PhysRevB.99.125127}
  {\bibfield  {journal} {\bibinfo  {journal} {Physical Review B}\ }\textbf
  {\bibinfo {volume} {99}} (\bibinfo {year} {2019}),\
  10.1103/PhysRevB.99.125127}\BibitemShut {NoStop}%
\bibitem [{\citenamefont {Granroth}\ \emph {et~al.}(2010)\citenamefont
  {Granroth}, \citenamefont {Kolesnikov}, \citenamefont {Sherline},
  \citenamefont {Clancy}, \citenamefont {Ross}, \citenamefont {Ruff},
  \citenamefont {Gaulin},\ and\ \citenamefont
  {Nagler}}]{Granroth2010SEQUOIA:SNS}%
  \BibitemOpen
  \bibfield  {author} {\bibinfo {author} {\bibfnamefont {G.~E.}\ \bibnamefont
  {Granroth}}, \bibinfo {author} {\bibfnamefont {A.~I.}\ \bibnamefont
  {Kolesnikov}}, \bibinfo {author} {\bibfnamefont {T.~E.}\ \bibnamefont
  {Sherline}}, \bibinfo {author} {\bibfnamefont {J.~P.}\ \bibnamefont
  {Clancy}}, \bibinfo {author} {\bibfnamefont {K.~A.}\ \bibnamefont {Ross}},
  \bibinfo {author} {\bibfnamefont {J.~P.}\ \bibnamefont {Ruff}}, \bibinfo
  {author} {\bibfnamefont {B.~D.}\ \bibnamefont {Gaulin}}, \ and\ \bibinfo
  {author} {\bibfnamefont {S.~E.}\ \bibnamefont {Nagler}},\ }\href {\doibase
  10.1088/1742-6596/251/1/012058} {\bibfield  {journal} {\bibinfo  {journal}
  {Journal of Physics: Conference Series}\ }\textbf {\bibinfo {volume} {251}}
  (\bibinfo {year} {2010}),\ 10.1088/1742-6596/251/1/012058}\BibitemShut
  {NoStop}%
\bibitem [{\citenamefont {Schmitt}\ and\ \citenamefont
  {Ouladdiaf}(1998)}]{Schmitt1998AbsorptionDiffraction}%
  \BibitemOpen
  \bibfield  {author} {\bibinfo {author} {\bibfnamefont {D.}~\bibnamefont
  {Schmitt}}\ and\ \bibinfo {author} {\bibfnamefont {B.}~\bibnamefont
  {Ouladdiaf}},\ }\href {\doibase 10.1107/S0021889898002672} {\bibfield
  {journal} {\bibinfo  {journal} {Journal of Applied Crystallography}\ }\textbf
  {\bibinfo {volume} {31}},\ \bibinfo {pages} {620} (\bibinfo {year}
  {1998})}\BibitemShut {NoStop}%
\bibitem [{\citenamefont {Laurita}(2017)}]{Laurita2017LowMagnets}%
  \BibitemOpen
  \bibfield  {author} {\bibinfo {author} {\bibfnamefont {N.~J.}\ \bibnamefont
  {Laurita}},\ }\emph {\bibinfo {title} {{Low Energy Electrodynamics of Quantum
  Magnets}}},\ \href@noop {} {Ph.D. thesis},\ \bibinfo  {school} {Johns Hopkins
  University} (\bibinfo {year} {2017})\BibitemShut {NoStop}%
\bibitem [{\citenamefont {Chauhan}\ \emph {et~al.}(2020)\citenamefont
  {Chauhan}, \citenamefont {Mahmood}, \citenamefont {Changlani}, \citenamefont
  {Koohpayeh},\ and\ \citenamefont {Armitage}}]{Chauhan2020TunableChain}%
  \BibitemOpen
  \bibfield  {author} {\bibinfo {author} {\bibfnamefont {P.}~\bibnamefont
  {Chauhan}}, \bibinfo {author} {\bibfnamefont {F.}~\bibnamefont {Mahmood}},
  \bibinfo {author} {\bibfnamefont {H.~J.}\ \bibnamefont {Changlani}}, \bibinfo
  {author} {\bibfnamefont {S.~M.}\ \bibnamefont {Koohpayeh}}, \ and\ \bibinfo
  {author} {\bibfnamefont {N.~P.}\ \bibnamefont {Armitage}},\ }\href {\doibase
  10.1103/PhysRevLett.124.037203} {\bibfield  {journal} {\bibinfo  {journal}
  {Physical Review Letters}\ }\textbf {\bibinfo {volume} {124}} (\bibinfo
  {year} {2020}),\ 10.1103/PhysRevLett.124.037203}\BibitemShut {NoStop}%
\bibitem [{\citenamefont {Er}\ \emph {et~al.}()\citenamefont {Er},
  \citenamefont {Cqrak}, \citenamefont {Garfunkel}, \citenamefont
  {Bsattkrthwaite},\ and\ \citenamefont {WExLER}}]{ErPH5K}%
  \BibitemOpen
  \bibfield  {author} {\bibinfo {author} {\bibfnamefont {N.~B.}\ \bibnamefont
  {Er}}, \bibinfo {author} {\bibfnamefont {W.~S.}\ \bibnamefont {Cqrak}},
  \bibinfo {author} {\bibfnamefont {M.~P.}\ \bibnamefont {Garfunkel}}, \bibinfo
  {author} {\bibfnamefont {C.}~\bibnamefont {Bsattkrthwaite}}, \ and\ \bibinfo
  {author} {\bibfnamefont {A.}~\bibnamefont {WExLER}},\ }\href@noop {} {\emph
  {\bibinfo {title} {{PH YSICAL REVIEW Atomic Heats of Copper, Silver, and Gold
  from 1'K to 5'K*}}}},\ \bibinfo {type} {Tech. Rep.}\BibitemShut {Stop}%
\bibitem [{\citenamefont {Meads}\ \emph {et~al.}()\citenamefont {Meads},
  \citenamefont {Forsythe}, \citenamefont {Giauque},\ and\ \citenamefont
  {63}}]{MeadsThe300K}%
  \BibitemOpen
  \bibfield  {author} {\bibinfo {author} {\bibfnamefont {P.~F.}\ \bibnamefont
  {Meads}}, \bibinfo {author} {\bibfnamefont {W.~R.}\ \bibnamefont {Forsythe}},
  \bibinfo {author} {\bibfnamefont {W.~F.}\ \bibnamefont {Giauque}}, \ and\
  \bibinfo {author} {\bibfnamefont {V.}~\bibnamefont {63}},\ }\href
  {https://pubs.acs.org/sharingguidelines} {\emph {\bibinfo {title} {{The Heat
  Capacities and Entropies of Silver and Lead from 15{${}^\circ$} to
  300{${}^\circ$}K}}}},\ \bibinfo {type} {Tech. Rep.}\BibitemShut {Stop}%
\bibitem [{\citenamefont {Bldg~mi}()}]{BldgmiLibraryLiterature}%
  \BibitemOpen
  \bibfield  {author} {\bibinfo {author} {\bibfnamefont {I.}~\bibnamefont
  {Bldg~mi}},\ }\href {\doibase 10.6028/NBS.MONO.21} {\emph {\bibinfo {title}
  {{Library, OCT 2 6 t of Standards Specific Heats and Entiialpies of J
  Technical Solids at Low Temperatures A Compilation From the Literature}}}},\
  \bibinfo {type} {Tech. Rep.}\BibitemShut {Stop}%
\bibitem [{\citenamefont {Knolle}\ \emph {et~al.}(2014)\citenamefont {Knolle},
  \citenamefont {Kovrizhin}, \citenamefont {Chalker},\ and\ \citenamefont
  {Moessner}}]{Knolle2014DynamicsFluxes}%
  \BibitemOpen
  \bibfield  {author} {\bibinfo {author} {\bibfnamefont {J.}~\bibnamefont
  {Knolle}}, \bibinfo {author} {\bibfnamefont {D.~L.}\ \bibnamefont
  {Kovrizhin}}, \bibinfo {author} {\bibfnamefont {J.~T.}\ \bibnamefont
  {Chalker}}, \ and\ \bibinfo {author} {\bibfnamefont {R.}~\bibnamefont
  {Moessner}},\ }\href {\doibase 10.1103/PhysRevLett.112.207203} {\bibfield
  {journal} {\bibinfo  {journal} {Physical Review Letters}\ }\textbf {\bibinfo
  {volume} {112}},\ \bibinfo {pages} {207203} (\bibinfo {year}
  {2014})}\BibitemShut {NoStop}%
\bibitem [{\citenamefont {Smith}\ \emph {et~al.}(2015)\citenamefont {Smith},
  \citenamefont {Knolle}, \citenamefont {Kovrizhin}, \citenamefont {Chalker},\
  and\ \citenamefont {Moessner}}]{Smith2015NeutronLiquid}%
  \BibitemOpen
  \bibfield  {author} {\bibinfo {author} {\bibfnamefont {A.}~\bibnamefont
  {Smith}}, \bibinfo {author} {\bibfnamefont {J.}~\bibnamefont {Knolle}},
  \bibinfo {author} {\bibfnamefont {D.~L.}\ \bibnamefont {Kovrizhin}}, \bibinfo
  {author} {\bibfnamefont {J.~T.}\ \bibnamefont {Chalker}}, \ and\ \bibinfo
  {author} {\bibfnamefont {R.}~\bibnamefont {Moessner}},\ }\href {\doibase
  10.1103/PhysRevB.92.180408} {\bibfield  {journal} {\bibinfo  {journal} {RAPID
  COMMUNICATIONS PHYSICAL REVIEW B}\ }\textbf {\bibinfo {volume} {92}},\
  \bibinfo {pages} {180408} (\bibinfo {year} {2015})}\BibitemShut {NoStop}%
\bibitem [{\citenamefont {Williams}\ \emph {et~al.}(2016)\citenamefont
  {Williams}, \citenamefont {Johnson}, \citenamefont {Freund}, \citenamefont
  {Choi}, \citenamefont {Jesche}, \citenamefont {Kimchi}, \citenamefont
  {Manni}, \citenamefont {Bombardi}, \citenamefont {Manuel}, \citenamefont
  {Gegenwart},\ and\ \citenamefont {Coldea}}]{Williams2016}%
  \BibitemOpen
  \bibfield  {author} {\bibinfo {author} {\bibfnamefont {S.~C.}\ \bibnamefont
  {Williams}}, \bibinfo {author} {\bibfnamefont {R.~D.}\ \bibnamefont
  {Johnson}}, \bibinfo {author} {\bibfnamefont {F.}~\bibnamefont {Freund}},
  \bibinfo {author} {\bibfnamefont {S.}~\bibnamefont {Choi}}, \bibinfo {author}
  {\bibfnamefont {A.}~\bibnamefont {Jesche}}, \bibinfo {author} {\bibfnamefont
  {I.}~\bibnamefont {Kimchi}}, \bibinfo {author} {\bibfnamefont
  {S.}~\bibnamefont {Manni}}, \bibinfo {author} {\bibfnamefont
  {A.}~\bibnamefont {Bombardi}}, \bibinfo {author} {\bibfnamefont
  {P.}~\bibnamefont {Manuel}}, \bibinfo {author} {\bibfnamefont
  {P.}~\bibnamefont {Gegenwart}}, \ and\ \bibinfo {author} {\bibfnamefont
  {R.}~\bibnamefont {Coldea}},\ }\href {\doibase 10.1103/PhysRevB.93.195158}
  {\bibfield  {journal} {\bibinfo  {journal} {Phys. Rev. B}\ }\textbf {\bibinfo
  {volume} {93}},\ \bibinfo {pages} {195158} (\bibinfo {year}
  {2016})}\BibitemShut {NoStop}%
\bibitem [{\citenamefont {Liu}\ \emph {et~al.}(2020)\citenamefont {Liu},
  \citenamefont {Chaloupka},\ and\ \citenamefont
  {Khaliullin}}]{Liu2020KitaevCompounds}%
  \BibitemOpen
  \bibfield  {author} {\bibinfo {author} {\bibfnamefont {H.}~\bibnamefont
  {Liu}}, \bibinfo {author} {\bibfnamefont {J.}~\bibnamefont {Chaloupka}}, \
  and\ \bibinfo {author} {\bibfnamefont {G.}~\bibnamefont {Khaliullin}},\
  }\href {\doibase 10.1103/PhysRevLett.125.047201} {\bibfield  {journal}
  {\bibinfo  {journal} {Physical Review Letters}\ }\textbf {\bibinfo {volume}
  {125}},\ \bibinfo {pages} {047201} (\bibinfo {year} {2020})}\BibitemShut
  {NoStop}%
\bibitem [{\citenamefont {Li}\ \emph {et~al.}(2021)\citenamefont {Li},
  \citenamefont {Winter}, \citenamefont {Kaib}, \citenamefont {Riedl},\ and\
  \citenamefont {Valent{\'{i}}}}]{Li2021ModifiedMagnets}%
  \BibitemOpen
  \bibfield  {author} {\bibinfo {author} {\bibfnamefont {Y.}~\bibnamefont
  {Li}}, \bibinfo {author} {\bibfnamefont {S.~M.}\ \bibnamefont {Winter}},
  \bibinfo {author} {\bibfnamefont {D.~A.}\ \bibnamefont {Kaib}}, \bibinfo
  {author} {\bibfnamefont {K.}~\bibnamefont {Riedl}}, \ and\ \bibinfo {author}
  {\bibfnamefont {R.}~\bibnamefont {Valent{\'{i}}}},\ }\href {\doibase
  10.1103/PhysRevB.103.L220408} {\bibfield  {journal} {\bibinfo  {journal}
  {Physical Review B}\ }\textbf {\bibinfo {volume} {103}},\ \bibinfo {pages}
  {L220408} (\bibinfo {year} {2021})}\BibitemShut {NoStop}%
\bibitem [{\citenamefont {Li}\ \emph {et~al.}(2020{\natexlab{a}})\citenamefont
  {Li}, \citenamefont {Rousochatzakis},\ and\ \citenamefont
  {Perkins}}]{Li2020Torque}%
  \BibitemOpen
  \bibfield  {author} {\bibinfo {author} {\bibfnamefont {M.}~\bibnamefont
  {Li}}, \bibinfo {author} {\bibfnamefont {I.}~\bibnamefont {Rousochatzakis}},
  \ and\ \bibinfo {author} {\bibfnamefont {N.~B.}\ \bibnamefont {Perkins}},\
  }\href {\doibase 10.1103/PhysRevResearch.2.033328} {\bibfield  {journal}
  {\bibinfo  {journal} {Phys. Rev. Research}\ }\textbf {\bibinfo {volume}
  {2}},\ \bibinfo {pages} {033328} (\bibinfo {year}
  {2020}{\natexlab{a}})}\BibitemShut {NoStop}%
\bibitem [{\citenamefont {Ducatman}\ \emph
  {et~al.}(2018{\natexlab{b}})\citenamefont {Ducatman}, \citenamefont
  {Rousochatzakis},\ and\ \citenamefont {Perkins}}]{Ducatman2018}%
  \BibitemOpen
  \bibfield  {author} {\bibinfo {author} {\bibfnamefont {S.}~\bibnamefont
  {Ducatman}}, \bibinfo {author} {\bibfnamefont {I.}~\bibnamefont
  {Rousochatzakis}}, \ and\ \bibinfo {author} {\bibfnamefont {N.~B.}\
  \bibnamefont {Perkins}},\ }\href
  {https://link.aps.org/doi/10.1103/PhysRevB.97.125125} {\bibfield  {journal}
  {\bibinfo  {journal} {Phys. Rev. B}\ }\textbf {\bibinfo {volume} {97}},\
  \bibinfo {pages} {125125} (\bibinfo {year} {2018}{\natexlab{b}})}\BibitemShut
  {NoStop}%
\bibitem [{\citenamefont {Li}\ \emph {et~al.}(2020{\natexlab{b}})\citenamefont
  {Li}, \citenamefont {Rousochatzakis},\ and\ \citenamefont
  {Perkins}}]{Li2020}%
  \BibitemOpen
  \bibfield  {author} {\bibinfo {author} {\bibfnamefont {M.}~\bibnamefont
  {Li}}, \bibinfo {author} {\bibfnamefont {I.}~\bibnamefont {Rousochatzakis}},
  \ and\ \bibinfo {author} {\bibfnamefont {N.~B.}\ \bibnamefont {Perkins}},\
  }\href {\doibase 10.1103/PhysRevResearch.2.013065} {\bibfield  {journal}
  {\bibinfo  {journal} {Phys. Rev. Research}\ }\textbf {\bibinfo {volume}
  {2}},\ \bibinfo {pages} {013065} (\bibinfo {year}
  {2020}{\natexlab{b}})}\BibitemShut {NoStop}%
\bibitem [{\citenamefont {Yang}\ \emph {et~al.}(2022)\citenamefont {Yang},
  \citenamefont {Wang}, \citenamefont {Rousochatzakis}, \citenamefont {Ruiz},
  \citenamefont {Analytis}, \citenamefont {Burch},\ and\ \citenamefont
  {Perkins}}]{Yang2022Signaturesbeta-Li_2IrO_3}%
  \BibitemOpen
  \bibfield  {author} {\bibinfo {author} {\bibfnamefont {Y.}~\bibnamefont
  {Yang}}, \bibinfo {author} {\bibfnamefont {Y.}~\bibnamefont {Wang}}, \bibinfo
  {author} {\bibfnamefont {I.}~\bibnamefont {Rousochatzakis}}, \bibinfo
  {author} {\bibfnamefont {A.}~\bibnamefont {Ruiz}}, \bibinfo {author}
  {\bibfnamefont {J.~G.}\ \bibnamefont {Analytis}}, \bibinfo {author}
  {\bibfnamefont {K.~S.}\ \bibnamefont {Burch}}, \ and\ \bibinfo {author}
  {\bibfnamefont {N.~B.}\ \bibnamefont {Perkins}},\ }\href
  {http://arxiv.org/abs/2202.00581} {\  (\bibinfo {year} {2022})}\BibitemShut
  {NoStop}%
\bibitem [{\citenamefont {Majumder}\ \emph
  {et~al.}(2019{\natexlab{b}})\citenamefont {Majumder}, \citenamefont {Freund},
  \citenamefont {Dey}, \citenamefont {Prinz-Zwick}, \citenamefont {B\"uttgen},
  \citenamefont {Skourski}, \citenamefont {Jesche}, \citenamefont {Tsirlin},\
  and\ \citenamefont {Gegenwart}}]{Majumder2019}%
  \BibitemOpen
  \bibfield  {author} {\bibinfo {author} {\bibfnamefont {M.}~\bibnamefont
  {Majumder}}, \bibinfo {author} {\bibfnamefont {F.}~\bibnamefont {Freund}},
  \bibinfo {author} {\bibfnamefont {T.}~\bibnamefont {Dey}}, \bibinfo {author}
  {\bibfnamefont {M.}~\bibnamefont {Prinz-Zwick}}, \bibinfo {author}
  {\bibfnamefont {N.}~\bibnamefont {B\"uttgen}}, \bibinfo {author}
  {\bibfnamefont {Y.}~\bibnamefont {Skourski}}, \bibinfo {author}
  {\bibfnamefont {A.}~\bibnamefont {Jesche}}, \bibinfo {author} {\bibfnamefont
  {A.~A.}\ \bibnamefont {Tsirlin}}, \ and\ \bibinfo {author} {\bibfnamefont
  {P.}~\bibnamefont {Gegenwart}},\ }\href {\doibase
  10.1103/PhysRevMaterials.3.074408} {\bibfield  {journal} {\bibinfo  {journal}
  {Phys. Rev. Materials}\ }\textbf {\bibinfo {volume} {3}},\ \bibinfo {pages}
  {074408} (\bibinfo {year} {2019}{\natexlab{b}})}\BibitemShut {NoStop}%
\bibitem [{\citenamefont {Stavropoulos}\ \emph {et~al.}(2018)\citenamefont
  {Stavropoulos}, \citenamefont {Catuneanu},\ and\ \citenamefont
  {Kee}}]{Stavropoulos2018Counter-rotatingModel}%
  \BibitemOpen
  \bibfield  {author} {\bibinfo {author} {\bibfnamefont {P.~P.}\ \bibnamefont
  {Stavropoulos}}, \bibinfo {author} {\bibfnamefont {A.}~\bibnamefont
  {Catuneanu}}, \ and\ \bibinfo {author} {\bibfnamefont {H.~Y.}\ \bibnamefont
  {Kee}},\ }\href {\doibase 10.1103/PhysRevB.98.104401} {\bibfield  {journal}
  {\bibinfo  {journal} {Physical Review B}\ }\textbf {\bibinfo {volume} {98}}
  (\bibinfo {year} {2018}),\ 10.1103/PhysRevB.98.104401}\BibitemShut {NoStop}%
\bibitem [{\citenamefont {Katukuri}\ \emph
  {et~al.}(2016{\natexlab{b}})\citenamefont {Katukuri}, \citenamefont {Yadav},
  \citenamefont {Hozoi}, \citenamefont {Nishimoto},\ and\ \citenamefont
  {van~den Brink}}]{Katukuri2016}%
  \BibitemOpen
  \bibfield  {author} {\bibinfo {author} {\bibfnamefont {V.~M.}\ \bibnamefont
  {Katukuri}}, \bibinfo {author} {\bibfnamefont {R.}~\bibnamefont {Yadav}},
  \bibinfo {author} {\bibfnamefont {L.}~\bibnamefont {Hozoi}}, \bibinfo
  {author} {\bibfnamefont {S.}~\bibnamefont {Nishimoto}}, \ and\ \bibinfo
  {author} {\bibfnamefont {J.}~\bibnamefont {van~den Brink}},\ }\href
  {http://dx.doi.org/10.1038/srep29585} {\bibfield  {journal} {\bibinfo
  {journal} {Sci. Rep.}\ }\textbf {\bibinfo {volume} {6}},\ \bibinfo {pages}
  {29585} (\bibinfo {year} {2016}{\natexlab{b}})}\BibitemShut {NoStop}%
\bibitem [{\citenamefont {Scheie}\ \emph {et~al.}(2019)\citenamefont {Scheie},
  \citenamefont {Dasgupta}, \citenamefont {Sanders}, \citenamefont {Sakai},
  \citenamefont {Matsumoto}, \citenamefont {Prisk}, \citenamefont {Nakatsuji},
  \citenamefont {Cava},\ and\ \citenamefont
  {Broholm}}]{Scheie2019HomogeneousAntiferromagnet}%
  \BibitemOpen
  \bibfield  {author} {\bibinfo {author} {\bibfnamefont {A.}~\bibnamefont
  {Scheie}}, \bibinfo {author} {\bibfnamefont {S.}~\bibnamefont {Dasgupta}},
  \bibinfo {author} {\bibfnamefont {M.}~\bibnamefont {Sanders}}, \bibinfo
  {author} {\bibfnamefont {A.}~\bibnamefont {Sakai}}, \bibinfo {author}
  {\bibfnamefont {Y.}~\bibnamefont {Matsumoto}}, \bibinfo {author}
  {\bibfnamefont {T.~R.}\ \bibnamefont {Prisk}}, \bibinfo {author}
  {\bibfnamefont {S.}~\bibnamefont {Nakatsuji}}, \bibinfo {author}
  {\bibfnamefont {R.~J.}\ \bibnamefont {Cava}}, \ and\ \bibinfo {author}
  {\bibfnamefont {C.}~\bibnamefont {Broholm}},\ }\href {\doibase
  10.1103/PhysRevB.100.024414} {\bibfield  {journal} {\bibinfo  {journal}
  {Physical Review B}\ }\textbf {\bibinfo {volume} {100}},\ \bibinfo {pages}
  {24414} (\bibinfo {year} {2019})}\BibitemShut {NoStop}%
\bibitem [{\citenamefont {Oshikawa}\ \emph {et~al.}(1999)\citenamefont
  {Oshikawa}, \citenamefont {Turnbull},\ and\ \citenamefont
  {Landee}}]{Oshikawa1999CharacterizationPresented}%
  \BibitemOpen
  \bibfield  {author} {\bibinfo {author} {\bibfnamefont {M.}~\bibnamefont
  {Oshikawa}}, \bibinfo {author} {\bibfnamefont {M.~M.}\ \bibnamefont
  {Turnbull}}, \ and\ \bibinfo {author} {\bibfnamefont {C.~P.}\ \bibnamefont
  {Landee}},\ }\href {\doibase 10.1103/PhysRevB.59.1008} {\bibfield  {journal}
  {\bibinfo  {journal} {Physical Review B - Condensed Matter and Materials
  Physics}\ }\textbf {\bibinfo {volume} {59}},\ \bibinfo {pages} {1008}
  (\bibinfo {year} {1999})}\BibitemShut {NoStop}%
\bibitem [{\citenamefont {Yamashita}\ \emph {et~al.}(2008)\citenamefont
  {Yamashita}, \citenamefont {Nakazawa}, \citenamefont {Oguni}, \citenamefont
  {Oshima}, \citenamefont {Nojiri}, \citenamefont {Shimizu}, \citenamefont
  {Miyagawa},\ and\ \citenamefont {Kanoda}}]{Yamashita2008ThermodynamicSalt}%
  \BibitemOpen
  \bibfield  {author} {\bibinfo {author} {\bibfnamefont {S.}~\bibnamefont
  {Yamashita}}, \bibinfo {author} {\bibfnamefont {Y.}~\bibnamefont {Nakazawa}},
  \bibinfo {author} {\bibfnamefont {M.}~\bibnamefont {Oguni}}, \bibinfo
  {author} {\bibfnamefont {Y.}~\bibnamefont {Oshima}}, \bibinfo {author}
  {\bibfnamefont {H.}~\bibnamefont {Nojiri}}, \bibinfo {author} {\bibfnamefont
  {Y.}~\bibnamefont {Shimizu}}, \bibinfo {author} {\bibfnamefont
  {K.}~\bibnamefont {Miyagawa}}, \ and\ \bibinfo {author} {\bibfnamefont
  {K.}~\bibnamefont {Kanoda}},\ }\href {\doibase 10.1038/nphys942} {\bibfield
  {journal} {\bibinfo  {journal} {Nature Physics}\ }\textbf {\bibinfo {volume}
  {4}},\ \bibinfo {pages} {459} (\bibinfo {year} {2008})}\BibitemShut {NoStop}%
\bibitem [{\citenamefont {Bourgeois-Hope}\ \emph {et~al.}(2019)\citenamefont
  {Bourgeois-Hope}, \citenamefont {Lalibert{\'{e}}}, \citenamefont
  {Lefran{\c{c}}ois}, \citenamefont {Grissonnanche}, \citenamefont {De~Cotret},
  \citenamefont {Gordon}, \citenamefont {Kitou}, \citenamefont {Sawa},
  \citenamefont {Cui}, \citenamefont {Kato}, \citenamefont {Taillefer},\ and\
  \citenamefont {Doiron-Leyraud}}]{taillefer}%
  \BibitemOpen
  \bibfield  {author} {\bibinfo {author} {\bibfnamefont {P.}~\bibnamefont
  {Bourgeois-Hope}}, \bibinfo {author} {\bibfnamefont {F.}~\bibnamefont
  {Lalibert{\'{e}}}}, \bibinfo {author} {\bibfnamefont {E.}~\bibnamefont
  {Lefran{\c{c}}ois}}, \bibinfo {author} {\bibfnamefont {G.}~\bibnamefont
  {Grissonnanche}}, \bibinfo {author} {\bibfnamefont {S.~R.}\ \bibnamefont
  {De~Cotret}}, \bibinfo {author} {\bibfnamefont {R.}~\bibnamefont {Gordon}},
  \bibinfo {author} {\bibfnamefont {S.}~\bibnamefont {Kitou}}, \bibinfo
  {author} {\bibfnamefont {H.}~\bibnamefont {Sawa}}, \bibinfo {author}
  {\bibfnamefont {H.}~\bibnamefont {Cui}}, \bibinfo {author} {\bibfnamefont
  {R.}~\bibnamefont {Kato}}, \bibinfo {author} {\bibfnamefont {L.}~\bibnamefont
  {Taillefer}}, \ and\ \bibinfo {author} {\bibfnamefont {N.}~\bibnamefont
  {Doiron-Leyraud}},\ }\href {\doibase 10.1103/PhysRevX.9.041051} {\bibfield
  {journal} {\bibinfo  {journal} {Physical Review X}\ }\textbf {\bibinfo
  {volume} {91}} (\bibinfo {year} {2019}),\
  10.1103/PhysRevX.9.041051}\BibitemShut {NoStop}%
\bibitem [{\citenamefont {Winter}\ \emph
  {et~al.}(2017{\natexlab{c}})\citenamefont {Winter}, \citenamefont {Riedl},
  \citenamefont {Maksimov}, \citenamefont {Chernyshev}, \citenamefont
  {Honecker},\ and\ \citenamefont {Valent{\'{i}}}}]{Winter2017BreakdownMagnet}%
  \BibitemOpen
  \bibfield  {author} {\bibinfo {author} {\bibfnamefont {S.~M.}\ \bibnamefont
  {Winter}}, \bibinfo {author} {\bibfnamefont {K.}~\bibnamefont {Riedl}},
  \bibinfo {author} {\bibfnamefont {P.~A.}\ \bibnamefont {Maksimov}}, \bibinfo
  {author} {\bibfnamefont {A.~L.}\ \bibnamefont {Chernyshev}}, \bibinfo
  {author} {\bibfnamefont {A.}~\bibnamefont {Honecker}}, \ and\ \bibinfo
  {author} {\bibfnamefont {R.}~\bibnamefont {Valent{\'{i}}}},\ }\href {\doibase
  10.1038/s41467-017-01177-0} {\bibfield  {journal} {\bibinfo  {journal}
  {Nature Communications}\ }\textbf {\bibinfo {volume} {8}} (\bibinfo {year}
  {2017}{\natexlab{c}}),\ 10.1038/s41467-017-01177-0}\BibitemShut {NoStop}%
\bibitem [{\citenamefont {Wang}\ \emph {et~al.}(2019)\citenamefont {Wang},
  \citenamefont {Normand},\ and\ \citenamefont {Liu}}]{Wang2019OneLattice}%
  \BibitemOpen
  \bibfield  {author} {\bibinfo {author} {\bibfnamefont {J.}~\bibnamefont
  {Wang}}, \bibinfo {author} {\bibfnamefont {B.}~\bibnamefont {Normand}}, \
  and\ \bibinfo {author} {\bibfnamefont {Z.~X.}\ \bibnamefont {Liu}},\ }\href
  {\doibase 10.1103/PhysRevLett.123.197201} {\bibfield  {journal} {\bibinfo
  {journal} {Physical Review Letters}\ }\textbf {\bibinfo {volume} {123}},\
  \bibinfo {pages} {197201} (\bibinfo {year} {2019})}\BibitemShut {NoStop}%
\bibitem [{\citenamefont {Aoki}\ \emph {et~al.}(2019)\citenamefont {Aoki},
  \citenamefont {Nakamura}, \citenamefont {Honda}, \citenamefont {Li},
  \citenamefont {Homma}, \citenamefont {Shimizu}, \citenamefont {Sato},
  \citenamefont {Knebel}, \citenamefont {Brison}, \citenamefont {Pourret},
  \citenamefont {Braithwaite}, \citenamefont {Lapertot}, \citenamefont {Niu},
  \citenamefont {Vali{\v{s}}ka}, \citenamefont {Harima},\ and\ \citenamefont
  {Flouquet}}]{Aoki2019UnconventionalUte2}%
  \BibitemOpen
  \bibfield  {author} {\bibinfo {author} {\bibfnamefont {D.}~\bibnamefont
  {Aoki}}, \bibinfo {author} {\bibfnamefont {A.}~\bibnamefont {Nakamura}},
  \bibinfo {author} {\bibfnamefont {F.}~\bibnamefont {Honda}}, \bibinfo
  {author} {\bibfnamefont {D.~X.}\ \bibnamefont {Li}}, \bibinfo {author}
  {\bibfnamefont {Y.}~\bibnamefont {Homma}}, \bibinfo {author} {\bibfnamefont
  {Y.}~\bibnamefont {Shimizu}}, \bibinfo {author} {\bibfnamefont {Y.~J.}\
  \bibnamefont {Sato}}, \bibinfo {author} {\bibfnamefont {G.}~\bibnamefont
  {Knebel}}, \bibinfo {author} {\bibfnamefont {J.~P.}\ \bibnamefont {Brison}},
  \bibinfo {author} {\bibfnamefont {A.}~\bibnamefont {Pourret}}, \bibinfo
  {author} {\bibfnamefont {D.}~\bibnamefont {Braithwaite}}, \bibinfo {author}
  {\bibfnamefont {G.}~\bibnamefont {Lapertot}}, \bibinfo {author}
  {\bibfnamefont {Q.}~\bibnamefont {Niu}}, \bibinfo {author} {\bibfnamefont
  {M.}~\bibnamefont {Vali{\v{s}}ka}}, \bibinfo {author} {\bibfnamefont
  {H.}~\bibnamefont {Harima}}, \ and\ \bibinfo {author} {\bibfnamefont
  {J.}~\bibnamefont {Flouquet}},\ }\href {\doibase 10.7566/JPSJ.88.043702}
  {\bibfield  {journal} {\bibinfo  {journal} {Journal of the Physical Society
  of Japan}\ }\textbf {\bibinfo {volume} {88}},\ \bibinfo {pages} {43702}
  (\bibinfo {year} {2019})}\BibitemShut {NoStop}%
\bibitem [{\citenamefont {Chen}\ \emph {et~al.}(2002)\citenamefont {Chen},
  \citenamefont {Ohara}, \citenamefont {Hedo}, \citenamefont {Uwatoko},
  \citenamefont {Saito}, \citenamefont {Sorai},\ and\ \citenamefont
  {Sakamoto}}]{Chen2002Observation2CoIn8}%
  \BibitemOpen
  \bibfield  {author} {\bibinfo {author} {\bibfnamefont {G.}~\bibnamefont
  {Chen}}, \bibinfo {author} {\bibfnamefont {S.}~\bibnamefont {Ohara}},
  \bibinfo {author} {\bibfnamefont {M.}~\bibnamefont {Hedo}}, \bibinfo {author}
  {\bibfnamefont {Y.}~\bibnamefont {Uwatoko}}, \bibinfo {author} {\bibfnamefont
  {K.}~\bibnamefont {Saito}}, \bibinfo {author} {\bibfnamefont
  {M.}~\bibnamefont {Sorai}}, \ and\ \bibinfo {author} {\bibfnamefont
  {I.}~\bibnamefont {Sakamoto}},\ }\href {\doibase 10.1143/JPSJ.71.2836}
  {\bibfield  {journal} {\bibinfo  {journal} {Journal of the Physical Society
  of Japan}\ }\textbf {\bibinfo {volume} {71}},\ \bibinfo {pages} {2836}
  (\bibinfo {year} {2002})}\BibitemShut {NoStop}%
\bibitem [{\citenamefont {Br{\"{u}}ning}\ \emph {et~al.}(2008)\citenamefont
  {Br{\"{u}}ning}, \citenamefont {Krellner}, \citenamefont {Baenitz},
  \citenamefont {Jesche}, \citenamefont {Steglich},\ and\ \citenamefont
  {Geibel}}]{Bruning2008CeFePO:Correlations}%
  \BibitemOpen
  \bibfield  {author} {\bibinfo {author} {\bibfnamefont {E.~M.}\ \bibnamefont
  {Br{\"{u}}ning}}, \bibinfo {author} {\bibfnamefont {C.}~\bibnamefont
  {Krellner}}, \bibinfo {author} {\bibfnamefont {M.}~\bibnamefont {Baenitz}},
  \bibinfo {author} {\bibfnamefont {A.}~\bibnamefont {Jesche}}, \bibinfo
  {author} {\bibfnamefont {F.}~\bibnamefont {Steglich}}, \ and\ \bibinfo
  {author} {\bibfnamefont {C.}~\bibnamefont {Geibel}},\ }\href {\doibase
  10.1103/PhysRevLett.101.117206} {\  (\bibinfo {year} {2008}),\
  10.1103/PhysRevLett.101.117206}\BibitemShut {NoStop}%
\bibitem [{\citenamefont {Majumder}\ \emph {et~al.}(2020)\citenamefont
  {Majumder}, \citenamefont {Prinz-Zwick}, \citenamefont {Reschke},
  \citenamefont {Zubtsovskii}, \citenamefont {Dey}, \citenamefont {Freund},
  \citenamefont {B\"uttgen}, \citenamefont {Jesche}, \citenamefont
  {K\'ezsm\'arki}, \citenamefont {Tsirlin},\ and\ \citenamefont
  {Gegenwart}}]{PhysRevB.101.214417}%
  \BibitemOpen
  \bibfield  {author} {\bibinfo {author} {\bibfnamefont {M.}~\bibnamefont
  {Majumder}}, \bibinfo {author} {\bibfnamefont {M.}~\bibnamefont
  {Prinz-Zwick}}, \bibinfo {author} {\bibfnamefont {S.}~\bibnamefont
  {Reschke}}, \bibinfo {author} {\bibfnamefont {A.}~\bibnamefont
  {Zubtsovskii}}, \bibinfo {author} {\bibfnamefont {T.}~\bibnamefont {Dey}},
  \bibinfo {author} {\bibfnamefont {F.}~\bibnamefont {Freund}}, \bibinfo
  {author} {\bibfnamefont {N.}~\bibnamefont {B\"uttgen}}, \bibinfo {author}
  {\bibfnamefont {A.}~\bibnamefont {Jesche}}, \bibinfo {author} {\bibfnamefont
  {I.}~\bibnamefont {K\'ezsm\'arki}}, \bibinfo {author} {\bibfnamefont {A.~A.}\
  \bibnamefont {Tsirlin}}, \ and\ \bibinfo {author} {\bibfnamefont
  {P.}~\bibnamefont {Gegenwart}},\ }\href {\doibase
  10.1103/PhysRevB.101.214417} {\bibfield  {journal} {\bibinfo  {journal}
  {Phys. Rev. B}\ }\textbf {\bibinfo {volume} {101}},\ \bibinfo {pages}
  {214417} (\bibinfo {year} {2020})}\BibitemShut {NoStop}%
\bibitem [{\citenamefont {Hammar}\ \emph {et~al.}(1999)\citenamefont {Hammar},
  \citenamefont {Stone}, \citenamefont {Reich}, \citenamefont {Broholm},
  \citenamefont {Gibson}, \citenamefont {Turnbull}, \citenamefont {Landee},\
  and\ \citenamefont {Oshikawa}}]{PhysRevB.59.1008}%
  \BibitemOpen
  \bibfield  {author} {\bibinfo {author} {\bibfnamefont {P.~R.}\ \bibnamefont
  {Hammar}}, \bibinfo {author} {\bibfnamefont {M.~B.}\ \bibnamefont {Stone}},
  \bibinfo {author} {\bibfnamefont {D.~H.}\ \bibnamefont {Reich}}, \bibinfo
  {author} {\bibfnamefont {C.}~\bibnamefont {Broholm}}, \bibinfo {author}
  {\bibfnamefont {P.~J.}\ \bibnamefont {Gibson}}, \bibinfo {author}
  {\bibfnamefont {M.~M.}\ \bibnamefont {Turnbull}}, \bibinfo {author}
  {\bibfnamefont {C.~P.}\ \bibnamefont {Landee}}, \ and\ \bibinfo {author}
  {\bibfnamefont {M.}~\bibnamefont {Oshikawa}},\ }\href {\doibase
  10.1103/PhysRevB.59.1008} {\bibfield  {journal} {\bibinfo  {journal} {Phys.
  Rev. B}\ }\textbf {\bibinfo {volume} {59}},\ \bibinfo {pages} {1008}
  (\bibinfo {year} {1999})}\BibitemShut {NoStop}%
\bibitem [{\citenamefont {Singh}\ \emph {et~al.}(2013)\citenamefont {Singh},
  \citenamefont {Tokiwa}, \citenamefont {Dong},\ and\ \citenamefont
  {Gegenwart}}]{Singh2013Spin8}%
  \BibitemOpen
  \bibfield  {author} {\bibinfo {author} {\bibfnamefont {Y.}~\bibnamefont
  {Singh}}, \bibinfo {author} {\bibfnamefont {Y.}~\bibnamefont {Tokiwa}},
  \bibinfo {author} {\bibfnamefont {J.}~\bibnamefont {Dong}}, \ and\ \bibinfo
  {author} {\bibfnamefont {P.}~\bibnamefont {Gegenwart}},\ }\href {\doibase
  10.1103/PhysRevB.88.220413} {\bibfield  {journal} {\bibinfo  {journal} {RAPID
  COMMUNICATIONS PHYSICAL REVIEW B}\ }\textbf {\bibinfo {volume} {88}},\
  \bibinfo {pages} {220413} (\bibinfo {year} {2013})}\BibitemShut {NoStop}%
\bibitem [{\citenamefont {Dender}\ \emph {et~al.}(1997)\citenamefont {Dender},
  \citenamefont {Hammar}, \citenamefont {Reich}, \citenamefont {Broholm},\ and\
  \citenamefont {Aeppli}}]{PhysRevLett.79.1750}%
  \BibitemOpen
  \bibfield  {author} {\bibinfo {author} {\bibfnamefont {D.~C.}\ \bibnamefont
  {Dender}}, \bibinfo {author} {\bibfnamefont {P.~R.}\ \bibnamefont {Hammar}},
  \bibinfo {author} {\bibfnamefont {D.~H.}\ \bibnamefont {Reich}}, \bibinfo
  {author} {\bibfnamefont {C.}~\bibnamefont {Broholm}}, \ and\ \bibinfo
  {author} {\bibfnamefont {G.}~\bibnamefont {Aeppli}},\ }\href {\doibase
  10.1103/PhysRevLett.79.1750} {\bibfield  {journal} {\bibinfo  {journal}
  {Phys. Rev. Lett.}\ }\textbf {\bibinfo {volume} {79}},\ \bibinfo {pages}
  {1750} (\bibinfo {year} {1997})}\BibitemShut {NoStop}%
\bibitem [{\citenamefont {Kenzelmann}\ \emph {et~al.}(2004)\citenamefont
  {Kenzelmann}, \citenamefont {Chen}, \citenamefont {Broholm}, \citenamefont
  {Reich},\ and\ \citenamefont {Qiu}}]{PhysRevLett.93.017204}%
  \BibitemOpen
  \bibfield  {author} {\bibinfo {author} {\bibfnamefont {M.}~\bibnamefont
  {Kenzelmann}}, \bibinfo {author} {\bibfnamefont {Y.}~\bibnamefont {Chen}},
  \bibinfo {author} {\bibfnamefont {C.}~\bibnamefont {Broholm}}, \bibinfo
  {author} {\bibfnamefont {D.~H.}\ \bibnamefont {Reich}}, \ and\ \bibinfo
  {author} {\bibfnamefont {Y.}~\bibnamefont {Qiu}},\ }\href {\doibase
  10.1103/PhysRevLett.93.017204} {\bibfield  {journal} {\bibinfo  {journal}
  {Phys. Rev. Lett.}\ }\textbf {\bibinfo {volume} {93}},\ \bibinfo {pages}
  {017204} (\bibinfo {year} {2004})}\BibitemShut {NoStop}%
\bibitem [{\citenamefont {Oshikawa}(2003)}]{doi:10.1143/JPSJS.72SB.36}%
  \BibitemOpen
  \bibfield  {author} {\bibinfo {author} {\bibfnamefont {M.}~\bibnamefont
  {Oshikawa}},\ }\href@noop {} {\bibfield  {journal} {\bibinfo  {journal}
  {Journal of the Physical Society of Japan}\ }\textbf {\bibinfo {volume}
  {72}},\ \bibinfo {pages} {36} (\bibinfo {year} {2003})}\BibitemShut {NoStop}%
\bibitem [{\citenamefont {Arnold}\ \emph {et~al.}(2014)\citenamefont {Arnold},
  \citenamefont {Bilheux}, \citenamefont {Borreguero}, \citenamefont {Buts},
  \citenamefont {Campbell}, \citenamefont {Chapon}, \citenamefont {Doucet},
  \citenamefont {Draper}, \citenamefont {Ferraz~Leal}, \citenamefont {Gigg},
  \citenamefont {Lynch}, \citenamefont {Markvardsen}, \citenamefont
  {Mikkelson}, \citenamefont {Mikkelson}, \citenamefont {Miller}, \citenamefont
  {Palmen}, \citenamefont {Parker}, \citenamefont {Passos}, \citenamefont
  {Perring}, \citenamefont {Peterson}, \citenamefont {Ren}, \citenamefont
  {Reuter}, \citenamefont {Savici}, \citenamefont {Taylor}, \citenamefont
  {Taylor}, \citenamefont {Tolchenov}, \citenamefont {Zhou},\ and\
  \citenamefont {Zikovsky}}]{Arnold2014MantidExperiments}%
  \BibitemOpen
  \bibfield  {author} {\bibinfo {author} {\bibfnamefont {O.}~\bibnamefont
  {Arnold}}, \bibinfo {author} {\bibfnamefont {J.~C.}\ \bibnamefont {Bilheux}},
  \bibinfo {author} {\bibfnamefont {J.~M.}\ \bibnamefont {Borreguero}},
  \bibinfo {author} {\bibfnamefont {A.}~\bibnamefont {Buts}}, \bibinfo {author}
  {\bibfnamefont {S.~I.}\ \bibnamefont {Campbell}}, \bibinfo {author}
  {\bibfnamefont {L.}~\bibnamefont {Chapon}}, \bibinfo {author} {\bibfnamefont
  {M.}~\bibnamefont {Doucet}}, \bibinfo {author} {\bibfnamefont
  {N.}~\bibnamefont {Draper}}, \bibinfo {author} {\bibfnamefont
  {R.}~\bibnamefont {Ferraz~Leal}}, \bibinfo {author} {\bibfnamefont {M.~A.}\
  \bibnamefont {Gigg}}, \bibinfo {author} {\bibfnamefont {V.~E.}\ \bibnamefont
  {Lynch}}, \bibinfo {author} {\bibfnamefont {A.}~\bibnamefont {Markvardsen}},
  \bibinfo {author} {\bibfnamefont {D.~J.}\ \bibnamefont {Mikkelson}}, \bibinfo
  {author} {\bibfnamefont {R.~L.}\ \bibnamefont {Mikkelson}}, \bibinfo {author}
  {\bibfnamefont {R.}~\bibnamefont {Miller}}, \bibinfo {author} {\bibfnamefont
  {K.}~\bibnamefont {Palmen}}, \bibinfo {author} {\bibfnamefont
  {P.}~\bibnamefont {Parker}}, \bibinfo {author} {\bibfnamefont
  {G.}~\bibnamefont {Passos}}, \bibinfo {author} {\bibfnamefont {T.~G.}\
  \bibnamefont {Perring}}, \bibinfo {author} {\bibfnamefont {P.~F.}\
  \bibnamefont {Peterson}}, \bibinfo {author} {\bibfnamefont {S.}~\bibnamefont
  {Ren}}, \bibinfo {author} {\bibfnamefont {M.~A.}\ \bibnamefont {Reuter}},
  \bibinfo {author} {\bibfnamefont {A.~T.}\ \bibnamefont {Savici}}, \bibinfo
  {author} {\bibfnamefont {J.~W.}\ \bibnamefont {Taylor}}, \bibinfo {author}
  {\bibfnamefont {R.~J.}\ \bibnamefont {Taylor}}, \bibinfo {author}
  {\bibfnamefont {R.}~\bibnamefont {Tolchenov}}, \bibinfo {author}
  {\bibfnamefont {W.}~\bibnamefont {Zhou}}, \ and\ \bibinfo {author}
  {\bibfnamefont {J.}~\bibnamefont {Zikovsky}},\ }\href {\doibase
  10.1016/j.nima.2014.07.029} {\bibfield  {journal} {\bibinfo  {journal}
  {Nuclear Instruments and Methods in Physics Research, Section A:
  Accelerators, Spectrometers, Detectors and Associated Equipment}\ }\textbf
  {\bibinfo {volume} {764}},\ \bibinfo {pages} {156} (\bibinfo {year}
  {2014})}\BibitemShut {NoStop}%
\bibitem [{\citenamefont {Seabold}\ and\ \citenamefont
  {Perktold}(2010)}]{Seabold2010Statsmodels:EconometricPython}%
  \BibitemOpen
  \bibfield  {author} {\bibinfo {author} {\bibfnamefont {S.}~\bibnamefont
  {Seabold}}\ and\ \bibinfo {author} {\bibfnamefont {J.}~\bibnamefont
  {Perktold}},\ }\href@noop {} {\enquote {\bibinfo {title}
  {{statsmodels:Econometric and statistical modeling with python}},}\ }
  (\bibinfo {year} {2010})\BibitemShut {NoStop}%
\end{thebibliography}%

\end{document}